\documentclass[prx,twocolumn,amsmath,amssymb,floatfix,footinbib,bibnotes,longbibliography]{revtex4-2}
\pdfoutput=1

\usepackage[colorlinks=true,citecolor=blue,linkcolor=magenta]{hyperref}
\usepackage{soul}
\usepackage{cleveref}
\usepackage{multirow}
\usepackage{amsmath}
\usepackage{bbm}
\usepackage{graphicx}
\usepackage{dsfont}
\usepackage{xcolor}
\usepackage{changepage}
\usepackage{fancyhdr}
\usepackage{amsthm, amssymb}
\usepackage{array}
\newcolumntype{P}[1]{>{\centering\arraybackslash}p{#1}}
\usepackage[T1]{fontenc}
\usepackage[latin9]{inputenc}
\usepackage[active]{srcltx}
\usepackage{xcolor}
\usepackage{amssymb}
\usepackage{esint}
\usepackage[version=4]{mhchem}
\usepackage{comment}
\usepackage{natbib}
\usepackage{float}

\usepackage{hyperref}

\graphicspath{ {Plots} }

\def \mr{\mathrm}

\newcommand{\mc}[1]{\mathcal{#1}}
\newcommand {\apgt} {\ {\raise-.5ex\hbox{$\buildrel>\over\sim$}}\ }
\newcommand {\aplt} {\ {\raise-.5ex\hbox{$\buildrel<\over\sim$}}\ }

\newcommand {\rem}[1]{}

\def  \w{\omega}
\def  \qinv{Q^{-1}}
\def  \G{\Gamma}
\def  \q0{\frac{\w_k^2}{\G}+z}

\newcommand{\ci}{\mathrm{i}}

\newcommand{\bs}[1]{\boldsymbol{#1}}

\def \titlename {Entanglement entropy of fermions in a strange metal}
\def \authorf{Santanu Singh$^1$}
\def \affiliationf{$^1$Centre for Condensed Matter Theory, Department of Physics, Indian Institute of Science, Bangalore 560012, India}
\def \authors{Surajit Bera$^2$}
\def \affiliations{$^2$JEIP, UAR 3573 CNRS, College de France, PSL Research University, Paris 75321, France}
\def \authort{Chenyuan Li$^{3}$}
\def \affiliationt{$^3$Rice Academy of Fellows, Rice University, Houston, Texas 77005, USA}
\def \authorfo{Subir Sachdev $^{4}$}
\def \affiliationfo{$^4$Department of Physics, Harvard University, Cambridge, Massachusetts 02138, USA}
\def \authorfi{Sumilan Banerjee$^1$}

\begin{document}
	\title{\titlename}
	\author{\authorf, \authors, \authort, \authorfo, \authorfi}
	\affiliation{\affiliationf,\affiliations,\affiliationt,\affiliationfo}
	\email{santanusingh@iisc.ac.in}
	\email{sumilan@iisc.ac.in}
	\date\today

\begin{abstract}
The subsystem-size dependence of ground-state entanglement entropy and its crossover to thermal entropy as a function of temperature are well understood for one-dimensional (1D) gapless systems described by conformal field theory (CFT), and for free fermions with a Fermi surface in any dimension. However, little is known about the entanglement entropy for gapless fermionic systems without quasi-particles, such as a strange metal. Here we study the entanglement entropy of fermions in a solvable large-$N$ 1D lattice model akin to the Yukawa-Sachdev-Ye-Kitaev (Yukawa SYK) model. In this model, two Fermi points are coupled to scalar bosons via spatially random Yukawa interactions, providing a solvable model of a strange metal when the bosons become critical at a quantum critical point. We exactly compute the second R\'{e}nyi entropy of fermions in a spatial subregion in this model. Our results unravel crucial role of intra-subregion entanglement between fermionic and bosonic degrees of freedom along with the inter-subregion entanglement in understanding the ground states of such strongly coupled fermion-boson systems. We show that the crossover from thermal entropy to entanglement entropy, is captured by a single scaling ansatz, that collapses the second R\'{e}nyi entropy of fermions for different subregion sizes and temperatures into a single universal curve. We further show that the universal scaling curve for the critical strange metal is well described by the standard CFT formula, albeit with an \emph{effective central charge} substantially larger than the non-interacting value.

\end{abstract}

\maketitle 


\section{Introduction}
Entanglement provides a unique window into fundamental correlations of quantum states and their characterization, e.g., gapped and gapless ground states of many-body systems with  symmetry broken and/or topological order \cite{LAFLORENCIE20161,KitaevPRL,Jiang2012}. Among the important measures of the entanglement of pure many-body quantum states are the von Neumann $S_A$ and $n$-th R\'{e}nyi entanglement entropies $S_A^{(n)}$ of a subsystem $A$. In gapless systems, the low-energy excitations that dominate low-temperature thermodynamics are also expected to control the ground-state entanglement, e.g., the scaling of $S_A$ with the subsystem size $L_A$, linking the entanglement and thermal entropy through a crossover with temperature \cite{Cardy,SwingleSenthil,Korepin}. The subsystem-size scaling of $S_A$ and the crossover are quite well understood for one-dimensional (1D) gapless bosonic or fermionic systems described by conformal field theory (CFT) \cite{Cardy,Korepin}, like a Tomonaga-Luttinger liquid (TLL) \cite{Giamarchi}, and non-interacting fermionic systems with Fermi surface in any dimension $d$ \cite{Klich,Swingle,Weifei2006,Swingle2012,Swingle2012a}. Also, there is some understanding of the entanglement scaling and crossover \cite{Swingle,SwingleSenthil,Swingle2012,Swingle2012a} for interacting Fermi liquids (FLs) in $d>1$, even though explicit calculations are rather scarce \cite{Tubmann,MuliDbosonization,bera_PRB}. In this context, even much less is known about $d\geq 1$ strongly interacting gapless fermionic systems that are not described by CFT or FL paradigms, such as non-Fermi liquids (NFLs) or strange metals \cite{Sachdev_QPT, Sachdev_review} without well-defined quasiparticles.

In recent years, zero-dimensional large-$N$ Sachdev-Ye-Kitaev (SYK) \cite{SachdevYe,Kitaev,SachdevPRX,Maldacena,Kitaev,Chowdhury} and related models \cite{BanerjeeAltman,EsterlisSchmalian,Aldape}, and their lattice and higher-dimensional generalizations \cite{Balents,Jian,Halder_VBS,Haldar_VBS_SB,Debanjan_Senthil} have provided a new avenue to study NFLs, interacting FLs and a plethora of other strongly correlated phenomena \cite{Haldar_VBS_SB,Debanjan_Senthil,Aldape,DiracScrambler} in a controlled manner. In the large-$N$ 2D Yukawa-SYK model, a large number ($N$) of fermion flavors with Fermi surface interact with $N$ flavors of quantum critical scalar bosons via spatially random Yukawa couplings \cite{patel_science,esterlisPRB,Li_2D_YSYK}. The NFL state in this model has been shown \cite{patel_science,esterlisPRB,Li_2D_YSYK,Guo2022,Guo2024,Patel2024,Patel2025,Lunts2025} to capture the phenomenology of strange metallic state in high-temperature cuprate superconductors \cite{Lee2006}. The 2D Yukawa-SYK model leads to a sharp Fermi surface without any quaisparticles, namely a critical Fermi surface \cite{patel_science,esterlisPRB, Li_2D_YSYK,Sachdev_review}. Here we study the subsystem-size scaling of entanglement entropy of fermions and entanglement-to-thermal-entropy crossover in a large-$N$ 1D Yukawa-SYK model. 

\begin{figure}[h!]
    \centering
    \includegraphics[width=\linewidth]{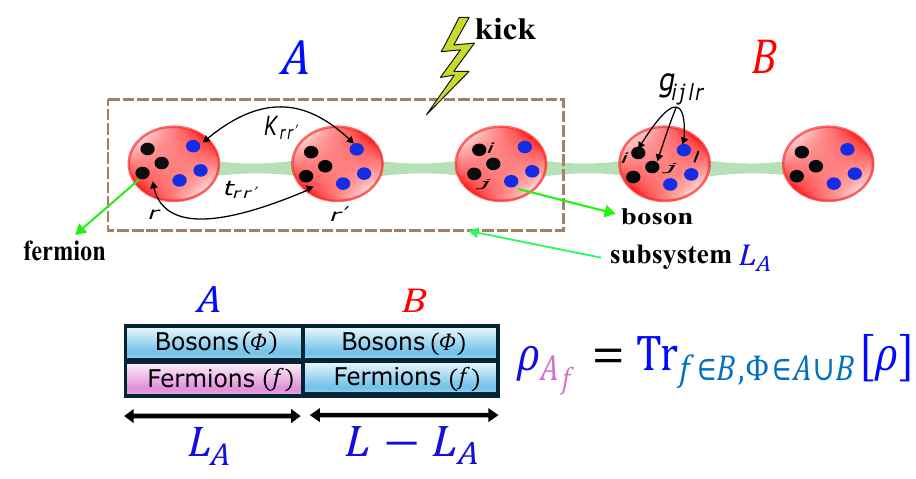}
    \caption{{\bf 1D Yukawa-SYK model and fermionic subsystem:} A one-dimensional model (sites
$r,r'=1,\cdots,L$, shown for $L=5$) of Yukawa-SYK \emph{quantum dots} having spinless fermions (black) and bosons (blue), each with $N$ (shown for $N=3)$ flavors (indexed by
$i,j,l$) at each site $r$. The fermion and boson flavors  within each dot interact via spatially random Yukawa interaction $g_{ijlr}$. Same flavor of fermions hop with amplitudes (black arrows) $t_{rr'}$ and bosonic fields are quadratically coupled via $K_{rr'}$, between the sites $r$ and $r'$. The one dimensional chain is divided into a spatial subsystem $A$ (dashed rectangle) of length $L_A$ and its complement $B$ of length $L-L_A$, shown here for $L_A=3$ and $L=5$. The entanglement kick acts only on the fermionic degrees of freedom within subsystem $A$ in the imaginary-time path integral. The lower panel illustrates the corresponding partition, where the fermionic subsystem consists of the fermions $(f)$ in region $A$ (magenta), while the fermions $(f)$ in region $B$ and all bosonic degrees $(\phi)$ of freedom are traced out to obtain the fermionic reduced density matrix $\rho_{{A}_f}$ in Eq.(\ref{eq:fermion_density_matrix}).}
\label{fig:model}
\end{figure}

\begin{figure}[h!]
    \centering
    \includegraphics[width=\linewidth]{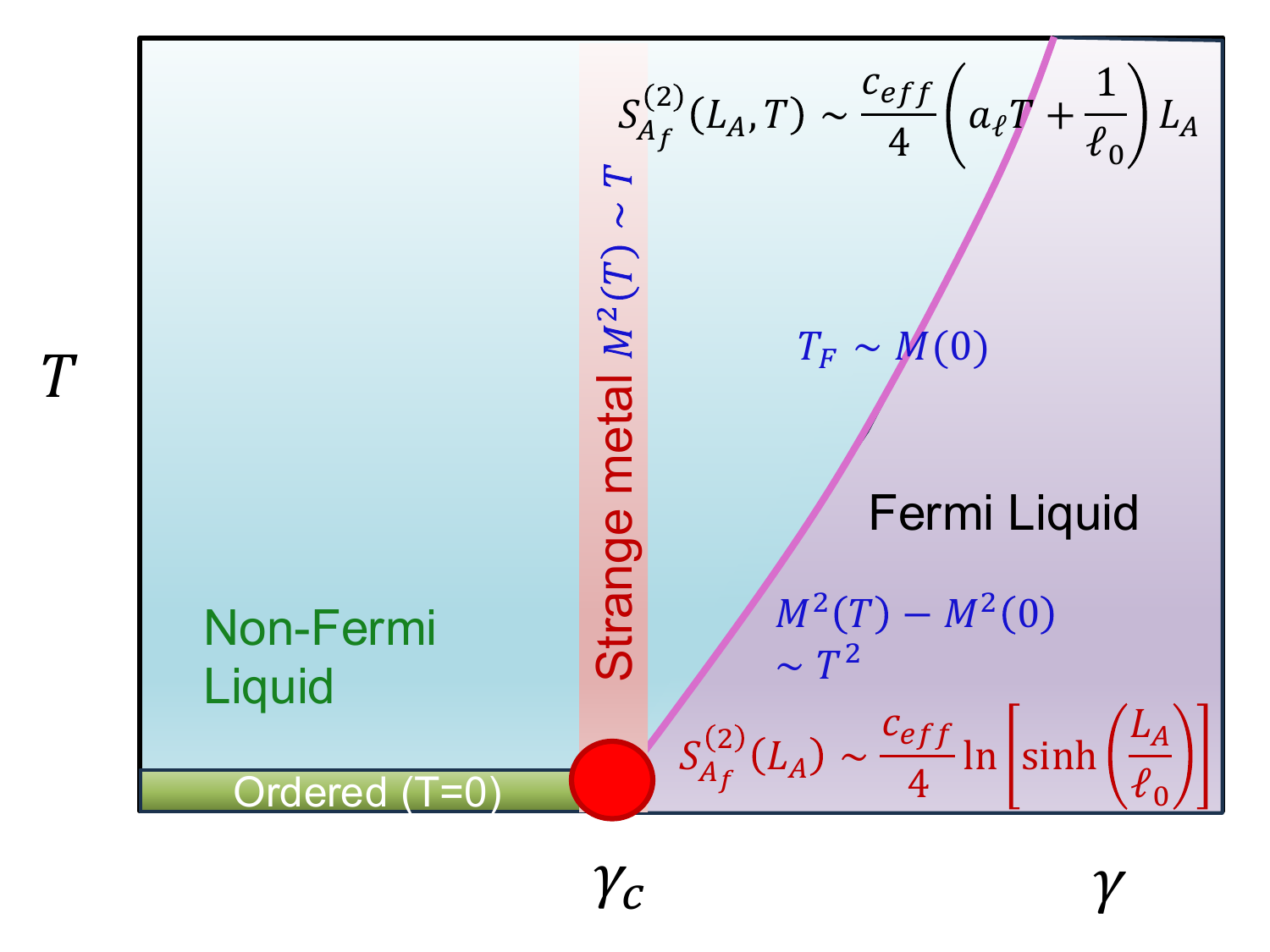}
    \caption{{\bf Phase diagram of 1D Yukawa-SYK model:} The schematic phase diagram of the 1D Yukawa-SYK model (Fig.\ref{fig:model}) as a function of temperature $T$ and a parameter $\gamma$ that controls the spherical constraint on the bosons. A quantum critical point (QCP) $\gamma=\gamma_c$ at $T=0$ separates a gapped disordered phase for $\gamma>\gamma_c$, with boson mass $M(0)$ and Fermi liquid (FL) ground state, and a gapless non-Fermi liquid ordered phase for $\gamma<\gamma_c$. The boson mass {$M^2(T)$}  increases as $\sim T^2$ in the FL phase, where the FL scale $T_\mr{F}\sim M(0)\to 0$ for $\gamma\to \gamma_c^+$. At the QCP, the coupling between fermions and critical bosons leads to a strange metal with critical Fermi points devoid of quasiparticles, and boson thermal mass $M(T)\sim \sqrt{T}$. For the strange metal and FL ground states ($\gamma\geq \gamma_c$), the second R\'{e}nyi entanglement entropy $S_{A_f}^{(2)}(L_A)$ of fermions in a spatial subregion of size $L_A$ (Fig.\ref{fig:model}) is well described by a conformal field theory (CFT)-like form with \emph{effective central charge} $c_{eff}> $ the non-interacting value ($c=1$), and an emergent fermion-boson entanglement length scale $\ell_0$. At higher temperatures, the entanglement entropy crosses over to a combination ($\propto L_A$) of thermal entropy ($\propto T$) and fermion-boson entanglement ($\propto 1/\ell_0$), that are also controlled by $c_{eff}$.  
    }
\label{fig:PhaseDiagram}
\end{figure}

In the 1D Yukawa-SYK model, spinless fermions with two Fermi points are coupled to scalar bosons via random Yukawa interactions (see Fig.\ref{fig:model}), providing a solvable model of a strange metal when the bosons become critical at a quantum critical point (QCP) (see Fig.\ref{fig:PhaseDiagram}). Through analytical low-energy solution of the large-$N$ saddle-point equations for the 1D Yukawa-SYK model, we show that the QCP occurs between zero-temperature $O(N)$ symmetry-broken ordered phase and a disordered phase when a tuning parameter $\gamma$ is varied across a critical value $\gamma=\gamma_c$, as shown in Fig.\ref{fig:PhaseDiagram}. The bosons become gapless with a boson thermal mass $M(T)$ vanishing as $\sqrt{T}$ as the temperature $T\to 0$ at the QCP. The zero-temperature ordered phase appears in the 1D model with continuous $O(N)$ symmetry due to long-range coupling in the time direction induced by Ohmic dissipation. The latter is generated self-consistently in the model from the coupling of the bosons with gapless fermions. The long-range coupling also leads to a 1D critical strange metallic state at QCP that is not described by a local CFT/TLL. The 1D strange metal has stronger NFL fermionic self energy $\sim \sqrt{\omega}$ at low energy $\omega\to 0$, compared to marginal Fermi liquid self-energy $\sim \omega \ln \omega$ in 2D Yukawa-SYK model \cite{patel_science,esterlisPRB,Li_2D_YSYK}. This facilitates potentially stronger signatures of NFL in the entanglement characteristics. We show that the model leads to NFL self energy $\sim \sqrt{\omega}$ even away from the QCP, in the $T=0$ ordered phase for $\gamma<\gamma_c$ where the thermal mass $M(T\to 0)\to 0$. In contrast, the disordered phase for $\gamma>\gamma_c$, where the bosons have a finite mass $M(0)$ at $T=0$, is shown to be a FL with imaginary part of self-energy $\sim \omega^2$. The zero-temperature mass $M(0)$ and the quaisparticle residue $Z$ for the fermions in the FL phase both vanish approaching the QCP. 
From the numerical solution of the large-$N$ saddle-point equations we find that the thermal entropy $S(T)$ varies linearly with $T$ at low temperatures for the strange metal at QCP and the FL phase. Even though the bosons are gapped in the FL phase, they still contribute to the linear-$T$ entropy coefficient, especially near the QCP, due to finite spectral weight induced by Ohmic dissipation at low energies \cite{hanggi2006,Hanggi_2008}. 

Using a recently developed imaginary-time path integral formalism \cite{Haldar}, we exactly compute the second R\'{e}nyi entropy $S_{A_f}^{(2)}$ of a \emph{fermionic subsystem} within a spatial subregion $A$ of length $L_A$ (Fig. \ref{fig:model}), for the large-$N$ 1D Yukawa SYK chain with total length $L$ in a thermal ensemble at $T$. To this end, we derive the large-$N$ saddle-point equations from the entanglement path integral and numerically solve them to obtain $S_{A_f}^{(2)}$ at the saddle point. The path integral method has been previously applied to compute entanglement properties of zero-dimensional SYK and related models \cite{Haldar}, as well as Hubbard model within dynamical mean-field theory (DMFT) \cite{bera_PRB}. The entanglement path integral formalism \cite{Chakraborty,Chakraborty_PRL, Haldar,Moitra,bera_PRB} replaces somewhat non-trivial space-time boundary conditions in the usual replica path integral for entanglement entropy \cite{Cardy} by auxiliary fermionic source fields. The latter are introduced through mathematical identities \cite{Haldar,Moitra,bera_PRB} involving reduced density matrix of a subsystem and fermionic displacement operators \cite{Cahill_Glauber}. 

Here, we employ the formalism to obtain the crossover from thermal entropy to entanglement in the 1D Yukawa-SYK model, and thus extract the scaling of entanglement entropy with $L_A$ by extrapolating to $T\to 0$ limit. As shown in schematic {Fig.\ref{fig:model}}, the entanglement entropy of fermionic subsystem $A_f$ in the region $A$ is expected to have contributions from -- (1) \emph{non-local} inter-subsystem entanglement of fermions in $A$ with the fermions and bosons in the rest of the system $B$, and (2) \emph{local} intra-subsystem entanglement of the fermions with bosons within the same spatial subregion $A$. The latter is expected to naturally lead to \emph{volume-law} entanglement $\sim L_A$ for fermions in the fermion-boson Yukawa-SYK model. We give an overview of our main results below.

\subsubsection*{Overview of the results}
In this work, we numerically obtain the following second R\'{e}nyi entropy of fermions (per flavor) for a spatial contiguous segment $A$ of length $L_A$  (Fig.\ref{fig:model}) in the large-$N$ 1D Yukawa-SYK model,
\begin{equation*}
     S_{A_{f}}^{(2)}(L_A,T)=-\frac{1}{N}\ln{\mathrm{Tr}_{A_{f}}\rho_{A_{f}}^{2}},
\end{equation*}
 where, $\rho_{A_f}=\mathrm{Tr}_{A_\phi,\, B_{f\phi}} \rho$ is the reduced density matrix of the fermion in $A$ and $\rho$ is the density matrix of the system at temperature $T$. Here $A_\phi$ refers to the bosonic degrees of freedom in $A$ and $B_{f\phi}$ both the fermionic and bosonic degrees of freedom in $B$.

\begin{figure*}
    \centering
    \includegraphics[width=0.9\linewidth]{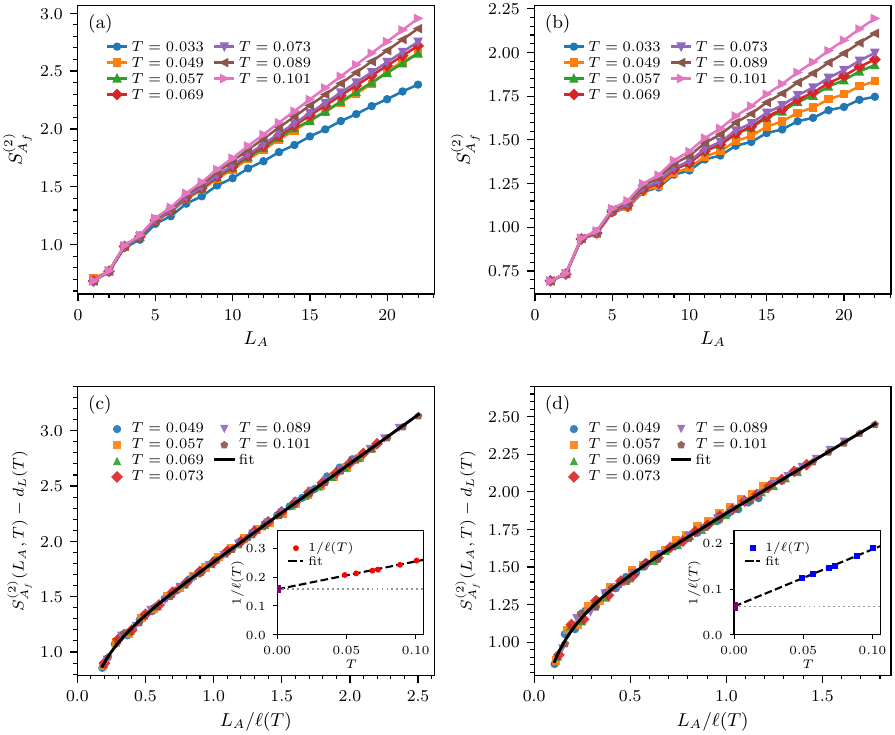}
    \caption{{\bf Fermionic-subsystem second R\'{e}nyi entropy and scaling collapse for 1D Yukawa-SYK model:} $S_{A_f}^{(2)}$ as a function of subsystem size $L_A$ for system size $L=50$ (see Fig.\ref{fig:model}) at different temperatures $T$, for (a) the strange metal at $\gamma/\gamma_c=1$ and (b) Fermi liquid at $\gamma/\gamma_c=1.84$. Corresponding scaling collapses in (c) and (d), where the scaling curves are fitted (solid black lines) with the CFT-like crossover formula in Eq.\eqref{eq:CFTCrossover}, yielding effective central charges, $c_{eff}=1.56\pm 0.02$ for $\gamma/\gamma_c=1$ and $c_{{eff}}=1.30 \pm 0.02$ for $\gamma/\gamma_c=1.84$. (Insets) The extracted inverse length scale $1/\ell(T)$, plotted as a function of $T$, exhibits a finite intercept at $T=0$, $1/\ell_0\simeq 0.16$ for $\gamma/\gamma_c=1$ and $1/\ell_0\simeq 0.06$ for $\gamma/\gamma_c=1.84$ (purple square). The details of the method to compute $S_{A_f}^{(2)}(L_A,T)$ in the large-$N$ 1D Yukawa-SYK model are discussed in Sec.\ref{sec:Ent_LargeN} and the analysis of the results in Sec.\ref{sec:EntYSYK}.} 
    \label{fig:SA2_Collpase_1DYSYK}
\end{figure*}

Fig.\ref{fig:SA2_Collpase_1DYSYK} presents the main results of this work. By studying $S_{A_f}^{(2)}(L_A,T)$ for the Yukawa-SYK chain of length $L$ [see Figs.\ref{fig:SA2_Collpase_1DYSYK}(a) and (b)], we obtain the following results for entanglement in the 1D Yukawa-SYK model.\\
(1) We numerically find that the crossover from thermal to entanglement entropy in $S_{A_f}^{(2)}$ in the strange metal, and the FL phase for $\gamma\geq \gamma_c$ is captured by a scaling ansatz 
\begin{align*}
S_{A_f}^{(2)}(L_A,T)&=f_L\left(\frac{L_A}{\ell(T)}\right)+\cdots,
\end{align*}
with a scaling function $f_L(x)$ and temperature-dependent length scale $\ell(T)$, for a sufficiently large system size $L$ and $L_A\lesssim L/2$. As a result, $S_{A_f}^{(2)}$ for different subsystem sizes $L_A$ and temperatures $T$ can be collapsed into a single universal curve, as shown in Figs.\ref{fig:SA2_Collpase_1DYSYK}(c) and (d). Analogous scaling ansatz for entropy-to-entanglement crossover has been discussed for generic NFLs with critical Fermi surface in Ref.\cite{SwingleSenthil}.\\
(2) The length $\ell(T)$, which controls the entropy-entanglement crossover, is found to follow 
\begin{align*}
\frac{1}{\ell(T)}=\frac{1}{\ell_1(T)}+\frac{1}{\ell_0}   
\end{align*}
with $1/\ell_0\neq 0$, and $\ell_1(T)\sim 1/T$ as $T\to 0$ for $\gamma\geq \gamma_c$ [see Figs.\ref{fig:SA2_Collpase_1DYSYK}(c) and (d)(insets)]. This, along with the extracted asymptotic behavior of the scaling function $f_L(x)\sim x$ for large $x$, i.e., $L_A\gg \ell(T)$, implies a (thermal) entropic contribution, $S_{A_f}^{(2)}/L_A\sim T$, at high temperatures, and a volume-law entanglement contribution, $S_{A_f}^{(2)}\sim L_A/\ell_0$, for $T\to 0$. We identify the latter with the intra-subsystem fermion-boson entanglement and $\ell_0$ with a length scale associated with such  entanglement. \\
(3) We further show that the universal scaling curve for $\gamma\geq \gamma_c$ is  well described by a CFT-like formula, 
\begin{align}
f\left(\frac{L_A}{\ell(T)}\right)= \frac{c_{eff}}{4}\ln{\left[\sinh\left(\frac{L_A}{\ell(T)}\right)\right]}+\cdots \label{eq:CFTCrossover}
\end{align} 
with a prefactor $c_{eff}$, denoted as an \emph{effective central charge} for reference. This implies a logarithmic scaling of the second R\'{e}nyi entanglement entropy 
\begin{align*}
S_{A_f}^{(2)}\simeq \frac{c_{eff}}{4}\ln\left(\frac{L_A}{\ell_0}\right)+\cdots 
\end{align*}
 for $L_A\ll \ell_0$ at $T=0$. Thus, the logarithmic scaling is cut off at the finite length scale $\ell_0$, and a volume-law entanglement scaling ensues for $L_A\gg \ell_0$.
\\
(4) We find that $c_{eff}$ and $1/\ell_0$ both decrease with $\gamma$ from the strange metal at QCP to the FL phase. At the QCP, $c_{eff}\simeq 1.6$, i.e., 60\% larger than the non-interacting or TLL value of the central charge $c=1$. Even for $\gamma>\gamma_c$, particularly for FL states close to QCP, we find that $c_{eff}>1$. \\
Due to numerical difficulty in dealing with a small thermal mass in the ordered phase ($\gamma<\gamma_c$) at low temperatures, we do not study the entanglement properties of the ordered phase in detail in this work. We also briefly discuss the effects of elastic scattering due to on-site one-body disorder, in addition to the random Yukawa coupling, on the entanglement properties. 

In summary, our main result is to demonstrate that the fermions predominantly get entangled locally with the bosons within the same subsystem in the strange metallic state of the 1D Yukawa-SYK model, which hosts strongly coupled fermions and bosons. Only for smaller subsystem sizes ($L_A\lesssim \ell_0$), a non-local entanglement dominates, varying logarithmically ($\sim \ln L_A$) with subsystem size, albeit with a prefactor $c_{eff}$ much larger than that of a 1+1D CFT or Luttinger liquid. 


The rest of the paper is organized as follows. In Sec.\ref{sec:Model}, we define the 1D Yukawa SYK model. Sec.\ref{sec:SaddlePoint} discusses the imaginary-time large-$N$ theory and corresponding saddle-point equations. The phase diagram of the model is characterized in terms of the boson thermal mass, order parameter, bosonic and fermionic self energies, and low-temperature thermodynamics in Sec.\ref{sec:PhaseDiagram}. In Sec.\ref{sec:Ent_LargeN}, we discuss our large-$N$ formalism to compute the second R\'{e}nyi entropy of fermions in a spatial subregion of the system. The results for the fermionic subsystem R\'{e}nyi entropy and entanglement properties of the strange metal and Fermi liquid states are presented in Sec.\ref{sec:Results}. We conclude in Sec.\ref{sec:Conclusion} with discussion of our results, their extensions and some future directions. The appendices provide details of the large-$N$ theory, low-energy analytical solutions of the large-$N$ equations and associated low-temperature expansions, finite-size effects, and the results for the effects of elastic scattering on the entanglement properties. The details of the numerical procedures for the solutions of the large-$N$ saddle-point equations in equilibrium and for the second R\'{e}nyi entropy are given in the Supplemental Material.

\section{One-dimensional (1D) Yukawa-SYK Model}\label{sec:Model}

As shown in Fig.\ref{fig:model}, the model consists of a 1D lattice of Yukawa-SYK \emph{quantum dots} \cite{EsterlisSchmalian} on $r=1,\cdots, L$ sites. Each dot contains $i=1,\cdots,N$ flavor of spinless fermions and real bosons, represented by canonical fermion operators $(c_{ir}^\dagger,c_{ir})$, and conjugate bosonic operators $(\phi_{ir},\pi_{ir})$, i.e, $[\phi_{ir},\pi_{jr'}]=\mathrm{i}\delta_{ij}\delta_{rr'}$
($\hbar=1$). The lattice spacing is set to one as the unit of length. The fermions and bosons are coupled via a spatially random Yukawa interaction. The Hamiltonian for the model is given by,
\begin{subequations}\label{eq:Model}
\begin{equation}
\mathcal{H} = \mathcal{H}_{f} + \mathcal{H}_{\phi} + \mathcal{H}_{f\phi},
\label{eq:H}
\end{equation}
where,
\begin{align}
\mathcal{H}_{f} &= \sum_{irr'} t_{rr'} c_{ir}^{\dagger} c_{ir'}
- \mu \sum_{ir} c_{ir}^{\dagger} c_{ir}, \label{eq:H_f}
 \\
\mathcal{H}_{\phi} &= \sum_{ir} \frac{\pi_{ir}^{2}}{2}
+ \frac{1}{2} \sum_{irr'} K_{rr'} \phi_{ir} \phi_{ir'}
+ \frac{1}{2} m_{b}^{2} \sum_{ir} \phi_{ir}^{2} \label{eq:H_phi}, \\
\mathcal{H}_{f\phi} &= \frac{1}{N} \sum_{ijlr}
g_{ijlr} c_{ir}^{\dagger} c_{jr} \phi_{lr}.
\label{eq:H_fphi}
\end{align}
\end{subequations}
The non-interacting dispersions of the fermions and bosons are defined through a nearest-neighbor hopping $t$ for fermions along $\nu=\pm \hat{\bs{x}}$ directions, i.e., $t_{rr'} = -t \sum_{\nu} \delta_{r,\, r'+\nu}$, a chemical potential $\mu$, a nearest-neighbor coupling,
 $K_{rr'} = K \big( 2 \delta_{r,r'} 
- \sum_{\nu} \delta_{r,\, r'+\nu} \big)$, for bosons, and a \emph{bare} boson mass $m_b$. Here $t$ is the fermion hopping, the coupling $K$ determines the dispersion of the boson, and $m_b$ is the bare boson mass. We set $t=1$ and choose it as the unit of energy. The Yukawa fermion-boson couplings $g_{ijlr}$ are complex Gaussian random variables that depend on both the spatial coordinate and flavor indices, with zero mean $\overline{g_{ijlr}}=0$  and variance $\overline{g_{ijlr}(g_{i'j'l'r'})^{*}}=g^{2}\delta_{ii'}\delta_{jj'}\delta_{ll'}\delta_{rr'}$ with  $g_{ijlr}=g_{jilr}^{*}$. Here $\overline{\cdots}$ denotes disorder averaging over realizations of the Yukawa coupling. 

We impose a spherical constraint \cite{patel_science,esterlisPRB, Li_2D_YSYK},
\begin{align}
\frac{1}{N}\sum_r\phi_{ir}^2=\frac{1}{\gamma}, \label{eq:SphericalConstraint}
\end{align}
on the bosonic field at each site $r$. The model is exactly solvable \cite{patel_science,esterlisPRB,EsterlisSchmalian,Li_2D_YSYK} in the large $N$ limit, i.e., for $N\to \infty$. Different phases of the model are accessed by tuning $\gamma$, while keeping the strength $g$ of the Yukawa coupling fixed. To this end, the bare boson mass $m_b$ is adjusted as a function of $T$ and $\gamma$ to satisfy the above constraint. As discussed later, this leads to renormalized boson thermal mass $M(\gamma,T)$ (Fig.\ref{fig:PhaseDiagram}) that vanishes approaching a QCP $\gamma_c$ from a disordered phase, i.e., $\gamma\to\gamma_c^+$ for $T\to 0$. At the QCP, original non-interacting Fermi points arising from $\mc{H}_f$ [Eq.\eqref{eq:H_f}] couples to gapless critical bosons and lead to a 1D strange metal with sharp critical Fermi points, albeit without any well-defined quasiparticles.

\section{Large-$N$ Saddle-Point Equations}\label{sec:SaddlePoint}
The model of Eq.\eqref{eq:Model} is studied by formulating the usual large-$N$ field theory \cite{patel_science,esterlisPRB,EsterlisSchmalian,Li_2D_YSYK} by writing the partition function $Z=\mr{Tr}(e^{-\beta \mc{H}})$ ($\beta=1/T$, $k_\mr{B}=1$) as an imaginary-time coherent-state path integral in terms of fermionic and bosonic fields $\bar{c}_{ir}(\tau),c_{ir}(\tau)$ and $\phi_{ir}(\tau)$, as discussed in detail in the Appendix \ref{app:LargeN}. To incorporate the possibility of the $O(N)$ symmetry breaking in the bosonic sector for $\gamma < \gamma_c$, we separate the ordered component from the transverse fluctuations by choosing the direction of the condensate along one component of the $N$-component field \cite{Sachdev_book}, i.e., we write
\begin{equation}
\boldsymbol{\phi}_{r}(\tau)=\Big(\sqrt{N}\,r_{0},\,\phi_{2r}(\tau),\,\phi_{3r}(\tau),\,\cdots,\phi_{Nr}(\tau)\Big),
\label{eq:O_N_symm}
\end{equation}
where $r_0$ denotes the condensate amplitude, while $\phi_{ir}(\tau)$ ($i=2,\cdots,N$) represents fluctuations transverse to the chosen ordered direction $i=1$. We perform the disorder average over the random Yukawa couplings $\{g_{ijlr}\}$ by introducing $m$ replicas in the imaginary-time path integral. The path integral for $\overline{Z^m}=\int \mc{D}(G,\Sigma,D,\Pi)e^{-m \mc{S}[G,\Sigma,D,\Pi]}$ can be written in terms of large-$N$ collective fields
\begin{subequations}
\begin{align}
 G_r(\tau,\tau')&=\frac{1}{N}\sum_i \bar{c}_{ir}(\tau')c_{ir}(\tau),\\ 
 D_r(\tau,\tau')&=\frac{1}{N}\sum_i \phi_{ir}(\tau)\phi_{ir}(\tau'),
\end{align}
\end{subequations}
and their conjugates $\Sigma_r(\tau,\tau')$ and $\Pi_r(\tau,\tau')$, after integrating out the fermionic and bosonic fields, and taking a replica symmetric and diagonal ansatz in the limit $m\to 0$. This leads to the large-$N$ action,
\begin{subequations}
\begin{align}\label{eq:LargeNAction}
\mc{S}&=-N\mr{Tr}\ln\left[-(G_0^{-1}-\Sigma)\right]+\frac{N-1}{2}\mr{Tr}\ln\left[D_0^{-1}-\Pi-\mr{i}\lambda\right]\nonumber\\
&+\frac{Nr_0^2}{2}\int_{r\tau,r'\tau'}\left[D_0^{-1}-\Pi-\ci \lambda\right]_{rr'}(\tau,\tau')+\frac{N}{2\gamma}\int_{r\tau}\ci \lambda_r(\tau)\nonumber \\
&+N\int_{\tau,\tau'}\sum_r\left[-\Sigma_r(\tau',\tau)G_r(\tau,\tau')+\frac{1}{2}\Pi_r(\tau',\tau)D_r(\tau,\tau')\right.\nonumber\\
&\left.+\frac{g^2}{2}G_r(\tau,\tau')G_r(\tau',\tau)D_r(\tau,\tau')\right].
\end{align}
In the above, we have used the shorthand notation $\int_{r\tau}=\int_0^\beta d\tau\sum_r$. Here 
\begin{align}
G_{0,rr'}^{-1}(\tau,\tau')&=-[(\partial_\tau-\mu)\delta_{rr'}-t_{rr'}]\delta(\tau-\tau'),  \label{eq:invG0}\\ 
\Sigma_{rr'}(\tau,\tau')&=\Sigma_r(\tau,\tau')\delta_{rr'},\\
D^{-1}_{0,rr'}(\tau,\tau')&=[(-\partial_\tau^2+m_b^2)\delta_{rr'}+K_{rr'}]\delta(\tau-\tau') \label{eq:invD0}\\
\Pi_{rr'}(\tau,\tau')&=\Pi_r(\tau,\tau')\delta_{rr'},\\
\lambda_{rr'}(\tau,\tau')&=\lambda_r(\tau)\delta_{rr'}\delta(\tau-\tau'). \label{eq:lambda_r}
\end{align}
\end{subequations}
The Lagrange multiplier field $\lambda_r(\tau)$ is introduced to impose the spherical constraint of Eq.\eqref{eq:SphericalConstraint}.

By extremizing the large-$N$ action $\mc{S}$ [Eq.\eqref{eq:LargeNAction}] with respect to $G_r(\tau,\tau'),~\Sigma_r(\tau,\tau'),~D_r(\tau,\tau'),~\Pi_r(\tau,\tau'),~\lambda_r(\tau)$ and $r_0$, and assuming translation invariance in space and time, e.g., $G_r(\tau,\tau')=G(\tau-\tau')$, $\lambda_r(\tau)=\lambda$, and similarly for $\Sigma,~D,~\Pi$, we obtain the large-$N$ self-consistent saddle-point equations. Using Fourier transforms for the momentum $(k,q)$ and (Matsubara) frequency space, the saddle-point equations are written as
\begin{subequations}\label{eq:SaddlePoint}
\begin{align}
G(\ci \omega_n)&=\frac{1}{L}\sum_k \frac{1}{\ci \omega_n+\mu-\epsilon_k-\Sigma(\ci \omega_n)}, \label{eq:G_iwn}\\
D(\ci \Omega_m)&=\frac{1}{L}\sum_q \frac{1}{\Omega_m^2+m_b^2+\omega_q^2-\Pi(\Omega_m)}+\beta \delta_{m,0}r_0^2,\label{eq:D_iwn}\\
\Sigma(\tau)&=g^2{D(\tau)}G(\tau), \label{eq:SaddleSigma}\\
\Pi(\tau)&=-g^2G(\tau)G(-\tau), \label{eq:Pi_tau}\\
\frac{1}{\beta L}\sum_{q,m}&\frac{1}{\Omega_m^2+m_b^2+\omega_q^2-\Pi(\ci \Omega_m)}+r_0^2=\frac{1}{\gamma},\label{eq:SaddleConstraint}\\
r_0M^2&(T)=0, \label{eq:r0}\\
M^2(T)&\equiv m_b^2-\Pi(\ci \Omega_m=0,T),\label{eq:ThermalMass}
\end{align}
\end{subequations}
where $\omega_n=(2n+1)\pi T$ and $\Omega_m=2m\pi T$ are the fermionic and bosonic Matsubara frequencies, with $n,m$ integers. In the above, we have absorbed $\lambda$ in $m_b^2$, i.e., $m_b^2-\ci \lambda \to m_b^2$. The non-interacting fermionic and bosonic dispersions are given by $\epsilon_k = -2t\cos k$, and $\omega_q^{2} = 2K(1-\cos q)$. The spatial randomness in the Yukawa coupling [Eq.\eqref{eq:H_fphi}] leads to local fermionic and bosonic self energies $\Sigma(\tau)$ $\Pi(\tau)$, respectively, after disorder averaging. These self energies are determined by the local Green's functions $G(\tau)$ and $D(\tau)$ for fermions and bosons in Eqs.\eqref{eq:SaddleSigma},\eqref{eq:Pi_tau}. In the next section, we obtain the phase diagram (Fig.\ref{fig:PhaseDiagram}) as a function of temperature $T$ and $\gamma$ by solving the above equations numerically, as well as analytically in the low-temperature/energy limit. We numerically solve the saddle-point equations [Eqs.\eqref{eq:SaddlePoint}] iteratively, both in imaginary-time, using Matsubara frequency representation, as well as by analytically continuing to real frequencies, $\ci \omega_n\to \omega+\ci 0^+$ and $\ci \Omega_m\to \Omega+\ci 0^+$. See Supplemental Material, Sec.\ref{app:numerical_saddle} for details. 


\section{Phase diagram}\label{sec:PhaseDiagram}
In this section, we discuss the phase diagram of the 1D Yukawa-SYK model based on the boson thermal mass $M(T)$, the order parameter $r_0$, and the fermionic and bosonic self energies at zero and low temperatures. We first analyze the phase diagram by solving the large-$N$ saddle-point equations [Eqs.\eqref{eq:SaddlePoint}] analytically at low energies. We verify the analytical results through numerical self-consistent solutions of the Eqs.\eqref{eq:SaddlePoint} later in the section. For the numerical solutions and for the rest of the paper we set the parameters $t=1,~\mu=0,~K=1$ and $g=1$ [Eq.\eqref{eq:Model}]. Here we take the bare boson mass $m_b$ and $t,~\mu$ to have dimension of energy ($E$), such that the bosonic field ($\phi$), $K$, $g$ and $\gamma$ have dimensions of $E^{-1/2}$, $E^2$, $E^{3/2}$, and $E$ respectively. We set $t$ as the unit of energy. The above choice of parameters leads to a QCP at $\gamma_c\simeq 1.36$.

\subsection{Zero-temperature phase diagram}
We first obtain the local fermionic Green's function $G(\ci \omega_n)$ at low energies at $T=0$ from Eq.\eqref{eq:G_iwn} by taking the continuum limit, $(1/L)\sum_k\to \int_{-\Lambda}^\Lambda dk/2\pi$, and linearizing the fermionic energy dispersion around the Fermi points $\pm k_\mr{F}$ as $\epsilon_k - \mu \simeq \pm v_\mr{F} k$,
where $v_\mr{F}$ is the Fermi velocity and $\Lambda$ is a momentum cutoff. As discussed in the Appendix \ref{app:SelfEnergy}, the Fermi points remain at the non-interacting value, even for the strange metal, since the Luttinger theorem holds across the phase diagram in the 1D Yukawa-SYK. In the low-energy limit, $\Lambda v_\mr{F}\gg |\omega_n|,|\Sigma(\ci \omega_n)|$ at $T=0$, we obtain (see Appendix \ref{app:LowEnergySol}),
\begin{align}
G(\ci \omega_n)\simeq -\frac{\ci}{v_\mr{F}} \mr{sgn}(\omega_n), \label{eq:G_iwn_LowE}
\end{align}
i.e., the same local Green's function as in the non-interacting limit ($g=0$). Using Eq.\eqref{eq:Pi_tau}, the above leads (Appendix \ref{app:LowEnergySol}) to boson self energy,
\begin{align}
\tilde{\Pi}(\ci\Omega_m)\simeq -\kappa |\Omega_m|  \label{eq:BosonPolarization}  
\end{align}
for $\Lambda v_\mr{F}\gg |\Omega_m|$, where $\tilde{\Pi}(\ci \Omega_m)=\Pi(\ci \Omega_m)-\Pi(\ci \Omega_m=0)$ and $\kappa=g^2/(\pi v_\mr{F}^2)$. As evident from the above expression, the bosons are damped with a dissipation strength $\kappa$. We verify the dissipative bosonic self energy from the numerical solutions of the large-$N$ saddle-point Eqs.\eqref{eq:SaddlePoint} in Fig.\ref{fig:self_energy}(a) for $t=1,~\mu=0,~K=1$ and $g=1$ [Eq.\eqref{eq:Model}]. The estimated dissipation strength $\kappa\simeq 0.08$ matches the slope of the retarded self-energy $\mr{Im}\Pi^R(\Omega)$ with frequency $\Omega$ ($\ci \Omega_m\to \Omega+\ci 0^+$). Using Eqs.\eqref{eq:BosonPolarization}, \eqref{eq:D_iwn},\eqref{eq:ThermalMass}, the local bosonic Matsubara propagator is given by
\begin{align}
D(\mathrm{i}\Omega_m) &= \frac{1}{2v_b}\frac{1}{\sqrt{\Omega_m^{2}+M^{2}(T)+\tilde{\Pi}(\mathrm{i}\Omega_m)}}{+\beta \delta_{m,0}r_0^2} .\label{eq:D_iwn_cont}
\end{align}
To obtain the above, we again take the continuum limit in Eq.\eqref{eq:D_iwn} by approximating the bosonic dispersion as $\omega_q^2\simeq v_b^2 q^2$ with a velocity $v_b=\sqrt{K}$. The bosonic thermal mass is calculated from the remaining saddle-point equations in Eq.\eqref{eq:SaddlePoint}. In particular, Eq.\eqref{eq:SaddleConstraint},
\begin{align}
D(\tau=0) &= r_0^2+\frac{T}{2v_b}\sum_{m}\frac{1}{\sqrt{\Omega_{m}^{2}+M^{2}(T)+\kappa|\Omega_{m}|}}=\frac{1}{\gamma}, \label{eq:Constraint_Cont}
\end{align}
along with $r_0^2M(T)=0$ [Eq.\eqref{eq:r0}], effectively describe \cite{Sachdev_book} a large-$N$ dissipative $O(N)$ model in $d=1$. At $T=0$, by converting the Matsubara summation into an integral, $T\sum_{m}\to\int_{-\Gamma}^\Gamma d\Omega/2\pi$, with a frequency cutoff $\Gamma$, we obtain the critical point at
\begin{subequations}\label{eq:ZeroT_Mr0}
\begin{align}
 \gamma_c\simeq \frac{2\pi v_b}{\ln\left(\frac{4\Gamma}{\kappa}\right)}, \label{eq:gammac} 
\end{align}
with
\begin{align}
M(0)&=0, ~~~~~r_0(0)=0,   \label{eq:Mr0_critical}
\end{align}
i.e., the mass and the order parameter both vanishes as $T=0$. In the disordered phase for $\gamma>\gamma_c$,
\begin{align}
M(0)&\simeq 2\Gamma e^{-2\pi v_b/\gamma}-\frac{\kappa}{2},~~~~~r_0(0)=0. \label{eq:Mr0_DisorderedPhase}
\end{align}
The above zero-temperature boson mass in the FL phase vanishes approaching the critical point, $\gamma\to \gamma_c^+$, as $M(0)\sim (\gamma-\gamma_c)$, in agreement with the low-temperature mass extracted from the numerical solution of the large-$N$ saddle-point Eqs.\eqref{eq:SaddlePoint}, as shown in Fig.\ref{fig:Mass_Z}(a) (Appendix \ref{app:SelfEnergy}).

In the $T=0$ ordered phase, $\gamma<\gamma_c$, we get
\begin{align}
M(0)&=0,~~~~~r_0^2(0)\simeq \frac{1}{\gamma}-\frac{1}{\gamma_c}. \label{eq:Mr0_OrderedPhase}
\end{align}
\end{subequations}
The ordered phase appears at zero temperature in this 1+1D model with continuous $O(N)$ symmetry due to dissipation-induced long-range coupling in the time direction when the boson mass vanishes.

\subsubsection{Fermion self-energy at $T=0$: Fermi and non-Fermi liquids}
We characterize the metallic state at zero temperature through the fermionic self-energy, which indicates whether there are Landau quasiparticles.

We obtain the zero-temperature fermionic self-energy from Eq.\eqref{eq:SaddleSigma}, using Eqs.\eqref{eq:G_iwn_LowE} and \eqref{eq:D_iwn_cont}, as
\begin{align}
\Sigma(\mathrm{i}\omega_n)&= -\frac{\mathrm{i}g^{2}}{4\pi v_{F}v_b}\int_{-\Gamma}^{\Gamma}\frac{\mathrm{sgn}(\omega+\Omega)}{\sqrt{\Omega^{2}+\kappa|\Omega|+M(0)^{2}}}d\Omega \nonumber \\
&-\frac{\mathrm{i}g^{2}r_{0}^{2}}{v_{F}}\mr{sgn}(\omega_n).
\label{eq:sigma_T_0}
\end{align}
As a result, using the $M(0)$ and $r_0(0)$ from Eqs.\eqref{eq:ZeroT_Mr0}, we evaluate the above expression (see Appendix \ref{app:SelfEnergy}), to obtain the Matsubara self energy $\Sigma(\ci \omega_n)$. Analytically continuing to real frequency through $\Sigma(\ci \omega_n\to \omega+\ci 0^+)=\Sigma^R(\omega)$ we obtain the retarded self energy $\Sigma^R(\omega)$. This leads to a FL self energy with the imaginary part $\mr{Im}\Sigma^R(\omega)\sim -\omega^2$ in the disordered phase for $\gamma>\gamma_c$, where $M(0)\neq 0$. Thus, we obtain a 1D FL in the disordered phase of the Yukawa-SYK model. 

In contrast, we obtain a NFL strange metallic self energy, 
\begin{align}
\Sigma^R(\omega)\simeq -f\sqrt{\omega}e^{\ci \pi/4}, \label{eq:NFL_Sigma_T0}
\end{align}
 at the critical point $\gamma=\gamma_c$ with $M(0)=0$, where $f=g^2/(\pi v_\mr{F}v_b\sqrt{\kappa})$. As a result, there are no low-energy quasiparticles at the Fermi points. The vanishing of mass at zero temperature in the ordered phase ($\gamma<\gamma_c$) also leads to a similar NFL self energy, $\mr{Im}\Sigma^R(\omega)\simeq -(g^2r_0^2/v_\mr{F})-f\sqrt{|\omega|}$, albeit with an additional effective elastic scattering that originates from the non-zero value of the order parameter $r_0$. This is expected since a non-zero static expectation value $\langle \phi_{lr}\rangle$ of the bosonic field in $\mc{H}_{f\phi}$ [Eq.\eqref{eq:H_fphi}] leads to an effective onsite disorder for fermions. We verify the above low-energy analytical form of the fermionic self energy in the FL ($\gamma>\gamma_c$) and NFL ($\gamma\leq \gamma_c)$ states from numerical solutions of the large-$N$ saddle-point equations [Eq.\eqref{eq:SaddlePoint}] at low temperatures, as discussed in the next section.

 Even in the absence of quasiparticles in the NFL state at the critical point, as evident from Eq.\eqref{eq:NFL_Sigma_T0}, the non-interacting Fermi points $\pm k_\mr{F}$ correspond to zero-energy poles of the retarded Green's function at $T=0$, i.e., $G_R^{-1}(\pm k_F,\omega=0)=0$, where $G_R^{-1}(k,\omega)=[\omega-\epsilon_k+\mu-\Sigma^R(\omega)]$. In appendix \ref{app:SelfEnergy} (Fig.\ref{fig:Mass_Z}), we numerically verify that the Luttinger count agrees with the fermion density, i.e., $n=\int_{-k_\mr{F}}^{k_\mr{F}}dk/(2\pi)$. Thus, the Fermi points demarcating the occupied and unoccupied regions in momentum space remain well-defined in the strange metal phase, even when quasiparticles are destroyed.

\begin{figure}[h!]
    \centering
    \includegraphics[width=0.9\linewidth]{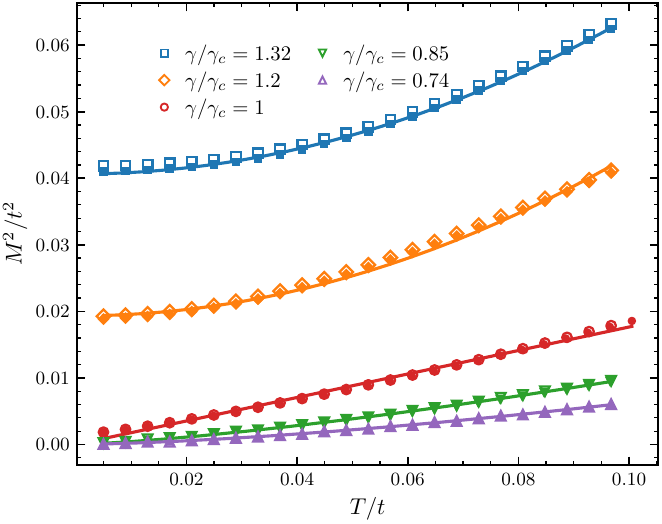}
   \caption{{\bf Boson thermal mass:} The thermal mass $M^2(T)$ for the 1D Yukawa-SYK model {(unfilled symbols)} with $t=1,~\mu=0,~K=1$ and $g=1$ is compared with that of the effective dissipative $O(N)$ model {(filled symbols)} with dissipation strength $\kappa=g^2/(\pi v_\mr{F}^2))\simeq 0.08$. For $\gamma>\gamma_c$, $M^2(T)-M^2(0)\sim T^2$, whereas $M^2(T)\sim T^\alpha$ for $\gamma\leq \gamma_c$. Up to logarithmic corrections, $\alpha=1$ at the QCP ($\gamma=\gamma_c$). For $\gamma<\gamma_c$, $M^2(T)$ shows a power law behavior with $\alpha=1.36$ for $\gamma/\gamma_c=0.85$ and $\alpha=1.51$ for $\gamma/\gamma_c=0.74$. The solid lines denote the power-law fits.}
\label{fig:MT_thermal_mass}
\end{figure}

\subsection{Finite-temperature phase diagram}
At finite but low temperature $T$, we obtain the bosonic thermal mass $M^2(T)$ analytically by evaluating the Matsubara sum in Eq.\eqref{eq:Constraint_Cont} using the Euler-Maclaurin expansion for the disordered phase ($\gamma > \gamma_c$) and at the critical point $\gamma=\gamma_c$ [see Appendix \ref{app:ThermalMass}]. For $\gamma> \gamma_c$, in the Fermi liquid regime, this low-temperature expansion converges self-consistently for $M^2(T)$ obtained from the leading-order terms. At the critical point $\gamma=\gamma_c$, the expansion is only marginally convergent, and hence there could be logarithmic corrections. To this end, in the disordered phase with Fermi liquid metallic state, for $T\ll (M^3/\kappa)^{1/2}$, we obtain
\begin{subequations}\label{eq:MT_LowT}
\begin{align}
M^2(T)-M^2(0)\sim T^2.    
\end{align}  
Similarly, at the strange metallic critical point we obtain
\begin{align}
M^2(T)\sim \kappa T,
\end{align}
modulo logarithmic corrections, at low temperature $T\ll \kappa$.
\end{subequations}
As shown in Fig.\ref{fig:MT_thermal_mass}, the above low-$T$ behaviors [Eqs.\eqref{eq:MT_LowT}] are consistent with $M^2(T)$ obtained from Eq.\eqref{eq:ThermalMass} by numerically solving saddle-point Eqs.\eqref{eq:SaddlePoint}. Numerically we access the phases and the critical region as a function of $\gamma$ and $T$, by adjusting the bare boson mass $m_b^2$ in Eqs.\eqref{eq:SaddlePoint} such that the spherical constraint [Eq.\eqref{eq:SaddleConstraint}] is satisfied. In Fig.\ref{fig:MT_thermal_mass}, we also compare the results of $M^2(T)$ for the 1D Yukawa-SYK model with that of an effective dissipative $O(N)$ model \cite{Sachdev_book} (Appendix \ref{app:LowEnergySol}) with dissipation strength $\kappa\simeq 0.08$ [Eq.\eqref{eq:BosonPolarization}], which corresponds to {$t=1,~\mu=0,~K=1$ and $g=1$} [Eq.\eqref{eq:Model}]. We find good agreement for $M^2(T)$ between the above two models in all the phases.

For $\gamma<\gamma_c$, the system is ordered only at $T=0$ [Eq.\eqref{eq:Mr0_OrderedPhase}, Fig.\ref{fig:PhaseDiagram}] with a finite thermal mass that approaches zero for $T\to 0$, as shown by numerically computed $M^2(T)$ in Fig.\ref{fig:MT_thermal_mass}. The leading-order low-temperature Euler-MacLaurin expansion (see Appendix \ref{app:ThermalMass}) suggests a temperature-dependence $M^2(T)\sim T^{4/3}$ closed to the critical point for $\gamma<\gamma_c$. However, this expansion is not convergent. One can also obtain a different temperature dependence [Appendix \ref{app:ThermalMass}], $M^2(T)\sim T^2$, by retaining only the \emph{classical} contribution from zeroth Matsubara frequency $\Omega_m=0$ in Eq.\eqref{eq:Constraint_Cont} for $T\ll \kappa v_b^2/\gamma^2$. The numerical results [Fig.\ref{fig:MT_thermal_mass}] for $M^2(T)$ in the 1D Yukawa-SYK and the dissipative $O(N)$ model are consistent with a power-law, $M^2(T)\sim T^\alpha$ at low temperature, where the exponent increases with decreasing $\gamma$ from $\alpha\approx 4/3$, close to the critical point ($\gamma\lesssim \gamma_c$), to $\alpha\approx 2$ deep in the $T=0$ ordered phase.

\begin{figure}[h!]
    \centering
    \includegraphics[width=\linewidth]{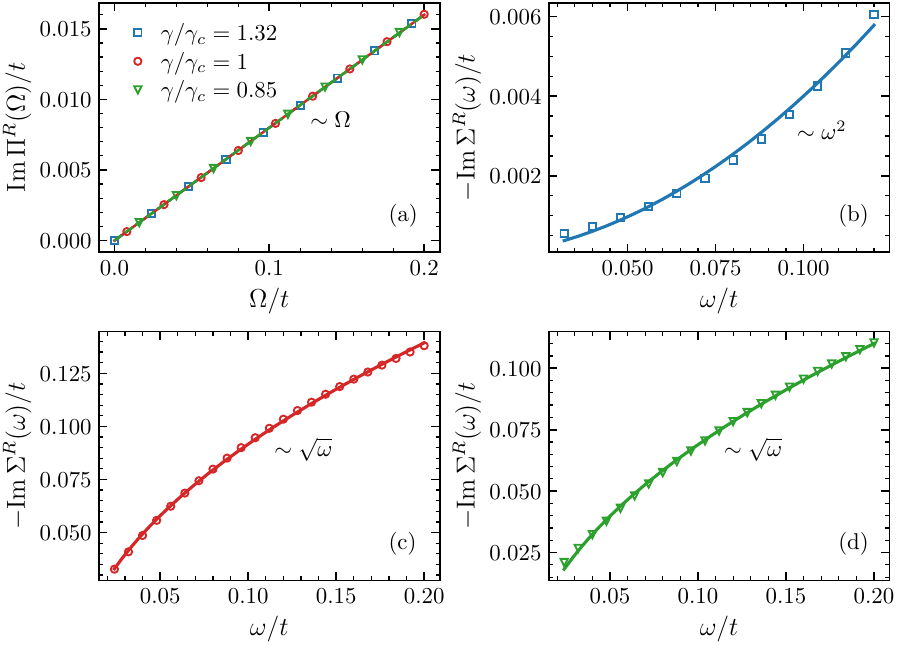}
   \caption{{\bf Bosonic and fermionic self energies:} (a) Imaginary part of the retarded bosonic self energy $\Pi^R(\Omega)$ for three values of $\gamma$ in the disordered ($\gamma/\gamma_c=1.32$), critical ($\gamma/\gamma_c=1$) and ordered ($\gamma/\gamma_c=0.85$) states with $t=1,~\mu=0,~K=1$ and $g=1$, at temperature $T=0.005$. (b) $\gamma/\gamma_c=1.32$, (c) $\gamma/\gamma_c=1$ and (d) $\gamma/\gamma_c=0.85$, show the corresponding imaginary part of the fermionic self-energies. The solid lines denote the power-law fits with the frequencies.
}
\label{fig:self_energy}
\end{figure}

\subsubsection{Fermion self energy at low temperatures}
We obtain the low-temperature fermion self-energy for $\gamma\geq \gamma_c$ through a low-temperature Euler-MacLaurin expansion in Appendix \ref{app:SelfEnergy}, using Eqs.\eqref{eq:G_iwn_LowE},\eqref{eq:D_iwn_cont} and low-temperature thermal mass $M^2(T)$ [Eqs.\eqref{eq:MT_LowT}] in Eq.\eqref{eq:SaddleSigma}. In the Fermi liquid phase for $\gamma>\gamma_c$ with $M(0)\neq 0$, we obtain,
\begin{align}
\Sigma^R(\omega)&\simeq -\frac{g^2}{2v_\mr{F}v_b}\left[\frac{\omega}{\pi M(0)}+\ci \frac{\kappa \omega^2}{4\pi M^3(0)}+\ci \frac{\pi \kappa T^2}{4M^3(0)}\right] \label{eq:FL_Sigma}
\end{align}
for $T\ll M(0)$ [Eq.\eqref{eq:Mr0_DisorderedPhase}]. As discussed in the Appendix \ref{app:SelfEnergy}, the above leads to an estimate of quasiparticle residue $Z=1/[1+g^2/(2\pi v_\mr{F}v_bM(0))]$, that vanishes for $\gamma\to \gamma_c^+$, {in agreement with the $Z$ extracted from numerically obtained fermion self-energy [see Fig.\ref{fig:Mass_Z}(b), Appendix \ref{app:SelfEnergy}]}. Fig.\ref{fig:self_energy}(b) shows the self-energy for $\gamma/ \gamma_c=1.32$, where we find a 
quadratic frequency dependence, i.e. $\mathrm{Im} \Sigma^{R}(\omega) \sim -\omega^2$ for $T=0.005\ll M(0)$, consistent with the above Fermi liquid behavior [Eq.\eqref{eq:FL_Sigma}].

In the finite-temperature quantum critical regime at $\gamma=\gamma_c$, for $T\ll M(T)\ll \omega$, we obtain the same NFL self energy of Eq.\eqref{eq:NFL_Sigma_T0}. Fig.\ref{fig:self_energy}(c) shows the numerically obtained self-energy at $\gamma=\gamma_c$, with $\mathrm{Im} \Sigma^{R}(\omega) \sim -\sqrt{|\omega|}$, for $T=0.005\ll \omega$, clearly indicating a non-Fermi liquid behavior with the absence of quasiparticle. On the ordered side for $\gamma/\gamma_c=0.85$, this non-Fermi liquid behavior persists for $T=0.005\ll \omega$ as shown in Fig.\ref{fig:self_energy}(d). 

\subsubsection{Low-temperature thermodynamics} \label{sec:LowTThermodynamics}
We now describe the thermodynamics at low temperatures, specifically the temperature dependence of thermal entropy $S(T)$ in the different phases. The (grand) free energy $F(T)$ (per site, per flavor) is obtained by evaluating the large-$N$ action $\mc{S}$ [Eq.\eqref{eq:LargeNAction}] using the saddle-point fermionic and bosonic Green's functions and their corresponding self-energies [Eq.\eqref{eq:SaddlePoint}]. Since the last two terms of the action in Eq.\eqref{eq:LargeNAction} cancel each other at the saddle point, the free energy can be written as $F=F_f+F_b$, in terms of a fermion ($F_f$) and a boson ($F_b$) contributions. The entropy is obtained from $S=-(\partial F/\partial T)=S_f+S_b$. As in the case of 2D Yukawa-SYK model \cite{esterlisPRB}, for any $\gamma$, the fermion contribution is unaffected by the interaction or the self-energy effects at low temperatures ($T\ll \Lambda v_\mr{F}$), and $S_f$ remains identical to the non-interacting value,
\begin{align}
S_f&\simeq \frac{\pi}{3v_\mr{F}}T.\label{eq:Sf}
\end{align}

For the bosonic contribution, we again perform an asymptotic low-temperature Euler-MacLaurin expansion of $\Omega(T)$ for $\gamma\geq \gamma_c$ [see Appendix \ref{app:Thermodynamics}]. In the Fermi liquid phase ($\gamma>\gamma_c$), we get linear-$T$ bosonic contribution to entropy,
\begin{align}
S_b&\simeq \frac{\pi \kappa}{6M(0)v_b}T,\label{eq:Sb_FL}
\end{align}
as expected due to low-energy excitations arising from Ohmic dissipation [Eq.\eqref{eq:BosonPolarization}], even when bosons are gapped \cite{hanggi2006,Hanggi_2008, Ford_PRB}. Thus, the linear-$T$ coefficient of total entropy in the FL phase is enhanced and the non-interacting value of Eq.\eqref{eq:Sf} is only recovered for large $\gamma$. This enhancement affects thermal entropy to entanglement crossover in the FL phase, as we discuss later in Sec.\ref{sec:EntYSYK}. In Figs.\ref{fig:ThermalEntropy} of Appendix \ref{app:Thermodynamics}, we compare our low-temperature analytical results for $S_f,~S_b$ and $S$ with those obtained from numerical solution of the saddle-point Eqs.\eqref{eq:SaddlePoint}.

At the QCP, the low-temperature expansion suggests a bosonic contribution to entropy, $S_b\sim \sqrt{T}$
along with linear-$T$ fermionic part [Eq.\eqref{eq:Sf}].
However, as shown in Fig.\ref{fig:ThermalEntropy}(a) of Appendix \ref{app:Thermodynamics}, our numerical results at $\gamma=\gamma_c$ show that the entropy $S$ or $S_b$ vanish linearly with $T$. For the temperature regime accessed in our calculations, we could not find signature of $\sim \sqrt{T}$ behavior in $S_b$. Since the low-temperature expansion is not convergent for $\gamma<\gamma_c$, we numerically obtain the entropy in the $T=0$ ordered phase, as shown in Fig.\ref{fig:ThermalEntropy}(b) of Appendix \ref{app:Thermodynamics}.




\section{Large-$N$ theory for entanglement in 1D Yukawa-SYK model} \label{sec:Ent_LargeN}

To understand the entanglement properties of the FL and NFL ground states in the 1D Yukawa-SYK model, we construct a large-$N$ theory for the second R\'{e}nyi entropy of a fermionic subsystem for the thermal state, described by the density matrix $\rho=e^{-\beta \mc{H}}/Z$ at temperature $T$. The subsystem R\'{e}nyi entropy is computed by numerically solving a set of large-$N$ saddle-point equations with entanglement replicas, as we discuss below. The large-$N$ theory is formulated based on a recent coherent-state path integral representation \cite{Haldar,Chakraborty_PRL} of the subsystem R\'{e}nyi entropy for fermions. 
The path integral representation has been earlier used to study entanglement in zero-dimensional SYK and other related large-$N$ models \cite{Haldar}, and in the FL and Mott insulating states of Hubbard model within DMFT approximation \cite{bera_PRB}. The readers are referred to Refs.\cite{Haldar,bera_PRB} for the details of the formalism. As in other imaginary-time path integral formulations \cite{QMC1,QMC2,Grover_QMC,Francesco_QMC,Broecker_QMC,Troyer_QMC,Fakher_QMC,Jonathan} implemented numerically, the ground-state entanglement typically cannot be directly accessed in this method. The information of the ground-state entanglement is extracted by systematically extrapolating the finite-$T$ numerical results to the $\beta \to \infty$ ($T\to 0$) limit through a scaling ansatz, as we discuss below.

 For a bi-partition of the system into subsystems $A$ and $B$, the $n$-th R\'{e}nyi entropy of the subsystem $A$ is defined by
 \begin{equation}
S_A^{(n)} = \frac{1}{1-n}\ln \mathrm{Tr}_A [\rho_A^n],
\label{eq:nth_renyi}
\end{equation} 
 where $n>1$ is an integer and the reduced density matrix for $A$, $\rho_A=\mathrm{Tr}_B\rho$, is obtained by taking partial trace of the density matrix $\rho$ of the system over the degrees of freedom of $B$. The analytical continuation to the limit $n\to 1$ leads to the subsystem von Neumann entropy $S_A=S_A^{(n\to 1)}$. However, a direct computation of the von Neumann entropy is not possible within our method or the standard replica field-theory approach \cite{Cardy}, unless $S_A^{(n)}$ can be obtained analytically as function of $n$ \cite{Cardy}. In this work, we only focus on the second R\'{e}nyi entropy $S_A^{(2)}$. Higher-order R\'{e}nyi entropies $S_A^{(n>2)}$ can be constructed similarly, but are numerically more challenging to compute. 
 
 In this work, we only consider the entanglement entropy of fermions in the 1D Yukawa-SYK model for a spatial contiguous segment $A$ of length $L_A$ with the rest of degrees of freedom in the system, which include fermions and bosons in segment $B$ of length $L-L_A$, as well as the bosons in the $A$ segment (see Fig.\ref{fig:model}). In the large-$N$ limit, we define,
 \begin{subequations}
\begin{equation}
     S_{A_{f}}^{(2)}=-\frac{1}{N}\ln{\mathrm{Tr}_{A_{f}}\rho_{A_{f}}^{2}},
     \label{eq:second_renyi}
\end{equation}
i.e., the second R\'{e}nyi entropy (per flavor) of the fermionic subsystem, where 
\begin{align} \label{eq:fermion_density_matrix}
 \rho_{A_f}&=\mathrm{Tr}_{A_\phi,\, B_{f\phi}} \rho   
\end{align}
\end{subequations}
is the reduced density matrix of the fermions in subsystem $A$. Here $A_\phi$ refers to the bosonic degrees of freedom in $A$ and $B_{f\phi}$ both the fermionic and bosonic degrees of freedom in $B$. Due to the local entanglement of the fermions with the bosons within $A$, we expect the entanglement entropy $S^{(2)}_{A_f}(L_A, T=0)$ to have a \emph{volume-law} contribution $\sim L_A$, in addition to the inter-subsystem entanglement between fermions in $A$ and the bosons and fermions in $B$. The latter quantifies the true non-local quantum correlations, and is expected to follow a sub-volumic dependence on $L_A$, e.g., $\sim \ln L_A$.

Following Refs.\cite{Haldar,bera_PRB}, we represent $\mathrm{Tr}_{A_f}\rho_{A_f}^2$ as
\begin{align}
e&^{-NS_{A_f}^{(2)}} 
= \mathrm{Tr}_{A_f} \rho_{A_f}^{2}\nonumber\\
&=\int d^2\xi f(\xi_{1},\xi_{2}) \mathrm{Tr}[\rho D_{A_f}(\xi_1)]\mathrm{Tr}[\rho D_{A_f}(\xi_2)],
\label{eq:expansion_in_disp_op}
\end{align}
in terms of normal-ordered fermionic displacement operators \cite{Cahill_Glauber}
\begin{align}
D(\xi_{\alpha})= e^{\sum_{i,r\in A}c_{ir}^{\dagger}\xi_{ir\alpha}}e^{-\sum_{i,r\in A}\bar{\xi}_{ir\alpha}c_{ir}}.
\label{eq:fermion_disp_op}
\end{align}
Here $\xi_\alpha=\{\bar{\xi}_{ir\alpha},\xi_{ir\alpha}\}$ denotes a set of auxiliary Grassmann variables with flavors $i=1,\cdots, N$ at site $r=1,\cdots,L_A$ in two entanglement replicas $\alpha=1,2$, and $d^2\xi=\prod_{i,r\in A,\alpha}d\bar{\xi}_{ir\alpha}d\xi_{ir\alpha}$. 
$\mathrm{Tr}$ denotes trace over the entire system. The Gaussian function
\begin{equation}
f(\xi_{1},\xi_{2})= 2^{NL_{A}}e^{-\frac{1}{2}\sum_{i,r\in A}\left(\bar{\xi}_{ir1}\xi_{ir1}+\bar{\xi}_{ir2}\xi_{ir2}-\bar{\xi}_{ir1}\xi_{ir2}+\bar{\xi}_{ir2}\xi_{ir1}\right)}
\label{eq:gaussian_factor}
\end{equation}
connects the two replicas. 

From Eq.\eqref{eq:expansion_in_disp_op}, we construct the coherent-state path integral
\begin{subequations}
\begin{align}
e^{-NS_{A_{f}}^{(2)}}&= \frac{Z_{A_f}^{(2)}}{Z^{2}}, \\
Z_{A_f}^{(2)}&=\int d^2\xi \mathcal{D}(\bar{c}_\alpha,c_\alpha)\mathcal{D}\phi_{\alpha}f(\xi_{1},\xi_{2})e^{-(\mathcal{S} + \mathcal{S}_{\xi})},
\label{eq:path_integral_2nd_renyi}
\end{align}
\end{subequations}
where,  
\begin{align}
\mathcal{S} &= \int_0^\beta d\tau\sum_\alpha  
[\sum_{ir} \{\bar c_{ir\alpha} (\partial_\tau - \mu) c_{ir\alpha}
+ \frac{1}{2} (\partial_\tau \phi_{ir\alpha})^2\}\nonumber\\
&+ \mathcal{H}(\bar c_\alpha, c_\alpha, \phi_\alpha)]
\label{eq:action_with_replica}
\end{align}
is the usual imaginary time action for the Hamiltonian $\mc{H}$ in Eq.\eqref{eq:Model}, but with two replicas $\alpha=1,2$. Here
\begin{equation}
    S_{\xi}=\int_{\tau}\sum_{i,r\in A,\alpha}\big[\bar{c}_{ir\alpha}(\tau)\delta(\tau-\tau_{0}^{+})\xi_{i\alpha r}-\bar{\xi}_{ir\alpha}\delta(\tau-\tau_{0})c_{ir\alpha}(\tau)\big]
    \label{eq:action_with_source_term}
\end{equation}
 is a source term involving the auxiliary Grassman source fields $\xi$ which act only on the fermionic fields in subsystem $A$ ($r=1,\cdots,L_A$) at an imaginary time $\tau_0$ on the thermal cycle $\tau\in[0,\beta)$. The time $\tau_0$ is arbitrary, but once chosen, the source term breaks the periodicity of imaginary time, along with the spatial translational symmetry due to the entanglement cut between $A$ and $B$. $Z$ is the thermal partition function discussed in Sec.\ref{sec:SaddlePoint}. For $Z_{A_{f}}^{(2)}$, we integrate out the auxiliary Grassmann fields $\xi$, to obtain 
\begin{equation}
Z_{A_f}^{(2)}= \int\mathcal{D}(\bar{c}_{\alpha},c_{\alpha})\mathcal{D}\phi_{\alpha}e^{-(\mathcal{S}+\mathcal{S}_{\text{kick}})}
\label{eq:path_int_kick_action}
\end{equation}
where 
{\small
\begin{equation}
    \mathcal{S}_{\text{kick}}=\int d\tau d\tau'\sum_{i,r\in A,\alpha\beta}\bar{c}_{ir\alpha}(\tau)M_{\alpha\beta}\delta(\tau-\tau_{0}^{+})\delta(\tau'-\tau_{0})c_{ir\beta}(\tau'),
    \label{eq:action_with_kick}
\end{equation}}
with the matrix
\begin{equation}
   M=  \left[\begin{array}{cc}
1 & 1\\
-1 & 1
\end{array}\right].
\label{eq:replica_matrix_M}
\end{equation}
 Thus $\mc{S}_{kick}$ couples the two entanglement replicas through a self-energy term that acts like a \emph{kick} at imaginary time $\tau=\tau_0$ on the fermionic subsystem $A_f$.

We obtain the disorder averaged second R\'{e}nyi entropy,
\begin{align}
N\overline{S_{A_{f}}^{(2)}}=  -\overline{\ln Z_{A_{f}}^{(2)}}+2\overline{\ln Z},
\label{eq:dis_av_2nd_renyi}
\end{align}
where $\overline{\cdots}$ represents disorder averaging over the random Yukawa couplings $\{g_{ijlr}\}$ [Eq.\eqref{eq:H_f}]. We evaluate each term in the above equation separately by introducing disorder replicas, as was done in Sec.\ref{sec:SaddlePoint} for thermal partition function $Z$. Thus, for $Z_{A_f}^{(2)}$ we also use the replica trick, 
\begin{subequations}
\begin{equation}
  \overline{\ln Z_{A_{f}}^{(2)}}=  \lim_{m\to0}\frac{\overline{\left(Z_{A_{f}}^{(2)}\right)^{m}}-1}{m} ,
\label{eq:enta_replica_trick}
\end{equation}
with
\begin{align}
 \overline{\left(Z_{A_{f}}^{(2)}\right)^{m}}=  \int\mathcal{D}(\bar{c}_{\alpha a},c_{\alpha a})\mathcal{D}\phi_{\alpha a}e^{-\mc{S}_{A_{f}}[\bar{c}_{\alpha a},c_{\alpha a},\phi_{\alpha a}]}   
\end{align}
\end{subequations}
Here $\mc{S}_{A_f}$ denotes the replicated action with disorder replica index $a = 1, \cdots, m$. Following steps similar to that for the thermal partition function in Appendix \ref{app:LargeN}, we average over $\{g_{ijlr}\}$ and introduce the large-$N$ bilocal in-time collective fields,
\begin{subequations}
\begin{align}
G_{r,\alpha a,\beta b} (\tau,\tau') 
&= \frac{1}{N} \sum_{i} 
\bar{c}_{ir\beta b}(\tau')\, c_{ir\alpha b}(\tau) , \\
D_{r,\alpha a,\beta b} (\tau,\tau') 
&= \frac{1}{N} \sum_{l} 
\phi_{l r\alpha a}(\tau)\, \phi_{l r\beta b}(\tau') ,
\end{align}
\label{eq:large_N_fields_replica}
\end{subequations}
together with their conjugate fields $\Sigma_{r,\beta b,\alpha a}(\tau',\tau)$ and $\Pi_{r,\beta b,\alpha a}(\tau',\tau)$, in the path integral for $\overline{(Z_{A_f}^{(2)})^m}$. 

In the presence of the entanglement kick term [Eq.\eqref{eq:action_with_kick}], which breaks space and time translation invariance, the order parameter $r_0$ [Eq.\eqref{eq:O_N_symm}] and the Lagrange multiplier field $\lambda$ [Eq.\eqref{eq:lambda_r}], or equivalently the bare boson mass $m_b$, should, in principle, become space and time dependent. Thus, these inhomogeneous parameters should be obtained self-consistently in the large-$N$ theory for subsystem R\'{e}nyi entropy. Such self-consistency becomes numerically very challenging. Thus, to make the numerical computation tractable, we approximate these parameters with their homogeneous and time-independent values obtained as a function of $\gamma$ and $T$ from the solution of the equilibrium saddle-point Eqs.\eqref{eq:SaddlePoint}. With this approximation, we integrate out the fermionic $(\bar{c},c)$ and bosonic $\phi$ fields in $\overline{(Z_{A_f}^{(2)})^m}$. We further take the symmetric and diagonal ansatz in disorder replicas as in Sec.\ref{sec:SaddlePoint}, obtain
\begin{subequations}
\begin{align}
\overline{\left(Z_{A_f}^{(2)}\right)^m}&=\int \mathcal{D}(G,\Sigma,D,\Pi)\,
e^{-m\mathcal{S}_{A_f}[G,\Sigma,D,\Pi]}.
\label{eq:replicated_partition}
\end{align}
The action above is given by
\begin{align}
\mathcal{S}&_{A_{f}}=  -N\mathrm{Tr}\ln\left[-\left(\tilde{G}_{0}^{-1}-\Sigma\right)\right]+\frac{N}{2}\mathrm{Tr}\ln\left[D_{0}^{-1}-\Pi\right] \nonumber \\
 & +N\int d\tau d\tau'\sum_{r,\alpha\beta}\Bigg[-\Sigma_{r,\beta\alpha}(\tau',\tau)G_{r,\alpha\beta}(\tau,\tau') \nonumber \\
 & +\frac{1}{2}\Pi_{r,\beta\alpha}(\tau',\tau)D_{r,\alpha\beta}(\tau,\tau') \nonumber \\
 & +\frac{g{}^{2}}{2}G_{r,\alpha\beta}(\tau,\tau')G_{r,\beta\alpha}(\tau',\tau)D_{r,\alpha\beta}(\tau,\tau')\Bigg].
\label{eq:final_action_entan_replica}
\end{align}
The self-energy kick term [Eq.\eqref{eq:action_with_kick}] enters above through the non-interacting fermionic Green's function,
\begin{align}
\tilde{G}^{-1}_{0, r\alpha, r'\beta}&(\tau,\tau') 
= -\left[(\partial_{\tau}-\mu)\delta_{rr'} + t_{rr'}\right]
\delta_{\alpha\beta}\delta(\tau-\tau') \nonumber \\
&\quad - M_{\alpha\beta}\,\delta(\tau - \tau_{0}^{+})
\delta(\tau' - \tau_{0})\,\delta_{r \in A}\delta_{rr'} ,
\label{eq:fermion_greenfn_kick}
\end{align}
with $\delta_{r\in A}=1$ for $r=1,\cdots,L_A$ and $\delta_{r\in A}=0$ otherwise. Here 
\begin{align}
\Sigma_{r\alpha,r'\beta}(\tau,\tau')&=\Sigma_{r,\alpha\beta}(\tau,\tau')\delta_{rr'},  \\
D^{-1}_{0,r\alpha,r'\beta}(\tau,\tau')&=D^{-1}_{0,rr'}(\tau,\tau')\delta_{\alpha\beta},\\
\Pi_{r\alpha,r'\beta}(\tau,\tau')&=\Pi_{r,\alpha\beta}(\tau,\tau')\delta_{rr'}.
\end{align}
\end{subequations}
The bosonic bare inverse Green's function $D_{0,rr'}^{-1}(\tau,\tau')$ is given in Eq.\eqref{eq:invD0}.

For $N \to \infty$, the Green's functions are obtained from the saddle point of the effective action $\mathcal{S}_{A_{f}}$ [Eq.\eqref{eq:final_action_entan_replica}], 
\begin{subequations}\label{eq:self_consis_entan_replica}
\begin{align}
G_{r\alpha,r'\beta}^{-1}(\tau,\tau')= & \tilde{G}_{0,r\alpha,r'\beta}^{-1}(\tau,\tau')-\Sigma_{r,\alpha\beta}(\tau,\tau')\delta_{rr'}, \label{eq:InvG_Ent}\\
D_{r\alpha,r'\beta}^{-1}(\tau,\tau')= & D_{0,r\alpha,r'\beta}^{-1}(\tau,\tau')-\Pi_{r,\alpha\beta}(\tau,\tau')\delta_{rr'}, \label{eq:InvD_Ent}\\
\Sigma_{r,\alpha\beta}(\tau,\tau')= & g^{2}G_{r,\alpha\beta}(\tau,\tau')D_{r,\alpha\beta}(\tau,\tau'),\label{eq:Sigma_Ent}\\
\Pi_{r,\alpha\beta}(\tau,\tau')= & -g^{2}G_{r,\alpha\beta}(\tau,\tau')G_{r,\beta\alpha}(\tau',\tau), \label{eq:Pi_Ent}
\end{align}
\end{subequations}
which are henceforth referred as large-$N$ second R\'{e}nyi saddle-point equations. The local Green's functions, that appear in the expressions for the self energies in Eqs.\eqref{eq:Sigma_Ent},\eqref{eq:Pi_Ent}, are obtained from the Green's functions $G$ and $D$, which are inverses of $G^{-1}$ and $D^{-1}$ in Eqs.\eqref{eq:InvG_Ent} and \eqref{eq:InvD_Ent}, respectively, i.e., $G_{r,\alpha\beta}(\tau,\tau')=G_{r\alpha,r\beta}(\tau,\tau')$ and $D_{r,\alpha\beta}(\tau,\tau')=D_{r\alpha,r\beta}(\tau,\tau')$.

The second term in Eq.~(\ref{eq:dis_av_2nd_renyi}), $\overline{\ln Z}$, is evaluated in a similar manner by computing the equilibrium saddle-point action $\mathcal{S}$ [Eq.\eqref{eq:LargeNAction}] using the solution of the saddle-point Eqs.\eqref{eq:SaddlePoint}, as discussed in Secs.\ref{sec:SaddlePoint} and \ref{sec:PhaseDiagram}. Finally, we express the disorder averaged fermionic subsystem second R\'{e}nyi entropy $\overline{S_{A_{f}}^{(2)}}$ [Eq.\eqref{eq:dis_av_2nd_renyi}] in terms of the difference between large-$N$ second R\'{e}nyi and equilibrium saddle-point actions,
\begin{equation}
\overline{S_{A_{f}}^{(2)}}=\frac{\mathcal{S}_{A_{f}}-2\mathcal{S}}{N}. 
\label{eq:final_renyi_entropy}
\end{equation}
We evaluate $\mc{S}_{A_f}$ above by numerically solving the second R\'{e}nyi saddle-point Eqs.\eqref{eq:self_consis_entan_replica} at finite temperatures $T$ with $m_b^2(\gamma,T)$ supplied from the self-consistent solutions of the equilibrium large-$N$ saddle-point Eqs.\eqref{eq:SaddlePoint} across the phase diagram (Fig.\ref{fig:PhaseDiagram}) as a function of $\gamma$ [Eq.\eqref{eq:SphericalConstraint}]. To this end, we solve Eqs.\eqref{eq:self_consis_entan_replica} by discretizing in imaginary time, $\beta=N_\tau\delta\tau$, and by iterating the equations till convergence within some numerical tolerance. Each numerical iteration, in principle, involves inverting large matrices of size $(2LN_\tau\times 2LN_\tau)$ in Eqs.\eqref{eq:InvG_Ent},\eqref{eq:InvD_Ent}, due to two entanglement replicas, $L$ sites and $N_\tau$ time points. However, since we only need the local Green's function to evaluate the self energies in Eqs.\eqref{eq:Sigma_Ent},\eqref{eq:Pi_Ent} at each iteration step, we use an efficient recursive Green's function method \cite{bera_PRB} to invert $G^{-1}$ and $D^{-1}$ in Eqs.\eqref{eq:InvG_Ent},\eqref{eq:InvD_Ent}. See Supplemental Material, Sec.\ref{app:renyi_numerical_saddle} for details. Finally, we numerically extrapolate the results for $S_{A_f}^{(2)}(\delta\tau)$ at finite discretization $\delta\tau$ for a fixed temperature $T=1/\beta$ to $\delta\tau\to 0$, i.e., $S_{A_f}^{(2)}=S_{A_f}^{(2)}(\delta\tau\to 0)$. 

At a finite temperature, in a gapless system like the 1D Yukawa-SYK model, we expect the subsystem fermionic second R\'{e}nyi entropy $S_{A_{f}}^{(2)}(L_A,T)$ to exhibit a crossover \cite{SwingleSenthil} as function of decreasing temperature, from thermal entropy to second R\'{e}nyi entanglement entropy as $T\to 0$. As discussed already, the fermionic subsystem R\'{e}nyi entanglement entropy $S_{A_f}^{(2)}(L_A,T=0)$ comprises of non-local inter-subregion entanglement and local intra-subregion fermion-boson entanglement. In Sec.\ref{sec:EntYSYK}, we show that this thermal to entanglement entropy crossover is captured by a scaling ansatz, which collapses data at different temperatures and subsystem sizes $L_A$ onto a single universal curve.

In the next section we first benchmark our time-discretized path integral method for subsystem second R\'{e}nyi entropy, the extrapolation to $\delta \tau\to 0$, and the scaling ansatz, for non-interacting fermions, where the subsystem second R\'{e}nyi entropy can be calculated efficiently using standard correlation-matrix method \cite{Casini_Huerta}. We then present our large-$N$ results for $S^{(2)}_{A_f}$ for Yukawa-SYK model, mainly focusing on the strange metallic state at QCP and the Fermi liquid phase for $\gamma \geq \gamma_c$.

\section{Results} \label{sec:Results}

\subsection{Benchmark for non-interacting fermions} \label{sec:NonIntBnchmark}
To benchmark our imaginary time path integral method and the $\delta\tau\to 0$ extrapolation, we first consider a non-interacting model of the nearest-neighbour tight-binding fermionic chain, which is given by $\mc{H}_f$ of Eq.\eqref{eq:H_f} with only one fermion flavor $N=1$. For such a non-interacting fermionic system, the second R\'{e}nyi entropy of a subsystem $A$ with $L_A$ sites can be obtained efficiently for large system sizes $L$ from the correlation matrix $C_{rr'}=\mr{Tr}[\rho c_r^\dagger c_{r'}]$ with $r,r'\in A$, where $\rho=\exp(-\beta \mc{H}_f)/Z_f$ with $Z_f=\mr{Tr}[e^{-\beta \mc{H}_f}]$. The second R\'{e}nyi entropy is then given by,
\begin{align}
S_A^{(2)}=-\mr{Tr}\ln\left[(\mathbb{I}-\mathbb{C})^2+\mathbb{C}^2\right]. \label{eq:SA2_NonInt_CorrMat}
\end{align}
Here $\mathbb{I}$ is the identity matrix, and the correlation matrix $\mathbb{C}$ is  computed using the single-particle eigenvalues and eigenfunctions of the single-particle Hamiltonian. 

\begin{figure}
    \centering
    \includegraphics[width=\linewidth]{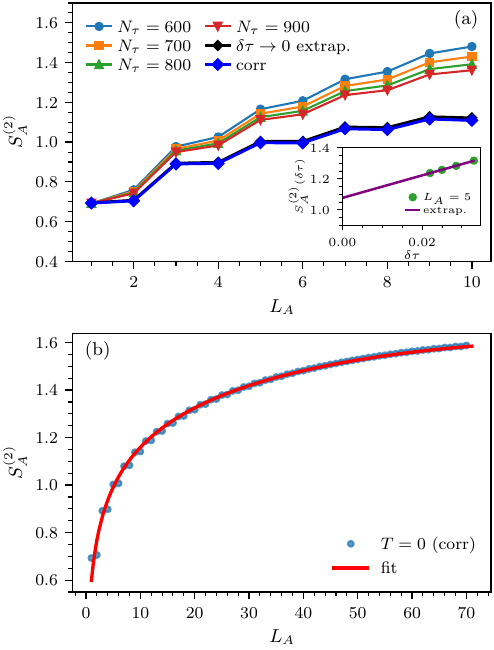}
    \caption{{\bf Subsystem second R\'{e}nyi entropy for non-interacting fermions:} (a) $S^{(2)}_{A_f}(\delta\tau)$ computed using the imaginary-time path integral method for nearest-neighbor tight-binding chain for different imaginary-time discretizations $\delta\tau$, the extrapolation to $\delta\tau \to 0$ and its comparison with the correlation matrix results. Inset: linear extrapolation to $\delta\tau\to 0$ for $L_A = 5$. 
    (b) $S^{(2)}_{A_f}$ obtained from the correlation matrix at $T=0$ for $L=200$, fitted to the conformal-field theory (CFT) formula [Eq. \eqref{eq:T_0_cft}] (solid line), yielding {$c = 0.97 \pm 0.01$}. 
    }
    \label{fig:SA2_NonInt_Benchmark}
\end{figure}

In comparison, the application of imaginary-time entanglement path integral formalism of Sec.\ref{sec:Ent_LargeN} to the non-interacting fermionic system \cite{Haldar} leads to
\begin{align}
 S^{(2)}_{A}&= -\ln{Z_{fA}^{(2)}}+2\ln{Z_f}\nonumber \\
 &= -\mathrm{Tr}\ln{\left(-\tilde{G}^{-1}_{0}\right)}+2\mathrm{Tr} \ln{\left(-G^{-1}_{0}\right)}, \label{eq:SA2_NonInt_PathInt}
\end{align}
where $Z_{fA}^{(2)}=\mr{Tr}_A\rho_A^2$ with $\rho_A=\mathrm{Tr}_B\rho$, and $G_0^{-1}$ and $\tilde{G}_0^{-1}$  are given in Eqs.\eqref{eq:invG0} and \eqref{eq:fermion_greenfn_kick}, respectively. To evaluate the above second R\'enyi entropy $S^{(2)}_{A}$, we discretize the imaginary-time interval $0 \leq \tau < \beta$ into $N_{\tau}$ time slices with step size $\delta\tau = \beta/N_{\tau}$. Using Eq.\eqref{eq:SA2_NonInt_PathInt}, we compute $S^{(2)}_{A}(\delta\tau)$ for a given discretization $\delta\tau$, by numerically evaluating the determinants of the $G_0^{-1}$ and $\tilde{G}_0^{-1}$, which are matrices in space ($r$), time ($\tau$) and/or replica indices ($\alpha=1,2$). Repetition of this calculation for different values of $\delta\tau$ and extrapolation to the limit $\delta\tau \to 0$ yield the result for $S^{(2)}_{A}$.
In Fig.\ref{fig:SA2_NonInt_Benchmark}(a), we plot $S^{(2)}_{A}(\delta\tau)$ for several values of $N_{\tau} = 600, 700, 800,$ and $900$, as a function of subsystem size $L_A$, for a system of size $L=30$ at temperature $T=0.05$. Then a linear extrapolation to $\delta\tau \to 0$ is done, as shown in Fig.\ref{fig:SA2_NonInt_Benchmark}(a).  Finally, we compare the extrapolated result $S^{(2)}_{A}(\delta\tau \to 0)$ with that obtained from the correlation matrix method [Eq.\eqref{eq:SA2_NonInt_CorrMat}]. The excellent agreement between the results from the time-discretized path integral and correlation matrix method can be seen in Fig.\ref{fig:SA2_NonInt_Benchmark}(a).

\subsubsection{Entanglement-to-entropy crossover, CFT formula and scaling ansatz} \label{sec:CFTCrossover_NonInt}

 Gapless 1D systems, e.g., a nearest-neighbor tight-binding fermionic chain, exhibit the logarithmic violation
of the area-law scaling of entanglement. These systems are
usually described by $1+1$D CFT characterized by some central charge $c$. For non-interacting fermions with two gapless chiral modes at the Fermi points, the R\'{e}nyi entropy $S^{(n)}_{A}$ at $T=0$ for a thermodynamically large system ($L\to\infty$) with periodic boundary condition is given by the CFT formula \cite{Cardy,Korepin} 
\begin{eqnarray}
    {S_{A}^{(n)}(L_A,T=0)=\dfrac{c}{6}\left(1+\dfrac{1}{n}\right) \ln{L_A} + b},
\label{eq:T_0_cft}
\end{eqnarray}
for $L_A\gg 1$. The logarithmic term above is universal with the central
charge $c$, and $b$ is a subleading nonuniversal constant originating from high-energy degrees of freedom.
At finite temperatures, the above CFT formula gets modified to
\begin{align}
    {S_{A}^{(n)}(L_A,T)=\dfrac{c}{6}\left(1+\dfrac{1}{n}\right) \ln\Big[\ell(T)\sinh{\left(\dfrac{ L_{A}}{\ell(T)}\right)}\Big] + b},
    \label{eq:finite_T_cft}
\end{align}
where $\ell(T)=v\beta/\pi$ is a thermal length (in units of lattice spacing) which diverges as $T\to0$ or equivalently, $1/\ell(T)$ vanishes as $T\to0$ limit. Here $v$ is a velocity, which is given by the Fermi velocity $v_\mr{F}$ for non-interacting fermions. At low temperature $T\to 0$, where $\ell(T)\gg L_A$, the expression of Eq.\eqref{eq:finite_T_cft}, reduces to the ground-state R\'{e}nyi entanglement entropy of Eq.\eqref{eq:T_0_cft}. On the other limit, $\ell(T)\ll L_A$, e.g., at high temperatures or very large subsystem sizes, one recovers the thermal R\'{e}nyi entropy,
\begin{align}
S_A^{(n)}(T)&\simeq \frac{c}{6\ell(T)}\left(1+\frac{1}{n}\right)L_A, \label{eq:SA2_Thermal}
\end{align}
that scales linearly with system size $L_A$.
The crossover function in Eq.\eqref{eq:finite_T_cft} therefore connects the universal part of the entanglement entropy to the low-temperature thermal entropy, since the same low-energy excitations control both in a gapless fermionic system with a Fermi surface \cite{SwingleSenthil,Korepin,Swingle2012}. Motivated by the CFT formula [Eq.\eqref{eq:finite_T_cft}] describing the entanglement-to-thermal entropy crossover \cite{SwingleSenthil},  we write the following general scaling ansatz for the subsystem second R\'{e}nyi entropy:
\begin{equation}
    S_{A}^{(2)}(L_A,T)=T^{\theta} f_{L}{\Bigg(\frac{L_A}{\ell(T)}\Bigg)}+d_{L}(T).
    \label{eq:scaling_ansatz}
\end{equation}
 Here $f_{L}(x)$ is the scaling function, which implicitly depends on the system size $L$; $d_{L}(T)$ represents the corrections to scaling, and $\ell(T)$ is the thermal length. Comparing with the CFT expression Eq.\eqref{eq:finite_T_cft} for $n=2$, we can identify $\theta=0$, $f_L(x)=(c/4)\ln[\sinh(x)]+b$, and $d_L(T)=(c/4)\ln\ell(T)$.
In our numerical scaling analysis, $f_{L}(x)$, $d_{L}(T)$ and $1/\ell(T)$ are taken as polynomials of their arguments for the range of finite temperatures ($T\neq 0$) that we consider. The coefficients of the polynomials are used as fitting parameters to collapse the data for $S_{A}^{(2)}(L_A,T)$ at different $L_A$ and $T$ into a single curve, and extract $f_L(x)$, $\ell(T)$ and $d_L(T)$ without relying on the CFT expression [Eq.\eqref{eq:final_renyi_entropy}]. As a result, the scaling ansatz of Eq.\eqref{eq:scaling_ansatz} is also used in the next section to analyze the fermionic second R\'{e}nyi entropy $S_{A_f}^{(2)}$ [Eq.\eqref{eq:second_renyi}] of the 1D Yukawa-SYK model, where there is no a priori validity of the CFT expression [Eq.\eqref{eq:finite_T_cft}]. Here, we note that $\ell(T)$ is only defined up to an overall scale factor $s_\ell$ within the argument of the scaling function due to its polynomial parameterization, where the scale factor $s_\ell$ can be absorbed in the redefinition of the coefficients of the polynomial.
The exponent $\theta$ can also be fixed more generally from the asymptotic behavior of the scaling function $f_{L}(x)$ for $L_A\gg \ell(T)$. In this limit, we must recover the thermal entropy, which scales linearly with $L_A$, implying $f_{L}(x \gg1) \sim x$. This implies $\theta=0$ since $S_A^{(2)}\propto 1/\ell(T)\propto T$ [Eq.\eqref{eq:SA2_Thermal}]. In the opposite limit, $\ell(T) \gg L_A$, i.e., $x \ll 1$), $f_{L}(x\ll 1) \sim \text{constant}$ modulo a logarithmic violation, $f_{L}(x\ll 1)\sim \ln{x}$ following Eq.\eqref{eq:T_0_cft}. 

\begin{figure}
    \centering
    \includegraphics[width=\linewidth]{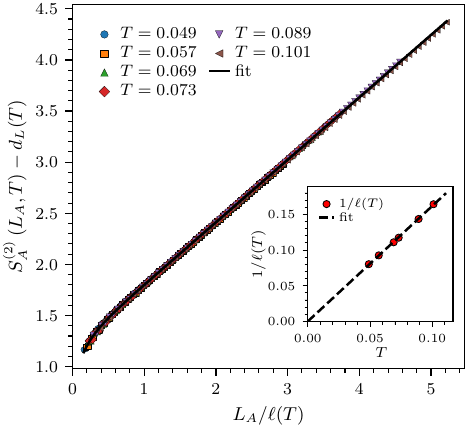}
    \caption{{\bf Scaling collapse of second Renyi entropy $S^{(2)}_{A}$ for non-interacting fermions:} $S_{A}^{(2)}(L_A,T)$ obtained from correlation matrix method for system size $L=200$ at different $T$ and subsystem size $L_A$ is collapsed into a single curve using the finite-size scaling ansatz [Eq.\eqref{eq:scaling_ansatz}]. We have subtracted the corrections to scaling from $S^{(2)}_{A_f}$ in the plot. The extracted scaling function is fitted (black solid line) with the CFT crossover formula [Eq.\eqref{eq:finite_T_cft}]. The extracted central charge for free fermions is {$c=0.97 \pm 0.01$}. (Inset) The inverse thermal length $1/\ell(T)$ is extracted from the scaling collapse as a function of $T$ and fitted with $1/\ell(T)=a_{\ell} T$.}
    \label{fig:SA2_Collpase_NonInt}
\end{figure}

To benchmark our scaling ansatz, we first perform a scaling analysis of $S^{(2)}_{A}$ computed using correlation matrix method [Eq.\eqref{eq:SA2_NonInt_CorrMat}] for $L=200$ for non-interacting tight-binding fermionic chain at low but finite temperatures ($T\neq 0$). The scaling ansatz of Eq.\eqref{eq:scaling_ansatz}
collapses the data for different subregion sizes $L_A$ and temperatures onto a single universal curve, as shown in Fig.\ref{fig:SA2_Collpase_NonInt}, thereby correctly capturing the crossover from entanglement entropy to thermal entropy for free fermions.
As shown in Fig.\ref{fig:SA2_Collpase_NonInt}, the resulting universal scaling curve is well fitted to the following CFT form,
\begin{align}
f_{L}\!\Bigg(\dfrac{L_A}{\ell(T)}\Bigg) &=S_A^{(2)}(L_A,T)-d_L(T)\nonumber \\
&=
\dfrac{c}{4}
\ln\!\Bigg[
\sinh\!\Big(s_\ell\dfrac{L_A}{\ell(T)}\Big)
\Bigg]
+
b,
\label{eq:eff_cft_fit}
\end{align}
where $\ell(T)$ extracted from the finite-size scaling collapse [Fig.\ref{fig:SA2_Collpase_NonInt}] is used. As a result, we account for the overall scale factor ($s_\ell$) in $\ell(T)$ by keeping $s_\ell$ as a fitting parameter along with $c$ and $b$ above. The scale factor has already been included in the thermal length shown in Fig.\ref{fig:SA2_Collpase_NonInt} (inset) through the redefinition, $\ell(T)/s_\ell\to \ell(T)$. The same procedure is followed for 1D Yukawa-SYK model in the next section.

From the fit to the universal scaling curve in Fig.\ref{fig:SA2_Collpase_NonInt}, we extract a central charge $c \simeq  0.97$, in agreement with $c$ extracted from the direct CFT fit [Eq.\eqref{eq:T_0_cft}] to the $T=0$ second R\'{e}nyi entanglement entropy for the same system size ($L=200$) in Fig.\ref{fig:SA2_NonInt_Benchmark}(b). The extracted $c$ is close to the expected CFT central charge of $c=1$ for $L\to\infty$. The deviation from $c=1$ could be attributed to the finite $L$. We have verified (not shown) that the extracted $c$ from CFT fit approaches $c\to 1$ with increasing $L$. {For the scaling collapse in Fig.\ref{fig:SA2_Collpase_NonInt}, we do not include the data of $S_A^{(2)}$ at $T=0$. This demonstrates that a $c$ consistent with the $T=0$ CFT fit [Fig.\ref{fig:SA2_NonInt_Benchmark}(b)] can be obtained from the finite-temperature entropy-to-entanglement crossover in $S_A^{(2)}(L_A,T)$ for sufficiently large $L$}. The extracted inverse thermal length $1/\ell(T)$ from the scaling analysis vanishes linearly with temperature $T$, as shown in the inset of Fig.\ref{fig:SA2_Collpase_NonInt}, following $\ell(T)=v_\mr{F}\beta/\pi$ in Eq.\eqref{eq:finite_T_cft}. 

With the above benchmarks established, in the following sections, we study the subsystem-size dependence and the universal crossover from entanglement-to-thermal entropy in the fermionic second R\'{e}nyi entropy for the 1D Yukawa-SYK model.


\subsection{Entanglement and its crossover to thermal entropy in 1D Yukawa SYK model}\label{sec:EntYSYK}

In this section, we present the results for fermionic subsystem second R\'{e}nyi entropy  $S^{(2)}_{A_f}$ [Eq.\eqref{eq:second_renyi}] in the 1D Yukawa-SYK model at the strange metallic quantum critical point (QCP), $\gamma/\gamma_c=1$ and away from criticality in the Fermi liquid regime, $\gamma>\gamma_c$ (Fig.\ref{fig:PhaseDiagram}). Due to difficulty with the numerical accuracy arising from a small thermal mass (Fig.\ref{fig:MT_thermal_mass}) in the self-consistency Eqs.\eqref{eq:self_consis_entan_replica} in the ordered phase ($\gamma<\gamma_c$) at low temperatures, we do not study the entanglement properties of the ordered phase in detail in this work. 

We solve the large-$N$ second R\'{e}nyi saddle-point equations [Eqs.\eqref{eq:self_consis_entan_replica}] self consistently by discretizing in imaginary time $\tau$ with a discretization step $\delta\tau=\beta/N_{\tau}$. See Supplemental Material, Sec.\ref{app:discretization} for details. To obtain $S^{(2)}_{A_f}$ from Eq.\eqref{eq:final_renyi_entropy}, we evaluate the second R\'{e}nyi saddle-point action $\mathcal{S}_{A_f}$ using the converged solutions of Eqs.\eqref{eq:self_consis_entan_replica}, and the equilibrium saddle-point action, $\mathcal{S}$ [Eq.\eqref{eq:LargeNAction}], from the corresponding solution of the equilibrium self-consistency Eqs.\eqref{eq:SaddlePoint}. Thus, we compute $S^{(2)}_{A_f}(\delta\tau)$ as a function of sub-system size $L_A$ for a given discretization $\delta\tau$. To access the continuum limit $\delta\tau \to 0$, we compute $S^{(2)}_{A_f}(\delta\tau)$ for several values of $\delta\tau$ and perform a linear extrapolation to $\delta\tau \to 0$. See Fig.\ref{fig:delta_tau}(a) in Appendix \ref{app:extrapolation}. In most of the calculations, we use three values of $\delta\tau$ in the range $0.017$--$0.03$ for this extrapolation. However, the linear extrapolation to $\delta\tau \to 0$ works well only for $\gamma \geq \gamma_c$. In contrast, for $\gamma < \gamma_c$, the extrapolation does not produce smooth curves for $S_{A_f}^{(2)}(L_A)$ in the $\delta\tau\to 0$ limit, presumably due to numerical issues related to small thermal mass (Fig.\ref{fig:MT_thermal_mass}) in the ordered phase. We show some results for $S^{(2)}_{A_f}(\delta\tau)$ for the smallest $\delta\tau$ accessed in Fig.\ref{fig:delta_tau}(b), Appendix \ref{app:extrapolation}, in the ordered phase.
For temperatures $T \lesssim 0.03$, computations with $\delta\tau\simeq 0.02$ become very demanding for system sizes $L\gtrsim 50$, as the size of the Green's function matrices in Eqs.\eqref{eq:self_consis_entan_replica}, $(2L N_{\tau} \times 2L N_{\tau})$, becomes very large (Supplemental Material, Sec.\ref{app:renyi_numerical_saddle}). Hence, we restrict our numerical computations to temperatures $T \gtrsim 0.03$. 
We implement an efficient recursive Green's function method for the solution of Eqs.\eqref{eq:self_consis_entan_replica} (Supplemental Material, Sec.\ref{app:recursive}) and to compute $S^{(2)}_{A_f}$ for system sizes $L=40-60$ and for the low-temperature range $T\simeq 0.05-0.12$. Most of our analysis, as discussed in this section, are for $L=50$. We show in Appendix \ref{app:system_size_c_l_0} that our main results do not depend on $L$ much for the range of $L=40-60$, accessed in our calculations.

\subsubsection{Scaling collapse of fermionic subsystem R\'{e}nyi entropy} 
We plot $S^{(2)}_{A_f}$ for $L=50$ as a function of $L_A<L/2$ at different temperatures in Figs.\ref{fig:SA2_Collpase_1DYSYK}(a) and (b) for the  strange metal at $\gamma/\gamma_c=1$ and the Fermi liquid at $\gamma/\gamma_c=1.84$. As evident from both of these figures, $S_{A_f}^{(2)}$ varies mostly linearly with $L_A$ at higher temperatures, indicating dominantly thermal entropic contribution. On the contrary, $S_{A_f}^{(2)}(L_A)$ deviates from linear behavior at lower temperatures, implying a crossover to entanglement.
We find that the data of $S^{(2)}_{A_f}$ for $\gamma\geq\gamma_c$ are well described by the scaling ansatz of Eq.\eqref{eq:scaling_ansatz} with $\theta \simeq 0$, with the asymptotic behaviors of the scaling function, $f_L(x\gg 1)\sim x$ and $f_L(x\ll 1)\sim \ln x$. The latter are evident in Figs.\ref{fig:SA2_Collpase_1DYSYK} (c) and (d), which show the corresponding scaling collapse of $S^{(2)}_{A_f}(L_A,T)$ using Eq.\eqref{eq:scaling_ansatz} in the strange metal and Fermi liquid regimes, respectively, with an inverse length scale $1/\ell(T)$ plotted in the insets. Remarkably, the extracted inverse length, $1/\ell(T)$, remains finite at $T = 0$, i.e., $1/\ell(T\to0)\to1/\ell_0$, for both non-Fermi liquid ($\gamma/\gamma_c=1$) and Fermi liquid ($\gamma/\gamma_c=1.84$) states. Since the $1/\ell(T)$ varies linearly with $T$, with a non-zero intercept $1/\ell_0$, we define
\begin{equation}
    \frac{1}{\ell(T)} = \frac{1}{\ell_1(T)} + \frac{1}{\ell_0},
    \label{eq:length_ysyk}
\end{equation}
where $1/\ell_1(T)= a_\ell T$ can be designated as an effective inverse thermal length related with the low-temperature thermodynamics of the 1D Yukawa-SYK model. The intercept $1/\ell_0$ can be interpreted as an inverse length scale associated with the local fermion-boson entanglement within the subregion $A$, as discussed below.

\subsubsection{Effective CFT-like description for crossovers between thermal entropy, and local and non-local entanglements}\label{sec:eff_cft_ysyk}

In Figs.\ref{fig:SA2_Collpase_1DYSYK} (c) and (d), the universal scaling curves $f_L(L_A/\ell(T))$ [Eq.\eqref{eq:scaling_ansatz}] for both the strange metal and Fermi liquid regimes are well fitted (solid black lines) by a CFT-like form of the scaling function
in Eq.\eqref{eq:eff_cft_fit}, but by replacing $c$ with a prefactor $c_{eff}$. The latter is denoted as \emph{effective central charge} for reference. Here $c_{eff}$, $s_\ell$ and $b$ are used as fitting parameters as in Sec.\ref{sec:CFTCrossover_NonInt}. From the fit of the universal scaling curves using the above form, we obtain $c_{\mathrm{eff}} \simeq 1.56$ for the strange metal ($\gamma/\gamma_c=1$), and $c_{\mathrm{eff}} \simeq 1.30$ for the Fermi liquid at $\gamma/\gamma_c=1.84$. The corresponding fermion-boson inverse length scales associated with intra-subregion entanglement are $1/\ell_0 \simeq 0.16$ and $1/\ell_0 \simeq 0.06$, respectively, as shown in Figs.\ref{fig:SA2_Collpase_1DYSYK}(c),(d) (insets). In these figures, $\ell(T)$ has been appropriately scaled, $\ell(T)/s_\ell\to\ell(T)$, with the $s_\ell$ extracted from the fit of the scaling curves [Figs.\ref{fig:SA2_Collpase_1DYSYK}(c) and (d)] to Eq.\eqref{eq:eff_cft_fit}. Motivated by the success of the CFT-like form [Eq.\eqref{eq:eff_cft_fit}] to describe the scaling curves in Figs.\ref{fig:SA2_Collpase_1DYSYK}(c) and (d), we have also directly fitted $S_{A_f}^{(2)}(L_A,T)$ curves in Figs.\ref{fig:SA2_Collpase_1DYSYK}(a) and (b) at a temperature $T$ using Eq.\eqref{eq:finite_T_cft} ($n=2$) with $c=c_{eff}$, $\ell$ and $b$ as fitting parameters. The CFT expression fits the data well, as discussed in Appendix \ref{app:CFTFit_YukawaSYK}. 

Given the numerically extracted form of $1/\ell(T)$ [Eq.\eqref{eq:length_ysyk}] from the scaling curves [Figs.\ref{fig:SA2_Collpase_1DYSYK}(c) and (d)], we can see that a single effective CFT-like form of Eq.\eqref{eq:eff_cft_fit} captures various crossovers among -- (1) non-local entanglement between fermions in $A$ subregion with fermionic and bosonic degrees of freedom in $B$ (Fig.\ref{fig:model}), (2) local intra-$A$ fermion-boson entanglement, and (3) thermal entropy. At zero temperature, in the limit $L_A\ll \ell_0$, Eq.\eqref{eq:eff_cft_fit} leads to a logarithmic scaling,
\begin{align}
S_{A_f}^{(2)}(L_A,T=0)\simeq \left(\frac{c_{eff}}{4}\right)\ln\left(\frac{L_A}{\ell_0}\right)+\cdots, \label{eq:SA2_Logarithmic}
\end{align}
presumably signifying non-local entanglement of the fermions in $A$ subsystem with the fermionic and bosonic degrees of freedom in the $B$ subsystem (Fig.\ref{fig:model}). However, evidently, this non-local entanglement is cut off by a length $L_A\simeq \ell_0$, and is obscured in the scaling curves [Figs.\ref{fig:SA2_Collpase_1DYSYK}(c) and (d)] by a large local volume-law entanglement between fermions and bosons in the same subregion $A$ for $L_A\gtrsim \ell_0$.

The local volumic entanglement can be directly inferred in the limit $L_A\gg \ell(T)$, where Eq.\eqref{eq:eff_cft_fit} leads to
\begin{align}
S_{A_f}^{(2)}(L_A,T)\simeq \frac{c_{eff}a_\ell}{4}TL_A+\frac{c_{eff}}{4\ell_0}L_A +\cdots. \label{eq:SA2_EntropyLocalEnt}
\end{align}
Here, the first term, that varies linearly with both $T$ and $L_A$, and dominates at higher temperatures, can be identified with the thermal second R\'{e}nyi entropy of the fermions in subsystem $A$. The second term above gives a linear-in $L_A$ contribution that remains non zero even at $T=0$. Thus, the second term can be recognized as the entanglement between fermions and bosons within the same subsystem $A$.

\begin{figure}
    \centering
    \includegraphics[width=\linewidth]{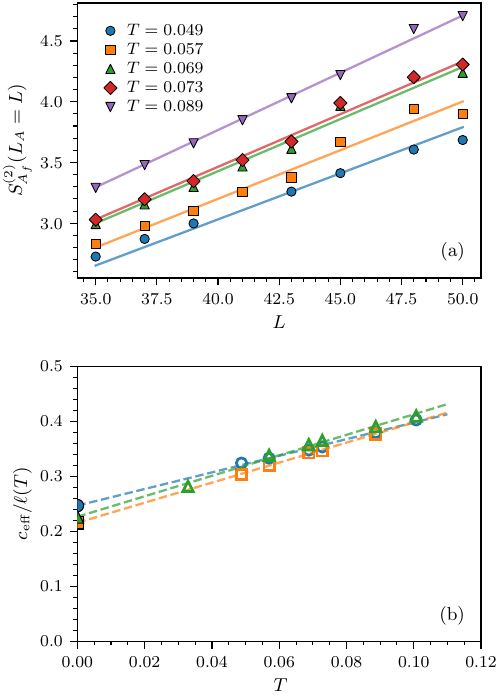}
    \caption{{\bf The second R\'{e}nyi entropy of fermions and fermion-boson entanglement:} (a) The second R\'{e}nyi entropy $S^{(2)}_f(L)=S_{A_f}^{(2)}(L_A=L)$ for strange metal ($\gamma/\gamma_c=1$) at different $T$ as a function of different system sizes $L$, fitted with $S^{(2)}_{A_f}(L,T)=[c_{eff}/4\ell(T)]L$. (b) The extracted $c_{eff}/\ell(T)$ ratio as a function of $T$ and its extrapolation $c_{eff}/\ell_0$ to $T=0$. We obtain $c_{eff}/\ell_0=0.215$ (filled orange square) from $S_f^{(2)}$, $c_{eff}/\ell_0=0.246$ (filled blue circle) from the fitting of scaling curves in Figs.\ref{fig:SA2_Collpase_1DYSYK}, and $c_{eff}/\ell_0=0.226$ (filled green triangle) from the direct CFT fits in Figs.\ref{fig:SA2_CFTFit}, Appendix \ref{app:CFTFit_YukawaSYK}.}
    \label{fig:SA2_EntropyLocalEnt}
\end{figure}

The fermion-boson entanglement can be deduced much more straightforwardly, without going through the route of fermionic subsystem R\'{e}nyi entropy $S_{A_f}^{(2)}(L_A,T)$ [Eq.\eqref{eq:second_renyi}] for $L_A<L/2$ and its scaling collapse (Figs.\ref{fig:SA2_Collpase_1DYSYK}). To this end, we compute the second R\'{e}nyi entropy of the fermions for the entire system of size $L$,
\begin{align}
S_f^{(2)}(L,T)=-\frac{1}{N}\mr{Tr}_f\rho_f^2, \label{eq:Thermal_Fermion2ndRenyi}
\end{align}
where $\rho_f=\mr{Tr}_\phi \rho$ is obtained by tracing out the bosonic degrees of freedom the entire system. By construction, $S_f^{(2)}(L,T)=S_{A_f}^{(2)}(L_A=L,T)$. Thus, $S_f^{(2)}(L,T)$ is computed using Eq.\eqref{eq:final_renyi_entropy} and the self-consistent solutions of the large-$N$ saddle-point Eqs.\eqref{eq:self_consis_entan_replica} for $L_A=L$. We expect $S_f^{(2)}(L,T)$ to follow Eq.\eqref{eq:SA2_EntropyLocalEnt} with $L_A$ replaced by $L$. We plot $S_{A_f}^{(2)}(L,T)$ as a function of $L$ for different temperatures in Fig.\ref{fig:SA2_EntropyLocalEnt}(a) {for the strange metal ($\gamma/\gamma_c=1$)}. Clearly, $S_{A_f}^{(2)}(L,T)$ varies linearly with $L$, as expected from Eq.\eqref{eq:SA2_EntropyLocalEnt}. We extract $c_{eff}/\ell(T)=c_{eff}(a_\ell T+1/\ell_0)$ from the slope of $S_{A_f}^{(2)}(L,T)$ with $L$ as a function of temperatures, as shown in Fig.\ref{fig:SA2_EntropyLocalEnt}(b). In the same graph, we also plot $c_{eff}/\ell(T)$ extracted from scaling analysis (Figs.\ref{fig:SA2_Collpase_1DYSYK}) and direct CFT fits (Figs.\ref{fig:SA2_CFTFit}, Appendix \ref{app:CFTFit_YukawaSYK}). We see that $c_{eff}/\ell(T)$, including its linear extrapolation $c_{eff}/\ell_0$ to $T\to 0$, obtained from $S_f^{(2)}(L,T)$ matches very well from those extracted from $S_{A_f}^{(2)}(L_A<L/2,T)$. Note that, in the latter, $c_{eff}$ and $\ell(T)$ can be separately estimated, whereas $S_f^{(2)}(L,T)$ only provides the combination $c_{eff}/\ell(T)$. Nevertheless, this analysis unambiguously establishes the contribution [the second term, Eq.\eqref{eq:SA2_EntropyLocalEnt}] of the volume-law entanglement between fermions and bosons to the fermionic-subsystem second R\'{e}nyi entropy $S_{A_f}^{(2)}(L_A,T)$. Similar results are expected for the Fermi liquid regime $\gamma>\gamma_c$. 

Thus, in summary, the second R\'{e}nyi entropy of fermions in a subsystem of 1D Yukawa SYK model exhibits crossovers between inter-subsystem logarithmic entanglement, intra-subsystem volume-law fermion-bosons entanglement and thermal entropy for both strange metal and Fermi liquid phase. Remarkably, the CFT-like formula of Eq.\eqref{eq:eff_cft_fit} with an effective central charge $c_{eff}$ and a length scale $\ell(T)$ with a finite value $\ell_0$ at $T=0$ captures all the above crossovers.

\begin{figure}
    \centering
    \includegraphics[width=\linewidth]{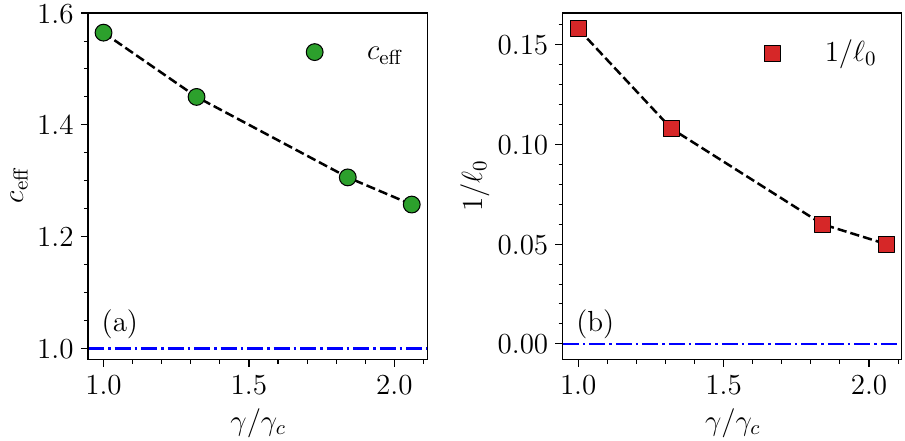}
    \caption{{\bf Effective central charge and fermion-boson entanglement length:} (a) The extracted effective central charge $c_{eff}$ from the universal scaling curves and their fits to CFT-like crossover formula as a function of $\gamma/\gamma_c$ for system size $L=50$. The CFT value $c=1$ for free fermions (blue dashed line) is also plotted for reference. (b) The extracted $T=0$ fermion-boson length $1/\ell_0$ as a function of $\gamma/\gamma_c$ for system size $L=50$. }
    \label{fig:ceff_ell0}
\end{figure}

We have carried out the above analyses for the two other values of $\gamma$ in the Fermi liquid phase and they give similar results. Summary of our main results are shown in Figs.\ref{fig:ceff_ell0}(a) and (b) in terms of $c_{eff}$ and $1/\ell_0$, extracted from scaling analysis of $S_{A_f}^{(2)}(L_A,T)$, as a function of $\gamma\geq \gamma_c$. We see that both $c_{eff}$ and $1/\ell_0$ monotonically decrease with $\gamma$ in the Fermi liquid phase. The increase of $1/\ell_0$, or the reduction of the length scale $\ell_0$, for $\gamma\to\gamma_c^+$ from the Fermi liquid side indicates that fermions and bosons get more strongly entangled locally increasing the intra-subsystem entanglement [Eq.\eqref{eq:SA2_EntropyLocalEnt}]. From the trend in Fig.\ref{fig:ceff_ell0}(b), we anticipate $1/\ell_0\to 0$ asymptotically for $\gamma\to \infty$, where the magnitude of the bosonic field vanishes due to the spherical constraint [Eq.\eqref{eq:SphericalConstraint}], and we expect to recover the non-interacting limit as in Fig.\ref{fig:SA2_Collpase_NonInt}(b)(inset).

The effective central charge $c_{eff}$ at $\gamma=\gamma_c$ is $\sim 60\%$ larger than CFT value of $c$ for the non-interacting fermions. It also remains substantially larger in the Fermi liquid phase, presumably approaching $c=1$ asymptotically for $\gamma\to\infty$. The larger effective central charge $c_{eff}$ indicates stronger fermion-boson entanglement. The enhanced $c_{eff}$ also suggests stronger non-local inter-subsystem entanglement which varies logarithmically with subsystem size [Eq.\eqref{eq:SA2_Logarithmic}], albeit for $L_A\lesssim \ell_0$. However, due to the later cut off, the logarithmic scaling, and hence $c_{eff}$, cannot be directly compared with logarithmic scaling of entanglement entropy in CFT [Eq.\eqref{eq:T_0_cft}] and its central charge. In CFT, the logarithmic behavior prevails for asymptotically large subsystem sizes $L_A$ at $T=0$. In contrast, the signature of such long-range entanglement in the fermionic-subsystem entanglement entropy $S_{A_f}^{(2)}(L_A,T=0)$ [Eq.\eqref{eq:second_renyi}]  in the Yukawa-SYK model is masked by the intra-subsystem volume-law fermion-boson entanglement beyond the length scale $\ell_0$. The length scale becomes quite short $\sim 10$ for $\gamma/\gamma_c\simeq 1$.

Nevertheless, a larger $c_{eff}$ in the 1D Yukawa-SYK model, even for the Fermi liquid states close to the QCP $\gamma=\gamma_c$, does indicate enhanced entanglement compared to the non-interacting limit or compared to the usual interacting 1D Luttinger liquids \cite{LAFLORENCIE20161}. As shown in {Fig.\ref{fig:system_size_c_l_0} in Appendix \ref{app:system_size_c_l_0}}, for the Fermi liquid state for $\gamma/\gamma_c=1.84$, we have verified that both $c_{eff}$ and $\ell_0$ more or less converge to the values shown in Figs.\ref{fig:ceff_ell0} ($L=50$) for the system sizes $L=40-60$. Thus, the enhanced $c_{eff}$, e.g., for a Fermi liquid state in the 1D Yukawa-SYK, model is presumably not a finite-size effect. We can qualitatively rationalize the increased $c_{eff}$ in the Fermi liquid regime based on the idea of thermal entropy-to-entanglement crossover, motivated by CFT. As discussed in Sec.\ref{sec:CFTCrossover_NonInt}, the central charge $c$ in the $T=0$ entanglement entropy [Eq.\eqref{eq:T_0_cft}] of CFT controls the linear-$T$ coefficient $\propto c a_\ell$ of the low-temperature entropy [Eq.\eqref{eq:SA2_Thermal}], where $a_\ell=\pi/v_\mr{F}$ for non-interacting fermions. 

In the same vein, as discussed in Sec.\ref{sec:LowTThermodynamics}, the low-temperature thermal entropy (per site, per flavor) in the Fermi liquid phase of the 1D Yukawa-SYK model is given by [Eqs.\eqref{eq:Sf},\eqref{eq:Sb_FL}],
\begin{align}
S(T)&\simeq \frac{1}{3}\left(\frac{\pi}{v_\mr{F}}+ \frac{\pi \kappa}{2M(0)v_b}\right)T\equiv \frac{1}{3}c_{eff}\tilde{a}_\ell T, \label{eq:LinearTCoeff_Entropy}
\end{align}
where the first term arises from the fermions and the second from (Ohmic) dissipative bosons, even when bosons are gapped [$M(0)\neq 0$]. The latter enhances the linear-$T$ coefficient of the entropy. We expect this enhancement to also reflect in $c_{eff}$ that controls the entanglement entropy. Thus, in Fig.\ref{fig:ThermalEntropy} of Appendix \ref{app:Thermodynamics}, we compare the coefficient $c_{eff}\tilde{a}_\ell$ extracted from thermal entropy with the $c_{eff}a_\ell$ [Eq.\eqref{eq:SA2_EntropyLocalEnt}] estimated from the fermionic-subsystem second R\'{e}nyi entropy [Fig.\ref{fig:SA2_Collpase_1DYSYK}(c),(d)]. We see that $c_{eff}\tilde{a}_\ell>\pi/v_\mr{F}$ is strongly enhanced compared to the non-interacting value $(\pi/v_\mr{F})$ approaching the QCP from the Fermi liquid side; $c_{eff}\tilde{a}_\ell$ tends towards the non-interacting value at larger $\gamma$, deep in the Fermi liquid regime. This is expected since the zero-temperature boson mass $M(0)$ [Eq.\eqref{eq:Mr0_DisorderedPhase}] increases with $\gamma$ in the Fermi liquid phase, reducing $c_{eff}\tilde{a}_\ell$ in Eq.\eqref{eq:LinearTCoeff_Entropy} towards the non-interacting value.
Curiously, $c_{eff}a_\ell$ [Eq.\eqref{eq:SA2_EntropyLocalEnt}] extracted from $S_{A_f}^{(2)}(L_A,T)$ remains close to the non-interacting value in Fig.\ref{fig:ThermalEntropy} (Appendix \ref{app:Thermodynamics}) for $\gamma\geq \gamma_c$, even though $c_{eff}$ is larger than the non-interacting value ($c=1$). In general, we do not expect $c_{eff}\tilde{a}_\ell$ and $c_{eff}a_\ell$ to match since thermal second R\'{e}nyi entropy of only the fermions $S_f^{(2)}(T)$ [Eq.\eqref{eq:Thermal_Fermion2ndRenyi}] and the thermal second R\'{e}nyi entropy $S^{(2)}(T)=-(1/N)\mathrm{Tr}(\rho^2)$ of the entire fermion-bosons system are not equal in general.


\section{Conclusions and discussions} \label{sec:Conclusion}

In this work, we have studied entanglement properties of fermions of a strange metal and Fermi liquids in the large-$N$ solvable 1D Yukawa-SYK model which hosts strongly coupled fermionic and bosonic degrees of freedom. The strange metallic state arises in this model due to the random Yukawa coupling of the gapless fermions at the Fermi points, and the bosons, which become critical at a QCP. The latter is tuned via a spherical constraint on the bosonic fields. We characterize the fermionic entanglement through the second R\'{e}nyi entropy of the fermions in a spatial subregion. The second R\'{e}nyi entropy of the fermionic subsystem is computed exactly in the large-$N$ limit through a imaginary-time coherent state path integral formalism for a system in a thermal ensemble at temperature $T$. We analytically and numerically obtain the spectral and thermodynamic properties of the fermions and bosons by solving a set of large-$N$ self-consistent saddle-point equations and relate some of these properties to the entanglement characteristics for $T\to 0$. 

We develop a universal scaling ansatz to analyze the fermionic second R\'{e}nyi entropy $S_{A_f}^{(2)}(L_A,T)$ as a function of subsystem size $L_A$ and $T$, and  extract the $T=0$ entanglement properties of fermions. Our main result is to unravel the significant role of the entanglement between fermions and bosons within the same spatial subregion in such a strongly coupled fermion-boson model. We find that a single CFT-like formula describes the universal scaling curve with a prefactor $c_{eff}$, and a length scale $\ell(T)$. The latter extrapolates to a non-zero value at $T=0$, unlike non-interacting fermions, and approaches a fermion-boson entanglement length scale $\ell_0$. The CFT-like formula captures all the crossovers between non-local inter-subregion entanglement $\sim c_{eff}\ln(L_A/\ell_0)$ for $L_A\lesssim \ell_0$, local intra-subregion fermion-boson entanglement $\sim c_{eff}L_A/\ell_0$ and thermal entropy $\sim c_{eff} TL_A $. For the strange metal and the Fermi liquid states close to the QCP, $c_{eff}$ is found to be substantially larger than the non-interacting value, the central charge $c=1$ of the 1+1 D CFT. The enhanced $c_{eff}$ can be rationalized based on the entanglement to entropy crossover, where the linear-$T$ coefficient of the thermal entropy acquires additional contribution from (Ohmic) dissipative bosons even when they are gapped in the Fermi liquid regime. The additional contribution also reflects in $c_{eff}$ which controls the entanglement. 

In the model considered throughout this work the disorder only enters through the interacting Yukawa-SYK term [Eq.\eqref{eq:H_fphi}], which leads to random inelastic scatterings. One could also incorporate on-site disorder \cite{patel_science,esterlisPRB, Li_2D_YSYK} by adding to $H_f$ [Eq.\eqref{eq:H_f}] a term $(1/\sqrt{N})\sum_{ir}v_{ijr}c_{ir}^\dagger c_{jr}$, where $v_{ijr}$ is a random complex intra-dot all-to-all flavor hopping at site $r$ with zero mean and finite variance. This preserves the large-$N$ solvability of the model. Such model has been extensively studied in 2D in the context of transport in strange metals \cite{patel_science,Li_2D_YSYK,esterlisPRB}. In that case, the random on-site disorder gives rise to an elastic scattering contribution to the fermion self-energy, whereas a marginal Fermi liquid term arises from the random Yukawa interactions. The elastic scattering is responsible for the $T=0$ residual resistivity, while the marginal Fermi-liquid self-energy leads to a linear-$T$ resistivity, in accordance with the phenomenology of strange metallic state in cuprates \cite{patel_science,Li_2D_YSYK,esterlisPRB,Sachdev2023,Guo2022,Guo2024,Patel2024,Patel2025,Lunts2025}. In our work for the 1D model here, we exclude the on-site disorder since we are primarily interested in the subsystem-size scaling of the entanglement entropy. As discussed in the Appendix \ref{app:diffusive_metal}, such disorder drives the system into a diffusive metallic phase in the large-$N$ limit, and gives rise to an elastic length scale \cite{potter,Haldar,Pouranvari2015AreaLaw} that cuts off the growth of the entanglement entropy with subsystem size $L_A$. As a result, even for gapless fermionic systems with a Fermi surface, the entanglement entropy approaches an area or boundary-law scaling $\sim L_A^{d-1}$ for large $L_A$.
 This contrasts, e.g., with a clean Fermi liquid, where a sharp Fermi surface gives rise to logarithmic violations of the boundary law. 

Explicit computation of the entanglement entropy of a strange metal and instances of deviation of the $c_{eff}$ from CFT central charge $c=1$ in 1D, or related coefficient in $d>1$ \cite{Klich,Swingle}, are rare in the literature. Small deviation of the prefactor to a $S_A\propto L_A \ln L_A$ variation in 2D from the coefficient expected from the Widom conjecture \cite{Widom,Klich,MathProof} for non-interacting fermions has been observed in variational quantum Monte Carlo (VQMC) computation of a Fermi liquid with Jastrow correlated wavefunction \cite{Tubman}. In contrast, the prefactor for a critical spin liquid with projected spinon Fermi sea has been found to be the same as that expected from Widom conjecture for a non-interacting Fermi sea \cite{ZhangGrover}. The Widom conjecture has been argued \cite{Swingle,SwingleSenthil,MuliDbosonization} to hold for interacting FL in $d>1$ based on the extension of 1+1D CFT arguments to higher dimensional Fermi surfaces. A large deviation ($\sim 100\%$) of the Widom coefficient has been found in the 2D non-Fermi liquid state with Fermi surface of composite fermions in electron liquids at high magnetic field with filling fraction $\nu=1/2$ \cite{Shao,Papic}. However, this deviation is presumably dependent on the microscopic details of the composite fermion wavefunction \cite{Motrunich,Papic}. 

The enhanced $c_{eff}$ in the fermionic subsystem R\'{e}nyi entropy $S_{A_f}^{(2)}$ is indicative of stronger non-local inter-subsystem entanglement $\sim c_{eff}\ln L_A$. However, in $S_{A_f}^{(2)}$ studied here, the logarithmic behavior is cut off for $L_A>>\ell_0$ by the fermion-boson entanglement length scale $\ell_0$. Hence, the logarithmic scaling and $c_{eff}$, cannot be directly compared with logarithmic scaling of entanglement entropy in CFT and its central charge. In CFT, the logarithmic behavior persists to asymptotically large subsystem sizes $L_A$ at $T=0$. In contrast, the signature of such long-range entanglement in the fermionic-subsystem entanglement entropy in the Yukawa-SYK model is obscured by the intra-subsystem volume-law fermion-boson entanglement beyond the length scale $\ell_0$. As a result, in future, it will be worthwhile to study the entanglement entropy of a full subsystem which contains both the fermionic and bosonic degrees of freedom in a spatial subregion and confirm that the logarithmic behavior $\sim c_{eff}\ln L_A$ ensues for asymptotically large subsystem sizes $L_A$. 

In the Yukawa-SYK model studied here, the fermions and bosons appear as separate degrees of freedom. However, often such models and other models of strange metals \cite{Lee2006,Sachdev_QPT,Sachdev2016Topological,Sachdev2016Topological,Watanabe2014Criterion} based on a QCP, are envisaged as effective low-energy models where the fermionic and bosonic degrees of freedom emerge from a purely fermionic models at higher energies, such as Hubbard or $tJ$ models in the context of high-temperature cuprate superconductors \cite{Lee2006,Norman2003Electronic}. It is thus interesting to understand the implications of the fermion-boson entanglement that arises at low energies, as in the 1D Yukawa-SYK model. It will be good to extend the large-$N$ entanglement computations to the marginal Fermi liquid ground state with critical Fermi surface in the 2D Yukawa-SYK model \cite{patel_science,esterlisPRB,Li_2D_YSYK}, which captures many salient phenomenologies of the strange metallic state around optimal doping in cuprates.

We end by noting that the measurements of entanglement entropy in solid-state materials remain an open challenge, even though the second R\'{e}nyi entanglement entropy has been measured in strongly interacting cold atomic optical lattice systems \cite{Islam2015}. Nevertheless, recently there have been exciting new developments \cite{Balut2025,Mazza2026} in the experimental measurements of entanglement witnesses, like quantum Fisher information \cite{Hauke2016}, for the putative strange metallic states in cuprates and heavy fermion systems. Thus, it would be fruitful to explore the possible signatures of emergent fermion-boson entanglement in such experimentally measurable entanglement witnesses. 

\begin{acknowledgements}
We acknowledge useful discussions with Sriram Ganeshan and Mohit Randeria. SB acknowledges support from ANRF, DST, India (File No. ANRF/ARG/2025/004045/PS).
SS was supported by the US National Science Foundation Grant DMR-2245246.
\end{acknowledgements}

\appendix
\section{Large-$N$ theory and saddle-point equations for 1D Yukawa-SYK model}\label{app:LargeN}
We write down the imaginary-time coherent-state path integral for the thermal partition function $Z=\mr{Tr}[e^{-\beta \mc{H}}]$ of the 1D Yukawa-SYK model [Eq.\eqref{eq:Model}] for a given realization of the random Yukawa-SYK couplings $\{g_{ijlr}\}$,
\begin{subequations}
\begin{align}
Z[g]&=\int \mc{D}(\bar{c},c)\mc{D}\phi e^{-\mc{S}[\bar{c},c,\phi]},
\end{align}
where
\begin{align}
\mathcal{S} &= \int_0^\beta d\tau  
[\sum_{ir} \{\bar{c}_{ir} (\partial_\tau - \mu) c_{ir}
+ \frac{1}{2} (\partial_\tau \phi_{ir})^2\}+ \mathcal{H}(\bar{c}, c, \phi)]
\end{align}
\end{subequations}
is the imaginary-time action. We perform the average over $\{g_{ijlr}\}$ using the usual replica trick, $\overline{\ln Z}=\lim_{m\to 0}(\overline{Z^m}-1)/m$. Following standard procedure in the large-$N$ SYK-type models (see, e.g., Refs.\cite{Gu2017SYKChain,BanerjeeAltman}), we further introduce the large-$N$ local-in-space and bi-local-in-time fields,
\begin{align}
 G_{r,ab} (\tau,\tau') &= \dfrac{1}{N} \sum_{i} \bar{c}_{ir b}(\tau') c_{ir a}(\tau) \\
    D_{r, a b} (\tau,\tau') &= \dfrac{1}{N} \sum_{l} \phi_{l r a}(\tau) \phi_{l r b}(\tau')  
\end{align}
and two conjugate fields $\Sigma$ and $\Pi$, to obtain disorder-averaged replicated partition function as
\begin{align}
\overline{Z^m}&\int\mc{D}(\bar{c},c)\mc{D}\phi \mc{D}(G,\Sigma,D,\Pi) e^{-S[\bar{c},c,\phi,G,\Sigma,D,\Pi]}. 
\end{align}
Here the replicated action is given by
\begin{widetext}
\begin{align}
\mathcal{S}= & \int d\tau d\tau'\sum_{irr',ab}\bar{c}_{ira}(\tau)\left[-\left(G_{0,rr'}^{-1}(\tau,\tau')\delta_{ab}-\Sigma_{r,ab}(\tau,\tau')\delta_{rr'}\right)\right]c_{ir'b}(\tau') \nonumber \\
&+\frac{1}{2}\int d\tau d\tau'\sum_{irr',ab}\phi_{ira}(\tau)\left[D_{0,rr'}^{-1}(\tau,\tau')\delta_{ab}-\Pi_{r,ab}(\tau,\tau')\delta_{rr'}\right]\phi_{ir'b}(\tau')\nonumber\\
 & +N\int d\tau d\tau'\sum_{r,ab}\Bigg[-\Sigma_{r,ba}(\tau',\tau)G_{r,ab}(\tau,\tau')+\frac{1}{2}\Pi_{r,ba}(\tau',\tau)D_{r,ab}(\tau,\tau') +\frac{g^{2}}{2}G_{r,ab}(\tau,\tau')G_{r,ba}(\tau',\tau)D_{r,ab}(\tau,\tau')\Bigg] \nonumber\\ & -\frac{N}{2}\int d\tau \sum_{ra}\mathrm{i}\lambda_{ra}(\tau)\left(\frac{1}{N}\sum_{i}\phi_{ira}^{2}(\tau)-\frac{1}{\gamma}\right),
\end{align}
\end{widetext}
where $a,b=1,\cdots,m$ are indices for $m$ replicas. The Lagrange multiplier field $\lambda_{ra}(\tau)$ enforces the spherical constraint [Eq.\eqref{eq:SphericalConstraint}] in the path integral. The non-interacting inverse propagators $G^{-1}_0$ and $D^{-1}_0$ are given in Eqs.\eqref{eq:invG0} and \eqref{eq:invD0}, respectively. We include the possibility of $O(N)$ symmetry breaking for the bosonoic field by separating the ordered component from the transverse fluctuations as in Eq.\eqref{eq:O_N_symm}. Subsequently, we integrate out the fermionic $(\bar{c},c)$ and transverse components of the bosonic $\phi$ fields and take a replica symmetric and diagonal ansatz, $G_{r,ab}(\tau,\tau')=G_r(\tau,\tau')\delta_{ab}$, and similarly for $\Sigma,~D,~\Pi$, with $\lambda_{ra}(\tau)=\lambda_r(\tau)$ to obtain $\overline{Z^m}=\int \mc{D}(G,\Sigma,D,\Pi)e^{-m\mc{S}[G,\Sigma,D,\Pi]}$ and the action of Eq.\eqref{eq:LargeNAction}. We extremize the action with respect to $G_r(\tau,\tau')$, $\Sigma_r(\tau,\tau')$, $D_r(\tau,\tau')$, $\Pi_r(\tau,\tau')$, $\lambda_r(\tau)$ and $r_0$ to obtain large-$N$ saddle-point equations, 
\begin{subequations}
\begin{align}
G^{-1}_{rr'}(\tau,\tau')&=G^{-1}_{0,rr'}(\tau,\tau')-\Sigma_r(\tau,\tau')\delta_{rr'},\\
D^{-1}_{rr'}(\tau,\tau')&=D^{-1}_{0,rr'}(\tau,\tau')-\Pi_r(\tau,\tau')\delta_{rr'},\\
\Sigma_r(\tau,\tau')&=g^2G_r(\tau,\tau')D_r(\tau,\tau'),\\
\Pi_r(\tau,\tau')&=-g^2G_r(\tau,\tau')G_r(\tau',\tau),\\
D_r(\tau,\tau)&+r_0^2=\frac{1}{\gamma},\\
r_0\int d\tau d\tau'&\sum_{rr'} D^{-1}_{rr'}(\tau,\tau')=0,
\end{align}
\end{subequations}
where $G_{rr}(\tau,\tau')=G_r(\tau,\tau')$ and $D_{rr}(\tau,\tau')=D_r(\tau,\tau')$ are the local propagators. Further, assuming translation invariance in space and time, we obtain the large-$N$ saddle-point Eqs.\eqref{eq:SaddlePoint}, that we solve numerically, and analytically at low energies in Sec.\ref{sec:SaddlePoint}.

\section{Low-energy solution of the large-$N$ saddle-point equations}\label{app:LowEnergySol}
We first obtain the local fermion Green's function from Eq.\eqref{eq:G_iwn} by converting the $k$ summation into an integral with cutoff $\Lambda$, and expanding the energy dispersion as $\epsilon_k-\mu=\pm v_\mr{F}k$ around the Fermi points $\pm k_\mr{F}$. This leads to,
\begin{align}
G(\mr{i}\omega_n)&\simeq 2\int_{-\Lambda}^{\Lambda}\frac{dk}{2\pi}\frac{1}{\mr{i}\omega_n-v_\mr{F}k-\Sigma(\mr{i}\omega_n)}\nonumber \\
&=-\frac{1}{v_\mr{F}}\ln{\frac{a+\mr{i}b-v_\mr{F}\Lambda}{a+\mr{i}b+v_\mr{F}\Lambda}}\nonumber \\
&\simeq -\frac{\ci}{v_\mr{F}}\mr{sgn}(\omega_n),~~~~~\Lambda v_\mr{F}\gg |\omega_n|,|\Sigma(\ci \omega_n)|. \label{eq:Glocal_iwn}
\end{align}
In the above, $a=-\mr{Re}\Sigma(\ci \omega_n)$ and $b=\omega_n-\mr{Im}\Sigma(\omega_n)$, with $\mr{sgn}(b)=\mr{sgn}(\omega_n)$ due to the analytical properties of the self-energy $\Sigma$. This is used to evaluate the bosonic self energy from Eq.\eqref{eq:Pi_tau}, which upon Fourier transformation gives $\Pi(\ci \Omega_n)=-g^2T\sum_n G(\ci \omega_n+\ci \Omega_m)G(\ci \omega_n)$. As a result, using Eq.\eqref{eq:Glocal_iwn}, we obtain at low temperatures $T\ll \Lambda v_\mr{F}$,
\begin{subequations}\label{eq:Pi_LowT}
\begin{align}
 \tilde{\Pi}&(\ci \Omega_m)=\Pi(\ci \Omega_m)-\Pi(\ci \Omega_m=0) \nonumber \\
 &\simeq \frac{g^2T}{v_\mr{F}^2}\sum_n \left[\mr{sgn}(\omega_n+\Omega_m)\mr{sgn}(\omega_n)-1\right],~~~|\Omega_m|\ll \Lambda v_\mr{F} \nonumber \\
 &=-\kappa |\Omega_m|, \\
 \Pi&(\ci \Omega_m=0)\equiv \Pi(0)=-g^2T\sum_n G(\ci \omega_n)^2\simeq \kappa \Gamma 
\end{align}
\end{subequations}
with $\kappa=g^2/\pi v_\mr{F}^2$. As a result, the bosonic propagator is given by $D(q,\ci \Omega_m)\equiv \tilde{D}(q,\ci \Omega_m)+\beta L\delta_{q,0}\delta_{m,0}r_0^2$, with the connected part,
\begin{align}
\tilde{D}(q,\ci \Omega_m)&=\frac{1}{\Omega_m^2+\omega_q^2+M^2(T)+\kappa |\Omega_m|}, 
\end{align}
and the boson thermal mass defined in Eq.\eqref{eq:ThermalMass}. The local bosonic propagator of Eq.\eqref{eq:D_iwn_cont} is obtained from Eq.\eqref{eq:D_iwn} by performing $q$ summation in the continuum approximation $\omega_q^2\simeq v_b^2 q^2$ for bosonic dispersion, i.e.,
\begin{align}
\tilde{D}(\ci \Omega_m)&=\int_{-\Lambda}^\Lambda \frac{dq}{2\pi} \frac{1}{\Omega_m^2+v_b^2q^2+M^2(T)+\kappa|\Omega_m|}\nonumber \\
&\simeq \frac{1}{2v_b}\frac{1}{\sqrt{\Omega_m^2+M^2(T)+\kappa|\Omega_m|}} \label{eq:LocalRegPropagator}
\end{align}
in the low-energy limit $v_b\Lambda\gg \sqrt{\Omega_m^2+M^2(T)+\kappa|\Omega_m|}$. The above bosonic propagator along with the Eq.\eqref{eq:Constraint_Cont} for spherical constraint and Eq.\eqref{eq:r0} for the order parameter correspond to an effective large-$N$ dissipative $O(N)$ model \cite{Sachdev_book}. The self-consistently generated dissipative self-energy term leads to long-range coupling in the time direction and lead to a $T=0$ ordered state with $r_0(T=0)\neq 0$ for $\gamma<\gamma_c$. We numerically solve Eqs.\eqref{eq:Constraint_Cont} and \eqref{eq:r0} self-consistently for $M^2(\gamma,T)$ and $r_0(\gamma,T)$ as a function of $\gamma$ and $T$ for a fixed dissipation strength $\kappa\simeq 0.08$ [Eq.\eqref{eq:BosonPolarization}], that corresponds to {$t=1,~\mu=0,~K=1$ and $g=1$} [Eq.\eqref{eq:Model}]. Since, $M^2(T)\neq 0$ at any finite $T$, $r_0(T\neq 0)=0$. The boson thermal mass obtained from the above effective dissipative $O(N)$ model agrees well [Fig.\ref{fig:MT_thermal_mass}] with the $M^2(T)$ obtained from the self-consistent solution of the full large-$N$ saddle-point Eqs.\eqref{eq:SaddlePoint} of the 1D Yukawa-SYK model.

\subsection{Boson mass and low-temperature phase diagram}\label{app:ThermalMass}

Since the effective dissipative $O(N)$ model provides a good description of the low-temperature properties of the bosonic degrees of freedom in the 1D Yukawa-SYK model, we first use Eqs.\eqref{eq:Constraint_Cont} and \eqref{eq:r0} for a fixed $\kappa$ to obtain the zero-temperature properties.  
At zero temperature $T=0$, we convert Matsubara summations in Eq.\eqref{eq:Constraint_Cont} to an integral
with a frequency cutoff $\Gamma$, 
and get 
\begin{align}
\tilde{D}(\tau=0)&=  \frac{1}{2v_b}\int_{-\Gamma}^{\Gamma}\frac{d\Omega}{2\pi}\frac{1}{\sqrt{\Omega^{2}+M^{2}(0)+\kappa|\Omega|}} \nonumber\\
&=  \frac{1}{2\pi v_b}\ln\left[\frac{\kappa+2\Gamma+2\sqrt{M^{2}+\kappa\Gamma+\Gamma^{2}}}{2M+\kappa}\right],\nonumber \\
\overset{\Gamma\gg M,\kappa}{\approx} & \frac{1}{2\pi v_b}\ln\left[\frac{4\Gamma}{2M(0)+\kappa}\right]=\dfrac{1}{\gamma}-r_0^2. \label{eq:LowEConstraintEq}
\end{align}
From the above and $r_0M^2(0)=0$, we obtain the $M(0)$ and $r_0$ of Eq.\eqref{eq:Mr0_DisorderedPhase} for $\gamma$ greater than a critical value $\gamma_c$, given in Eq.\eqref{eq:gammac}, and at the QCP ($\gamma=\gamma_c$) where both $M(0)$ and $r_0(0)$ vanish [Eq.\eqref{eq:Mr0_critical}]. For $\gamma<\gamma_c$, we get
\begin{align}
r_{0}^{2}+\frac{1}{2\pi v_b}\ln\left[\frac{4\Gamma}{\kappa}\right]\simeq  \frac{1}{\gamma}
\end{align}
with $M(0)=0$, corresponding to the zero-temperature ordered phase, as in Eq.\eqref{eq:Mr0_OrderedPhase}. 

Next, we compute boson thermal mass $M(T)\neq 0$ [Eq.\eqref{eq:ThermalMass}] at finite temperatures, where $r_0(T)=0$ for all $\gamma$. After performing the $q$ summation through integration as in Eq.\eqref{eq:LocalRegPropagator}, we rewrite the constraint of Eq.\eqref{eq:SaddleConstraint} as
\begin{equation}
\begin{aligned}
D(\tau=0)= & \frac{T}{2v_b}\sum_{m}\frac{1}{\sqrt{\Omega_{m}^{2}+\kappa|\Omega_{m}|+M^{2}(T)}}=\dfrac{1}{\gamma}. \label{eq:FiniteT_Constraint}
\end{aligned}
\end{equation} 
At low $T$, we approximate the Matsubara summation above using the Euler-MacLaurin low-temperature expansion,
\begin{subequations}
\begin{align}
\sum_{m=a}^{b}f(m)\approx & \int_{a}^{b}dm f(m)+\frac{1}{2}\left(f(a)+f(b)\right) \nonumber \\ 
&+\sum_{k=1}^{\infty} \frac{B_{2k}}{(2k)!}\left(f^{(2k-1)}(b)-f^{(2k-1)}(a)\right), \label{eq:EulerMaclaurin}
\end{align}
where $B_{2k}$s are the Bernoulli numbers, e.g., $B_2=1/6$, and $f^{(2k-1)}(m)$ refers to the $(2k-1)$-th derivative of the function $f(m)$, which is given by $f(m)=(T/2v_b)(\Omega_m^2+\kappa|\Omega_m|+M^2(T))^{-1/2}$ in Eq.\eqref{eq:FiniteT_Constraint}. The same result can be obtained by using Poisson summation formula in Eq.\eqref{eq:FiniteT_Constraint} and making a low-temperature expansion. The low-temperature expansion of Eq.\eqref{eq:EulerMaclaurin} leads to,
\begin{align}
D(\tau=0)\overset{\Gamma\gg M,\kappa}{\simeq} & \frac{1}{2\pi v_b}\ln\left(\frac{4\Gamma}{2M+\kappa}\right)+\frac{\pi\kappa}{12M^{3}c}T^{2}+\cdots \nonumber \\
&=\frac{1}{\gamma}. \label{eq:LowTConstraintEq}
\end{align}
\end{subequations} 
The first term on the right hand side of the first line corresponds to the zero-temperature contribution obtained earlier in Eq.\eqref{eq:LowEConstraintEq}, while the $T^{2}$ term represents the leading thermal correction.
For $\gamma>\gamma_{c}$, in the disordered phase where $M(0)\neq0$, we get 
\begin{align}
M^{2}(T)\simeq & M(0)^{2}+\left(M(0)+\frac{\kappa}{2}\right)\frac{\pi^{2}\kappa}{3M(0)^{2}}T^{2}. \label{eq:MT_DisorderedPhase}
\end{align}
As evident from above, the above expansion is valid below a temperature scale $T\ll (M^{3}/\kappa)^{1/2}$.

At the critical point, $\gamma=\gamma_{c}$ [Eq.\eqref{eq:gammac}], we obtain
\begin{equation}
    M^{2}(T)\simeq  \frac{\pi\kappa}{2\sqrt{3}}T. \label{eq:MT_Critical}
\end{equation}
The above is valid below a temperature scale $T\ll \kappa$. For $\gamma<\gamma_{c}$, which corresponds zero-temperature ordered phase with $M(0)=0$,
using Eq.\eqref{eq:LowTConstraintEq}, we can estimate a thermal mass $M^2(T)\sim T^{4/3}$, close to the critical point ($\gamma\lesssim \gamma_c$). However, this does not correspond to a self-consistent solution, since the Euler-MacLaurin expansion of Eq.\eqref{eq:EulerMaclaurin} is not convergent for $M^2(T)\sim T^{4/3}$. This can be deduced by looking at the higher order term in the Euler-MacLaurin expansion of Eq.\eqref{eq:EulerMaclaurin}, which diverges as $\sim T^{2k}/M^{4k-1}$ for $M\to 0$. Thus, for the thermal mass, $M(T)\sim T^\alpha$, vanishing as a power law with exponent $\alpha$ as $T\to 0$, one needs $\alpha\leq 2k/(4k-1)$ for all $k$, i.e., $\alpha\leq 1/2$. This consideration implies that the low-temperature expansion [Eq.\eqref{eq:EulerMaclaurin}] is convergent for the disordered phase ($\gamma>\gamma_c$), where $M(T)\sim T^0$ [Eq.\eqref{eq:MT_DisorderedPhase}], and is marginal at the critical point ($\gamma=\gamma_c$) with $M(T)\sim T^{1/2}$ [Eq.\eqref{eq:MT_Critical}]. However, the low-order expansion in Eq.\eqref{eq:LowTConstraintEq} produces a non-convergent solution, $M(T)\sim T^{2/3}$, for $\gamma<\gamma_c$. 

\begin{figure}
     \centering
     \includegraphics[width=\linewidth]{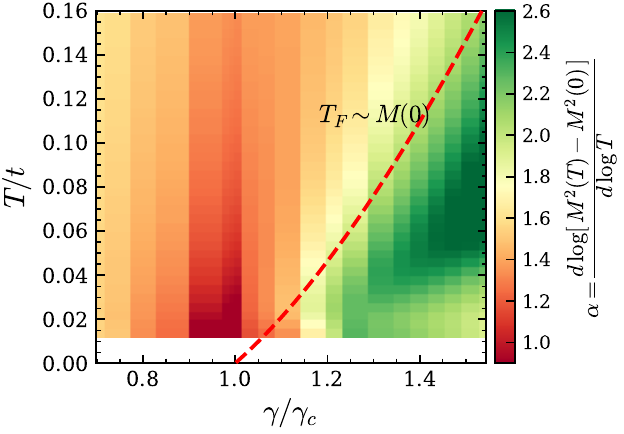}
     \caption{\textbf{Phase diagram of dissipative $O(N)$ model:} The $\gamma-T$ phase diagram of the effective 1D dissipative $O(N)$ represented by the logarithmic derivative $\alpha\equiv d\ln[M^2(T)-M^2(0)]/d\ln T$ (color). The analytically estimated Fermi liquid scale $T_F\sim M(0)$ [Eq.\eqref{eq:Mr0_DisorderedPhase}] is shown as a reference.}
     \label{fig:O_N_model_phase}
 \end{figure}

From Eq.\eqref{eq:FiniteT_Constraint}, we can also obtain a \emph{classical-like} solution by keeping only the contribution from the zeroth Matsubara frequency ($\Omega_m=0$). This leads to 
\begin{align}
M(T)\simeq \frac{\gamma T}{2 v_b}.
\end{align}
Considering the neglected contribution from non-zero Matsubara frequencies, we can see that the above can be a good approximation for $T\ll 8\pi v_b^2 \kappa/3\gamma^2$. Our numerical results for $M^2(T)$ in the 1D Yukawa-SYK and the dissipative $O(N)$ models in Fig.\ref{fig:MT_thermal_mass} are consistent with the analytically estimated $M^2(T)$ for $\gamma\geq \gamma_c$ [Eqs.\eqref{eq:fermion_density_matrix}, \eqref{eq:MT_Critical}]. The numerical results also suggest a power-law, $M^2(T)\sim T^\alpha$, for $\gamma<\gamma_c$ at low temperatures. The exponent increases with decreasing $\gamma$ from $\alpha\approx 4/3$, close to the critical point ($\gamma\lesssim \gamma_c$), to $\alpha\approx 2$ deep in the $T=0$ ordered phase. As an underpinning of the schematic phase diagram of Fig.\ref{fig:PhaseDiagram}, and to provide an overall picture of the exponent associated with $M^2(T)$ across the $\gamma-T$ phase diagram, in Fig.\ref{fig:O_N_model_phase}, we plot the numerical logarithmic derivative $d\ln[M^2(T)-M^2(0)]/d\ln T$ as an approximation for the exponent $\alpha$ for the effective $O(N)$ model. Here $M^2(0)$ has been estimated from the numerically obtained $M^2$ at the lowest temperature $T=0.005$ accessed. As evident, $\alpha\approx 2$ at low temperature in the Fermi liquid phase and $\alpha$ approaches $\approx 1$ for $\gamma\to \gamma_c$, consistent with the analytical expectations [Eqs.\eqref{eq:MT_DisorderedPhase}, \eqref{eq:MT_Critical}]. For $\gamma<\gamma_c$, $\alpha$ varies from $\approx 1.2-1.3$ close to the QCP to $\approx 1.8$ deeper ($\gamma/\gamma_c\sim 0.7$) inside the $T=0$ ordered phase.
\subsection{Fermionic Self energy}\label{app:SelfEnergy}
From Eq.\eqref{eq:SaddleSigma}, the fermion self energy can be written as 
\begin{equation}
    \Sigma(\mathrm{i}\omega_n)= g^{2}r_{0}^{2}G(\mathrm{i}\omega_{n})+g^{2}\frac{1}{\beta}\sum_{m}G(\mathrm{i}\omega_{n}+\mathrm{i}\Omega_{m})\tilde{D}(\mathrm{i}\Omega_{m}). \label{eq:Sigma_Matsubara}
\end{equation}
At $T=0$, for $\gamma>\gamma_c$, using Eqs.\eqref{eq:Glocal_iwn} and \eqref{eq:LocalRegPropagator}, we obtain
\begin{subequations}
\begin{align}
\Sigma&(\mathrm{i}\omega)\simeq  -\frac{\mathrm{i}g^{2}}{4\pi v_{F}v_b}\int_{-\Gamma}^{\Gamma}\frac{\mathrm{sgn}(\omega+\Omega)}{\sqrt{\Omega^{2}+\kappa|\Omega|+M(0)^{2}}}d\Omega\nonumber \\
= & \frac{\mathrm{-i}g^{2}}{2\pi v_{F}v_b}\ln\left[\frac{\kappa+2\omega+2\sqrt{\omega^{2}+\kappa\omega+M(0)^{2}}}{2M(0)+\kappa}\right]\mathrm{sgn}(\omega) \label{eq:Sigma_ZeroT} \\
\overset{\omega\to0}{\simeq} & -\frac{\mathrm{i}g^{2}}{2\pi v_{F}v_b}\left(\frac{\omega}{M(0)}-\frac{\kappa}{4M(0)^{3}}\omega^{2}\right)\mathrm{sgn}(\omega)
\end{align}
\end{subequations}
Analytically continuing to the real frequency $\mr{i}\omega \to \omega+\mr{i}0^+$, we obtain the retarded self energy at low energies,
\begin{equation}
    \Sigma^{R}(\omega)
\simeq-\frac{g^{2}}{2\pi v_{F}v_b}\left(\frac{\omega}{M(0)}+\mathrm{i}\frac{\kappa}{4M(0)^{3}}\omega^{2}\right), \label{eq:Sigma_FL_ZeroT}
\end{equation}
which corresponds to a Fermi liquid.

At the critical point ($\gamma=\gamma_c$) with $M(0)=0$, we obtain from Eq.\eqref{eq:Sigma_ZeroT},
\begin{align}
\Sigma(\mathrm{i}\omega)\simeq & -\frac{\mathrm{i}g^{2}}{2\pi v_{F}v_b}\ln\left[\frac{\kappa+2\omega+2\sqrt{\omega^{2}+\kappa\omega}}{\kappa}\right]\mathrm{sgn}(\omega) \\
\overset{\omega\to0}{\approx} & -\frac{\mathrm{i}g^{2}}{\pi v_{F}v_b\sqrt{\kappa}}\sqrt{|\omega|}\mathrm{sgn}(\omega).
\end{align}
The above leads to upon analytic continuation,
\begin{equation}
    \Sigma^{R}(\omega)\simeq -\frac{g^{2}}{\pi v_{F}v_b\sqrt{\kappa}}\sqrt{\omega}e^{\mathrm{i}\pi/4}.\label{eq:Sigma_ZeroTCritical}
\end{equation}
a non-Fermi liquid self energy correponding to the strange metalic state at QCP in the 1D Yukawa-SYK model.

The self energy has a similar form in the zero-temperature ordered phase, $\gamma<\gamma_c$, where also $M(0)=0$,
\begin{equation}
  \Sigma^{R}(\omega)\simeq  -\frac{\mathrm{i}g^{2}r_{0}^{2}}{v_{F}}-\frac{g^{2}}{\pi v_{F}v_b\sqrt{\kappa}}\sqrt{\omega}e^{\mathrm{i}\pi/4},\label{eq:Sigma_ZeroTOrdered}
\end{equation}
except the first term on the right hand side of Eq.\eqref{eq:Sigma_Matsubara} that leads to an additional effective elastic scattering due to the non-zero order parameter $r_0\neq 0$.
The non-zero order parameter leads to an effective elastic scattering at $T=0$.

\begin{figure}
    \centering
\includegraphics[width=\linewidth]{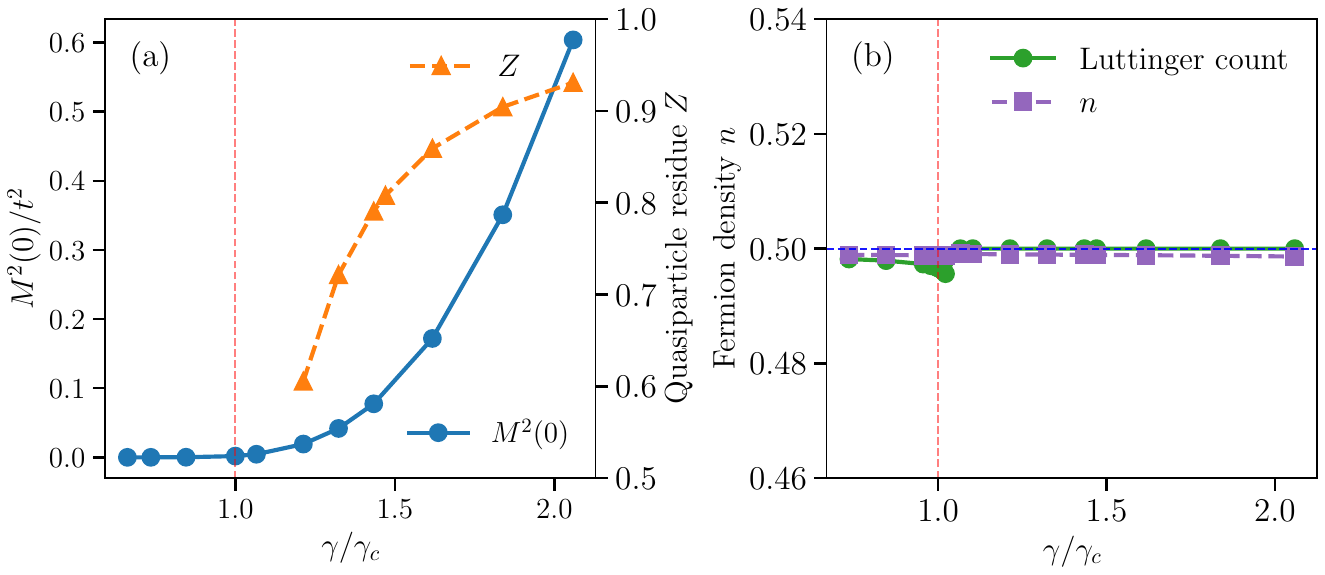}
    \caption{{\bf Boson mass, qusiparticle residue and Luttinger theorem:} (a)  (Left axis) the $T=0$ renormalized mass of the boson $M^2(0)$ as a function of $\gamma/\gamma_c$. The mass vanishes at the critical point $\gamma/\gamma_c=1$, and remains zero for $\gamma<\gamma_c$. In Fermi-liquid regime ($\gamma>\gamma_c$), $M^2(0)$ becomes finite. (Right axis) the quasiparticle residue $Z$, extracted from the fermionic self-energy, plotted as a function of  $\gamma/\gamma_c$ in the Fermi liquid regime. (b) The Luttinger count is compared with the fermion density $n$ at $T=0.005$ and chemical potential $\mu=0$ as function of $\gamma/\gamma_c$.}
    \label{fig:Mass_Z}
\end{figure}

The Fermi liquid states for $\gamma>\gamma_c$ can be characterized in terms of the quasiparticle residue $0 < Z \leq 1$, defined as 
\begin{equation}
    Z =  \dfrac{1}{1 - \left.\partial_\omega \mathrm{Re}\left[\Sigma^R(\omega)\right]\right|_{\omega = 0}} .
\end{equation}
In practice, we compute $Z$ numerically by evaluating the derivative of self-consistently [Eqs.\eqref{eq:SaddlePoint}] obtained $\mathrm{Re}\,\Sigma^R(\omega)$ at $\omega=0$. 
We also analytically estimate $Z$ from Eq.\eqref{eq:Sigma_FL_ZeroT} as
\begin{equation}
Z = \frac{1}{1 + \dfrac{g^2}{2\pi v_b v_\mr{F} M(0)}},
\end{equation}
which directly relates $Z$ to the boson mass $M(0)$ for $\gamma>\gamma_c$ [Eq.\eqref{eq:Mr0_DisorderedPhase}]. As the system approaches the quantum critical point (QCP), the boson mass vanishes as $M^2(0) \sim (\gamma-\gamma_c)^2$. As a result, the quasiparticle residue vanishes as $Z \sim (\gamma - \gamma_c)$. 
Away from the QCP, in the Fermi-liquid phase ($\gamma > \gamma_c$), $M(0)$ grows with increasing $\gamma$, resulting in a corresponding increase of the quasiparticle residue, with $Z \to 1$ deep in the Fermi-liquid regime. To substantiate this behavior, we compute $M^2(0)$ at the lowest numerically accessible temperature $T = 0.005$. As shown in Fig.\ref{fig:Mass_Z}(a), our numerical results for $M^2(0)$ and $Z$ are consistent with the above analytical expectations.  

As evident from the analytical expressions of the self-energies in Eqs.\eqref{eq:Sigma_FL_ZeroT}, \eqref{eq:Sigma_ZeroTCritical} and \eqref{eq:Sigma_ZeroTOrdered} across all $\gamma$, the non-interacting Fermi points $\pm k_\mr{F}$ corresponds to zero-energy poles of the retarded Green's function $G_R(k,\omega)=(\omega-\epsilon_k-\Sigma^R(\omega))^{-1}$, i.e., $G^{-1}_R(\pm k_\mr{F},0)=0$. Here we have neglected the imaginary part of the self-energy due to elastic scattering in Eq.\eqref{eq:Sigma_ZeroTOrdered} for $\gamma<\gamma_c$. We expect the zero-energy poles at the non-interacting Fermi points $\pm k_\mr{F}$ to sharply delineate the occupied region of the momentum space and lead to a Luttinger theorem. We thus verify that the Luttinger count, 
\begin{equation}
n_\mr{L} = 2 \int_{0}^{k_F} \frac{dk}{2\pi} = \frac{k_F}{\pi}.
\end{equation}
agrees with the numerically computed fermion density,
\begin{equation}
n = \int d\omega \, \rho_f(\omega) \, n_F(\omega),
\end{equation}
where $n_F(\omega)$ is the Fermi-Dirac distribution function and $\rho_f(\omega)=-(1/\pi L)\sum_k \mr{Im}G_R(k,\omega)$ single-particle density of states of fermions obtained from the numerical solution of the large-$N$ saddle-point Eqs.\eqref{eq:SaddlePoint}.
We compare the above Luttinger count $n_\mr{L}$ with fermion density $n$ for $T=0.005$ and chemical potential $\mu=0$ in Fig.\ref{fig:Mass_Z}(b), that demonstrates the existence of the Luttinger theorem for Fermi and non-Fermi liquid states, across all $\gamma$ in the 1D Yukawa-SYK model. 

We further obtain the fermion self-energy at low $T$ for $\gamma\geq \gamma_c$ using the Euler-MacLaurin expansion [Eq.\eqref{eq:EulerMaclaurin}] with $M=M(T)$
\begin{widetext}
\begin{align}
\Sigma(\mathrm{i}\omega_{\pm n})\simeq & -\frac{\mathrm{i}g^{2}}{2v_{F}v_b}T\sum_{m}\frac{\mathrm{sgn}(\omega_{\pm n}+\Omega_{m})}{\sqrt{\Omega_{m}^{2}+\kappa|\Omega_{m}|+M^{2}}},\hspace{1em}\hspace{1em}+n=0,1,\cdots,\hspace{1em}-n=-1,-2,\cdots\\
\simeq & -\frac{\mathrm{i}g^{2}}{2\pi v_{F}v_b}\mathrm{sgn}(\omega_{\pm n})\left[\ln\left(\frac{\kappa+2(|\omega_{\pm n}|-\pi T)+2\sqrt{(|\omega_{\pm n}|-\pi T)^{2}+\kappa(|\omega_{\pm n}|-\pi T)+M^{2}}}{2M+\kappa}\right)\right.\nonumber \\
 & \left.+\frac{\pi T}{\sqrt{(|\omega_{\pm n}|-\pi T)^{2}+\kappa(|\omega_{\pm n}|-\pi T)+M^{2}}}-\frac{\pi^2T^2}{6}\left(\frac{2(|\omega_{\pm n}|-\pi T)+\kappa}{[(|\omega_{\pm n}|-\pi T)^{2}+\kappa(|\omega_{\pm n}|-\pi T)+M^{2})]^{3/2}}-\frac{\kappa}{M^{3}}\right)\right] \label{eq:Sigma_EulerMacLaurin}
\end{align}
\end{widetext}
The above expansion is valid below a temperature scale $T\ll (M^{3}(T)/\kappa)^{1/2}$.
To analyze the above self-energies at low energies, we first assume
$\kappa\gg \omega=|\omega_{\pm n}|,~T$. We can subsequently expand the self energy in $\omega$
and $T$ in two regimes, $T,\omega\ll M(T)$
and $T\ll M(T)\ll\omega$. 

For $\gamma>\gamma_c$, $T, \omega \ll M(T)$ is the relevant regime to diagnose Fermi liquid characteristics. In this regime, upon analytic continuation to real frequencies, we obtain
\begin{align}
\Sigma^{R}(\omega)\simeq  -\frac{g^{2}}{2v_{F}v_b}\left[\frac{\omega}{\pi M(T)}+\mathrm{i}\frac{\kappa \omega^{2}}{4\pi M^{3}(T)}+\mathrm{i}\frac{\pi\kappa T^{2}}{4M^{3}(T)}\right]
\end{align}
Using $M^{2}(T)-M^{2}(0)\sim T^{2}$ [Eq.\eqref{eq:MT_DisorderedPhase}] for the Fermi liquid case, the above leads to
\begin{equation}
\Sigma^{R}(\omega)\simeq  -\frac{g^{2}}{2v_{F}v_b}\left[\frac{\omega}{\pi M(0)}+\mathrm{i}\frac{\kappa \omega^{2}}{4\pi M^{3}(0)}+\mathrm{i}\frac{\pi\kappa T^{2}}{4M^{3}(0)}\right],
\end{equation}
which is applicable for $T,\omega\ll M(0),(M^3(0)/\kappa)^{1/2}$.

At the QCP, $\gamma=\gamma_c$, where $M(0)=0$, $r_{0}=0$, $M^{2}(T)\sim T$ [Eqs.\eqref{eq:Mr0_critical}, \eqref{eq:MT_Critical}], $T\ll M(T)\ll\omega$ is the relevant frequency regime for characterizing non-Fermi liquid self energy in the quantum critical regime. Here, we obtain
\begin{equation}
\Sigma^{R}(\omega)\simeq  -\frac{g^{2}}{2v_{F}v_b}\left[\mathrm{i}\left(\frac{\pi\kappa T^{2}}{6M^3(T)}-\frac{2M(T)}{\pi\kappa}\right)+\frac{2\sqrt{\omega} e^{\mathrm{i}\pi/4}}{\pi\sqrt{\kappa}}\right],
\end{equation}
which reduces to the zero-temperature self energy of Eq.\eqref{eq:Sigma_ZeroTCritical} at the QCP for $T\ll M(T)\ll \omega$. The low-$T$ Euler-MacLaurin expansion in Eq.\eqref{eq:Sigma_EulerMacLaurin} for the self energy is not convergent for $\gamma<\gamma_c$, as in the case of boson thermal mass in Appendix \ref{app:ThermalMass}.

\section{Low-temperature thermodynamics}\label{app:Thermodynamics}

In this section, we evaluate the free energy and entropy of the 1D Yukawa-SYK model. In the large-$N$ limit, the low-temperature ($T\neq 0$) free energy ${F}$ (per flavor, per site) is obtained by evaluating the action [Eq.\eqref{eq:LargeNAction}] at the saddle point, 
\begin{align}
\beta LF &= -\mathrm{Tr} \ln \left[ -\left( G_0^{-1} - \Sigma \right) \right] 
+ \frac{1}{2} \mathrm{Tr} \ln \left[ D_0^{-1} - \Pi -\ci \lambda \right] \nonumber\\
&+ \int_{0}^{\beta} d\tau \sum_r \left[ -\Sigma_r(-\tau) G_r(\tau) 
+ \frac{1}{2} \Pi_r(-\tau) D_r(\tau)\right. \nonumber \\
&\left. + \frac{g^2}{2} G_r(\tau) G_r(-\tau) D_r(\tau) \right] -  \int_{0}^{\beta} d\tau \sum_r\dfrac{\mathrm{i}\lambda_r(\tau)}{2\gamma}. \label{eq:F_Action}
\end{align}
As a result, the free energy can be written as
\begin{subequations}
\begin{align}
F&=  F_{f}+F_{b},\\
F_{f}&= -T\sum_{n}\int\frac{dk}{2\pi}\ln\left[\frac{\epsilon_{k}-\mu-\mathrm{i}\omega_{n}+\Sigma(\mathrm{i}\omega_{n})}{\epsilon_{k}-\mu-\mathrm{i}\omega_{n}}\right] \nonumber \\
&-T\sum_{n}G(\mathrm{i}\omega_{n})\Sigma(\mathrm{i}\omega_{n})-T\int\frac{dk}{2\pi}\ln\left(1+e^{-\beta(\epsilon_{k}-\mu)}\right), \label{eq:Ff}\\
F_{b}&=\frac{T}{2}\sum_{m}\int\frac{dq}{2\pi}\ln\left[\frac{\Omega_{m}^{2}+\omega_{q}^{2}+m_{b}^{2}-\Pi(\mathrm{i}\Omega_{m})}{\Omega_{m}^{2}+\omega_{q}^{2}+m_{b}^{2}}\right]\nonumber \\
&-\frac{m_{b}^{2}}{2\gamma}+T\int\frac{dq}{2\pi}\ln\left[1-e^{-\beta\sqrt{m_{b}^{2}+\omega_{q}^{2}}}\right]\nonumber \\
&+\frac{1}{2}\int\frac{dq}{2\pi}\sqrt{m_{b}^{2}+\omega_{q}^{2}}, \label{eq:Fb}
\end{align}
\end{subequations}
where we have subtracted and added the non-interacting fermion and
boson contributions to make the Matsubara summations convergent. The Lagrange multiplier field $\lambda_r(\tau)=\lambda$ in Eq.\eqref{eq:LargeNAction} imposes the spherical constraint [Eq.\eqref{eq:SphericalConstraint}] as a function of $\gamma$ and $T$ at the saddle point and has been absorbed in $m_b^2$ above, i.e., $m_b^2-\ci \lambda\to m_b^2(\gamma,T)$. We also omit a temperature-independent constant term in Eq.\eqref{eq:Fb}.

At low temperatures, we compute the fermion determinant part using $\epsilon_{k}-\mu\approx\pm v_{F}k$
around the two Fermi points to obtain
\begin{align}
&\int dk \ln\left[\frac{\epsilon_{k}-\mu-\mathrm{i}\omega_{n}+\Sigma(\mathrm{i}\omega_{n})}{\epsilon_{k}-\mu-\mathrm{i}\omega_{n}}\right] \nonumber \\
&\simeq  2\int_{-\Lambda}^{\Lambda}dk\ln\left[\frac{k-\frac{\mathrm{i}\omega_{n}-\Sigma(\mathrm{i}\omega_{n})}{v_{F}}}{k-\frac{\mathrm{i}\omega_{n}}{v_{F}}}\right]
\overset{\Lambda\to \infty}{\simeq} -  2\pi G(\mathrm{i}\omega_{n})\Sigma(\mathrm{i}\omega_{n}).
\end{align}
As a result, the first and the second terms on the right hand side of Eq.\eqref{eq:Ff} cancel each other, leaving only the free-fermion contribution for the fermionic part $F_f$ of the free energy.
\begin{align}
F_{f}&\simeq  
    -T\int\frac{dk}{2\pi}\ln\left(1+e^{-\beta(\epsilon_{k}-\mu)}\right) \nonumber \\
    &\simeq  -2T\int_{-\Lambda}^{\Lambda}\frac{dk}{2\pi}\ln\left(1+e^{-\beta v_{F}k}\right)
\overset{\Lambda\to\infty}{\simeq}  \mathrm{const.}-\frac{\pi}{6v_{F}}T^{2}
\end{align}
Consequently, the fermionic contribution to the entropy is given by
\begin{align}
S_{f}= & -\frac{\partial F_f}{\partial T}\simeq \frac{\pi}{3v_{F}}T, \label{eq:Sf_LowT}
\end{align}
as quoted in Eq.\eqref{eq:Sf}.

Now, we compute the bosonic contribution with  $\bar{\Pi}(\mathrm{i}\Omega_{m})=-\kappa|\Omega_{m}|$ [Eqs.\eqref{eq:BosonPolarization},\eqref{eq:Pi_LowT}]  and $m_{b}^{2}(T)=  M^{2}(T)+\Pi(0)\simeq M^2(T)+\kappa\Gamma$ [Eq:\eqref{eq:ThermalMass}] at low temperatures $T\to 0$, such that
\begin{subequations}
\begin{align}
\frac{1}{2}&\int_{-\Lambda}^{\Lambda}\frac{dq}{2\pi}\sqrt{m_{b}^{2}+v_b^{2}q^{2}} \nonumber \\
&\simeq  \mathrm{const.}+\frac{1}{4\pi v_b}\left[\ln\left(\frac{\Lambda v_b+\sqrt{\Lambda^{2}v_b^{2}+m_{0}^{2}}}{m_{0}}\right)\right]\delta M^{2}(T),
\end{align}
\begin{equation}
T\int_{-\Lambda}^{\Lambda}\frac{dq}{2\pi}\ln\left(1-e^{-\beta\sqrt{m_{0}^{2}+\delta M^{2}(T)+v_b^{2}q^{2}}}\right)\approx  -T\frac{\Lambda}{\pi}e^{-\beta m_{0}}
\end{equation}
where $m_{0}^{2}=M^{2}(0)+\kappa\Gamma$, $\delta M^2(T)=M^2(T)-M^2(0)$, and
\begin{align}
&\frac{T}{2}\sum_{m}\int_{-\Lambda}^{\Lambda}\frac{dq}{2\pi}\ln\left[\frac{\Omega_{m}^{2}+v_b^{2}q^{2}+M^{2}(0)+\delta M^{2}(T)+\kappa|\Omega_{m}|}{\Omega_{m}^{2}+v_b^{2}q^{2}+m_{0}^{2}+\delta M^{2}(T)}\right] \nonumber \\
&\overset{\Lambda\to\infty}{\simeq}  \frac{T}{2v_b}\sum_{m}\left(\sqrt{\Omega_{m}^{2}+M^{2}(0)+\delta M^{2}(T)+\kappa|\Omega_{m}|}\right. \nonumber \\
 & \left.-\sqrt{\Omega_{m}^{2}+m_{0}^{2}+\delta M^{2}(T)}\right)
\end{align}
\end{subequations}
We again perform Euler-MacLaurin expansion for the above term for $\gamma\geq \gamma_c$
\begin{align}
&\frac{T}{2v_b}\sum_{m}\left(\sqrt{\Omega_{m}^{2}+M^{2}(0)+\delta M^{2}(T)+\kappa|\Omega_{m}|}\right. \nonumber \\
& \left.-\sqrt{\Omega_{m}^{2}+m_{0}^{2}+\delta M^{2}(T)}\right) \nonumber \\
&\simeq\mathrm{const.}+\frac{1}{4\pi v_b}\left(\ln\left[\frac{2\Gamma+\kappa+2\sqrt{M^{2}(0)+\Gamma(\Gamma+\kappa)}}{2M(0)+\kappa}\right]\right. \nonumber \\
& \left.-\ln\left[\frac{\Gamma+\sqrt{\Gamma^2+m_0^2}}{m_0}\right]\right)\delta M^{2}(T)-\frac{\pi T^{2}\kappa}{12M(T)v_b}
\end{align}
Combining all the above in Eq.\eqref{eq:Fb}, and relating the frequency and momentum cutoffs, i.e., $\Gamma\simeq\Lambda v_b$, we get
\begin{align}
F_{b}&\simeq  \mr{const.}-\frac{\pi T^{2}\kappa}{12M(T)v_b} \nonumber \\
&+\left(\frac{1}{4\pi v_b}\ln\left[\frac{2\Gamma+\kappa+2\sqrt{M^{2}(0)+\Gamma(\Gamma+\kappa)}}{2M(0)+\kappa}\right]-\frac{1}{2\gamma}\right) \nonumber \\
&\times \delta M^{2}(T)-T\frac{\Lambda}{\pi}e^{-\beta m_{0}} \nonumber 
\end{align}
The coefficient of the $\delta M^2(T)$ above in the brackets vanish due to the spherical constraint [Eq.\eqref{eq:LowEConstraintEq}] with $r_0^2(T\neq 0)=0$. The last term is negligible at low temperatures. Thus, we finally get at low temperatures for $\gamma\geq \gamma_c$,
\begin{align}
F_b\simeq -\frac{\pi T^2\kappa}{12 M(T)v_b}+\mr{const.}   
\end{align}

\begin{figure}[h!]
    \centering
    \includegraphics[width=\linewidth]{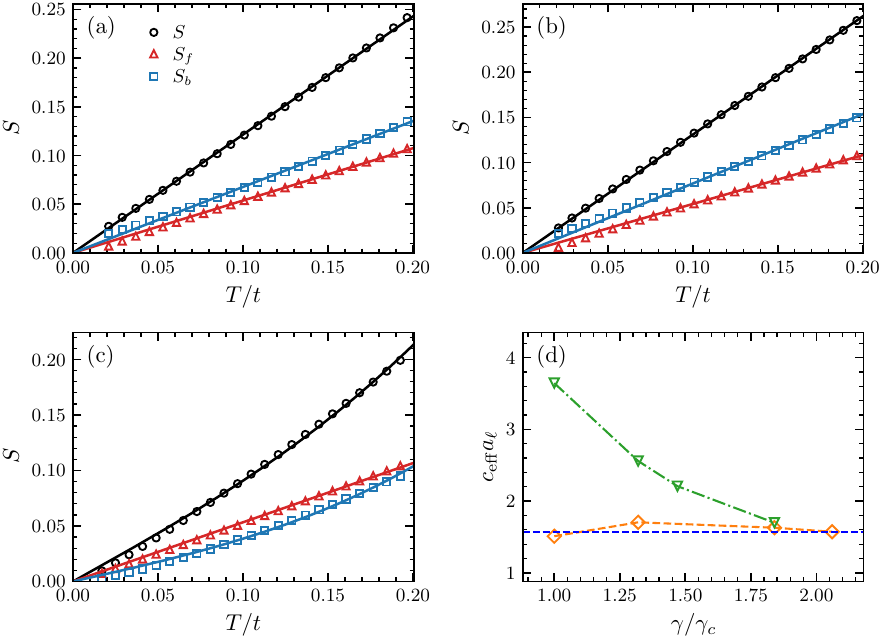}
    \caption{\textbf{Thermal entropy of the 1D Yukawa--SYK model:} Total thermal entropy $S$, fermionic contribution $S_{f}$, and bosonic contribution $S_{b}$ as functions of temperature $T$ for (a) the strange metal at
$\gamma/\gamma_{c}=1$, (b) $\gamma/\gamma_{c}=0.85$, and (c) the Fermi liquid at $\gamma/\gamma_{c}=1.32$. For all values of $\gamma/\gamma_{c}$, the fermionic entropy is fitted (solid lines) to a linear-$T$ form with a coefficient close to the non-interacting value $\pi/(3v_{F})\approx 0.524$. The bosonic contribution remains linear in $T$
for $\gamma\leq\gamma_{c}$, whereas in the Fermi liquid regime $S_{b}$ is better fitted (solid line) to $S_{b}\simeq f_{1}T+f_{2}T^{3}$. (d) Comparison between the coefficient of the fermionic subsystem second
R\'{e}nyi entropy, $c_{\mathrm{eff}}a_{\ell}$ (orange diamonds), defined in Eq.~\eqref{eq:SA2_EntropyLocalEnt} and obtained from Figs.\ref{fig:SA2_Collpase_1DYSYK},\ref{fig:SA2_EntropyLocalEnt}, and the coefficient of the thermal entropy, $c_{\mathrm{eff}}\tilde{a}_{\ell}$
(green downward triangles), defined in Eq.~\eqref{eq:LinearTCoeff_Entropy} and obtained from the fit to the thermal entropy. The non-interacting value of $ca_\ell=\pi/v_\mr{F}\approx1.57$ is shown as a reference (blue dashed line)}.
\label{fig:ThermalEntropy}
\end{figure}

In the disordered or Fermi liquid phase $\gamma>\gamma_c$ with $M^2(T)-M^2(0)\sim T^2$ [Eq.\eqref{eq:MT_DisorderedPhase}], we obtain the following free energy and entropy for $T\to 0$,
\begin{subequations}\label{eq:FbSb_FL}
\begin{align}
F_b&\simeq -\frac{\pi \kappa T^2}{12 M(0)v_b}+\mr{const.}, \\
S_b&=-\frac{\partial F_b}{\partial T}\simeq \frac{\pi\kappa}{6 M(0)v_b}T
\end{align}
\end{subequations}
as mentioned in Eq.\eqref{eq:Sb_FL}. Thus, combining the fermionic [Eq.\eqref{eq:Sf_LowT}] and the bosonic contributions, we obtain the total entropy of Eq.\eqref{eq:LinearTCoeff_Entropy} in the Fermi liquid phase. The entropy varies linearly with $T$, albeit with an enhanced coefficient due to the (Ohmic) dissipative bosons \cite{hanggi2006,Hanggi_2008,Ford_PRB}, even when bosons are gapped ($M(0)\neq 0$) for $\gamma>\gamma_c$. 

At the strange metallic QCP ($\gamma=\gamma_c$), using $M^2(T)\sim T$ [Eq.\eqref{eq:MT_Critical}], we obtain $F_b\sim -T^{3/2}$, and $S_b\sim T^{1/2}$. However, we could not detect the signature of this $\sqrt{T}$ behavior in $S_b$ in our numerical calculations, as shown in Fig.\ref{fig:ThermalEntropy}(a). This could be due to the temperature regime accessed not being sufficiently low, or the additional corrections to the employed Euler-MacLaurin low-$T$ expansion [Eq.\eqref{eq:EulerMaclaurin}] for $M^2(T)$ and $F(T)$. The Euler-MacLaurin expansion at the critical point $\gamma=\gamma_c$ is only marginally convergent for $T\to 0$.

In Figs.\ref{fig:ThermalEntropy}(a) and (c), we compare the analytical low-$T$ expansion results for $S_f$ [Eq.\eqref{eq:Sf_LowT}], $S_b$ [Eq.\eqref{eq:FbSb_FL}] and $S=S_f+S_b$ with those obtained from Eq.\eqref{eq:F_Action} using the numerical solution of the large-$N$ saddle-point Eqs.\eqref{eq:SaddlePoint} for (a) the strange metal at $\gamma/\gamma_c=1$ and (c) the Fermi liquid at $\gamma/\gamma_c=1.32$. We also show the numerical results for $\gamma/\gamma_c=0.85$, corresponding to the $T=0$ ordered phase [Fig.\ref{fig:PhaseDiagram}], in Fig.\ref{fig:ThermalEntropy}(b). For all values of $\gamma$, the fermionic contribution can be fitted well with a linear-$T$ variation with coefficient close to the expected non-interacting value $\pi/(3v_\mr{F})\approx 0.52$ [Eq.\eqref{eq:Sf_LowT}]. As shown in Fig.\ref{fig:ThermalEntropy}(a), at the QCP ($\gamma/\gamma_c=1$), $S_b$, and hence the total entropy $S$, exhibit a linear-$T$ behavior for the temperature range accessed, unlike the expected $\sqrt{T}$ variation from the low-temperature expansion, as discussed above. A linear temperature dependence is also observed for $S_b$ at $\gamma/\gamma_c=0.85$, as shown in Fig.\ref{fig:ThermalEntropy}(b). 

In Fig.\ref{fig:ThermalEntropy}(c) for the Fermi liquid ($\gamma/\gamma_c=1.32$), $S_b\propto T$ at low temperatures, but the data is better fitted with $S_b\simeq f_1T+f_2 T^3$ form over a larger temperature range. The $T^3$ correction can be rationalized from the low-temperature expansion in Eqs.\eqref{eq:FbSb_FL} by keeping next order term. We perform the same analysis for a few other values of $\gamma>\gamma_c$ in the Fermi liquid phase and extract $c_{eff}\tilde{a}_\ell$ of Eq.\eqref{eq:LinearTCoeff_Entropy} from the linear-$T$ coefficient of entropy $S$. The result of $c_{eff}\tilde{a}_\ell$ for $\gamma\geq \gamma_c$ is plotted as function of $\gamma$ in Fig.\ref{fig:ThermalEntropy}(d), and is compared with $c_{eff}a_\ell$ of Eq.\eqref{eq:SA2_EntropyLocalEnt} extracted from the fermionic subsystem R\'{e}nyi entropy $S_f^{(2)}(L,T)=S_{A_f}^{(2)}(L_A=L,T)$ [Eq.\eqref{eq:Thermal_Fermion2ndRenyi}] in Fig.\ref{fig:SA2_EntropyLocalEnt}(b). The non-interacting value $(\pi/v_\mr{F})$ from the CFT crossover formula of Eq.\eqref{eq:finite_T_cft} for these quantities is also shown as a reference in Fig.\ref{fig:ThermalEntropy}(d).

\begin{figure}[h!]
    \centering
    \includegraphics[width=\linewidth]{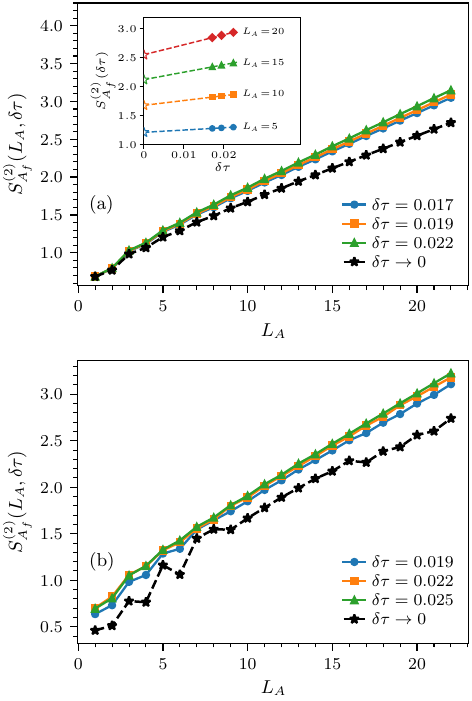}
\caption{{\bf Extrapolation of $S_{A_f}^{(2)}(L_A,\delta \tau)$ to $\delta \tau\to 0$:} Fermionic second R\'{e}nyi entropy $S^{(2)}_{A_{f}}(L_{A},\delta\tau)$
as a function of $L_{A}$ for three values of the discretization step $\delta\tau=\beta/N_{\tau}$, together with the continuum-limit result
$S^{(2)}_{A_{f}}(L_{A},\delta\tau\to 0)\equiv S^{(2)}_{A_{f}}(L_{A})$ (black
dashed curve), for (a) the strange metal ($\gamma/\gamma_{c}=1$) at $T\simeq 0.069$ and (b) the ordered phase ($\gamma/\gamma_{c}=0.85$) at $T\simeq 0.057$, both at system size $L=50$. In the ordered phase, the linear extrapolation does not yield a smooth continuum-limit curve for $S^{(2)}_{A_{f}}(L_{A})$. Inset of (a): $S^{(2)}_{A_{f}}(\delta\tau)$ as a
function of $\delta\tau$ for several subsystem sizes, with the corresponding linear extrapolations (dashed lines) and extrapolated values at $\delta\tau\to 0$ (open stars).}
    \label{fig:delta_tau}
\end{figure}

\section{The extrapolation of $S^{(2)}_{A_f}(\delta\tau)$ to $\delta\tau \to 0$}\label{app:extrapolation}
As explained in Appendix~\ref{app:discretization}, we solve the large-$N$ saddle-point self-consistency equation [Eqs.~\eqref{eq:self_consis_entan_replica}]
for the fermionic second R\'{e}nyi entropy by discretizing it in imaginary time with a step $\delta\tau=\beta/N_{\tau}$. To reach the continuum limit $\delta\tau\to 0$, we compute $S^{(2)}_{A_{f}}(\delta\tau)$ at three values of $\delta\tau$ lying between $0.017$ and $0.03$, and then extrapolate linearly to
$\delta\tau\to 0$. This works reliably for $\gamma\geq\gamma_{c}$; in the ordered state, by contrast, the extrapolated curve for $S^{(2)}_{A_{f}}(L_{A})$ is no longer smooth. Staying within this range of $\delta\tau$ also becomes increasingly costly for $T\lesssim 0.03$, since the size $2NN_{\tau}\times 2NN_{\tau}$ of the Green's function matrix grows with $\beta=1/T$. For this reason, we confine most of our large-$N$ saddle-point calculations for fermionic subsytem second R\'{e}nyi entropy [Eq.\eqref{eq:final_renyi_entropy}] to the
temperature window $T\simeq 0.05$--$0.12$. 

In Fig.~\ref{fig:delta_tau}, we show $S^{(2)}_{A_{f}}(L_{A},\delta\tau)$ as a
function of the subsystem size $L_{A}$ for the smallest values of $\delta\tau$ accessible in our calculations, together with the continuum-limit result $S^{(2)}_{A_{f}}(L_{A},\delta\tau\to 0)\equiv S^{(2)}_{A_{f}}(L_{A})$, for (a) the strange metal ($\gamma/\gamma_{c}=1$) and (b) the orderd phase
($\gamma/\gamma_{c}=0.85$). In contrast to the strange metal regime, in the ordered state the linear extrapolation of $S^{(2)}_{A_f}(\delta\tau)$ does not yield a smooth curve of the $S^{(2)}_{A_f}$, presumably due to numerical issue related to the smaller boson thermal mass [Fig.~\ref{fig:MT_thermal_mass}]. In the inset of Fig.~\ref{fig:delta_tau} we further show $S^{(2)}_{A_{f}}(\delta\tau)$ as a function of $\delta\tau$, along with the linear extrapolation, for a few subsystem sizes in the strange metal. Repeating this procedure for all subsystem sizes $L_{A}$ and temperatures $T$ yields a smooth continuum-limit curve of $S^{(2)}_{A_{f}}(L_{A})$ for $\gamma\geq \gamma_c$, shown in Figs.~\ref{fig:SA2_Collpase_1DYSYK}(a) and \ref{fig:SA2_Collpase_1DYSYK}(b) for the strange metal and the Fermi liquid, respectively.


\begin{figure*}
    \centering
    \includegraphics[width=\linewidth]{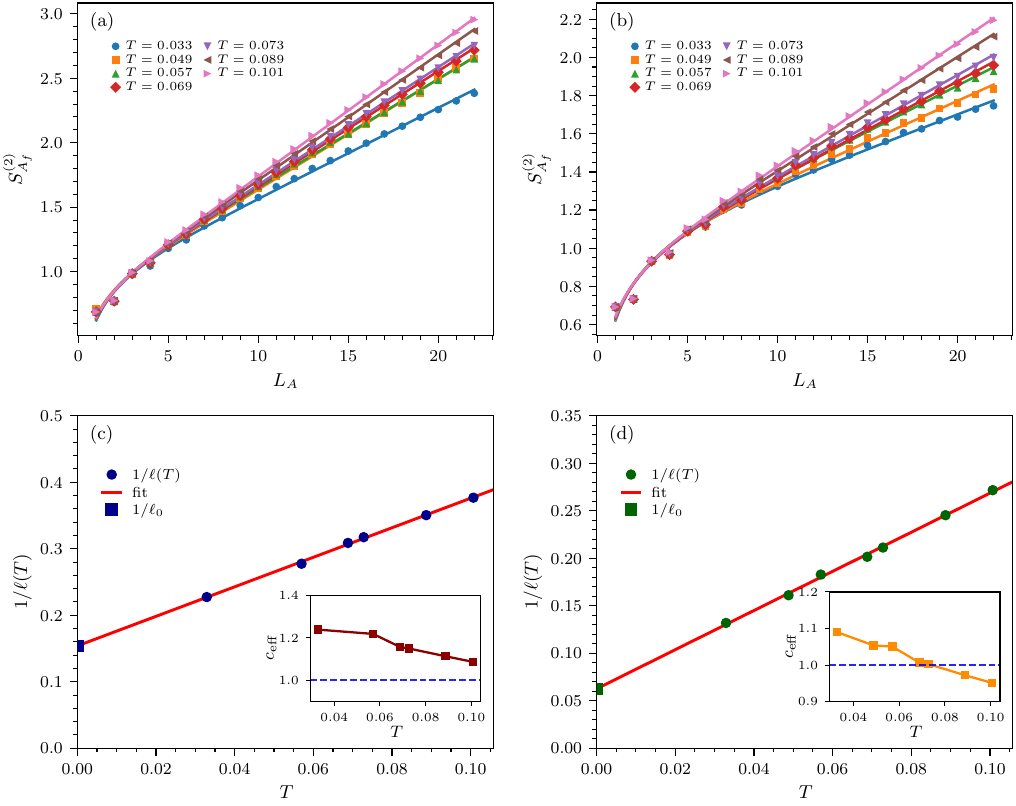}
    \caption{{\bf Fermionic-subsystem second R\'{e}nyi entropy of 1D Yukawa-SYK model and CFT fits:} (a) $\gamma/\gamma_c=1$ (strange metal) and (b) $\gamma/\gamma_c=1.84$ (Fermi liquid) -- The fermionic subsystem second R\'{e}nyi entropy $S^{(2)}_{A_f}$ as a function of subsystem size $L_A$ for system size $L=50$, shown for different temperatures and fitted (solid lines) with CFT crossover function of Eq.\eqref{eq:finite_T_cft}. The extracted inverse length $1/\ell(T)$ plotted as a function of $T$ from the CFT fits, with finite intercepts at $T=0$, $1/\ell_0\simeq 0.15$ in (c) $\gamma/\gamma_c=1$, and $1/\ell_0=0.06$ in (d) $\gamma/\gamma_c=1.84$. The insets show the corresponding effective central charge $c_{eff}$ as a function of $T$, compared with the the CFT value (blue dashed line) $c=1$. }\label{fig:SA2_CFTFit}
\end{figure*}

\section{CFT fitting to the $S_A^{(2)}(L_A,T)$ in the 1D Yukawa-SYK model} \label{app:CFTFit_YukawaSYK}
Here, we directly fit the results [Fig.\ref{fig:SA2_Collpase_1DYSYK}(a) and Fig.\ref{fig:SA2_Collpase_1DYSYK}(b)] for the fermionic subsystem second R\'enyi entropy $S^{(2)}_{A_f}(L_A)$ [Eq.\eqref{eq:second_renyi}] at a temperature $T$ in the 1D Yukawa-SYK model using the CFT crossover formula in Eq.\eqref{eq:finite_T_cft} with the parameters $c=c_{eff}$, $1/\ell(T)$, and $b$. We fit $S^{(2)}_{A_f}(L_A)$ at different temperatures for $\gamma/\gamma_c = 1$ and $\gamma/\gamma_c = 1.84$, as shown by the solid lines in Figs.\ref{fig:SA2_CFTFit}(a) and (b), respectively. As evident, the CFT crossover function provides a good description of the data in both the non-Fermi liquid and Fermi-liquid regimes.
From the fits, we extract the inverse thermal length $1/\ell(T)$, which follows the form of Eq.\eqref{eq:length_ysyk}, with $1/\ell_1(T)\propto T$. The results are shown in Figs.\ref{fig:SA2_CFTFit}(c) and (d) for $\gamma/\gamma_c = 1$ and $\gamma/\gamma_c = 1.84$, respectively. The inverse thermal length remains finite as $T \to 0$, yielding a non-zero fermion-boson entanglement length scales $1/\ell_0 \simeq 0.15$ and $1/\ell_0 \simeq 0.06$, in close agreement with the values obtained from the scaling collapse in Fig.\ref{fig:SA2_Collpase_1DYSYK}(c) and Fig.\ref{fig:SA2_Collpase_1DYSYK}(d) (insets).
The extracted  effective central charge $c_{\mathrm{eff}} > 1$ in the quantum critical regime and also in Fermi-liquid regime. The effective central charge $c_{\mathrm{eff}}$, obtained from direct CFT fits, is shown as a function of $T$ in the insets of Figs.\ref{fig:SA2_CFTFit}(c) and (d) (insets). $c_{eff}$ decreases with temperature in both cases and is compared with the free-fermion CFT value $c=1$ (blue dashed line). 

The temperature dependence of $c_{\mathrm{eff}}$ is a consequence of fitting $S_{A_f}^{(2)}(L_A)$ curve at each temperature independently with $c_{\mathrm{eff}}$ as a free parameter, unlike $c_{eff}$ extracted using Eq.\eqref{eq:eff_cft_fit} from the universal scaling curves [Figs.\ref{fig:SA2_Collpase_1DYSYK}(c) and (d)]. Thus, the scaling analysis presumably provides much more robust way to analyze the $S_{A_f}^{(2)}(L_A,T)$. In any case, compared to the temperature-independent $c_{eff}$ extracted from the scaling analysis, we obtain somewhat lower values of $c_{eff}$, $c_{eff}\simeq 1.3$ for the strange metal at $\gamma/\gamma_c=1$ and $c_{eff}\simeq 1.1$ for the Fermi liquid at $\gamma/\gamma_c=1.84$, at the lowest temperature $T\simeq 0.03$ from the direct CFT fit. The temperature-dependence of the length scale $\ell(T)$ extracted from the direct CFT fit also follows behavior similar to Eq.\eqref{eq:length_ysyk} with $1/\ell_1(T)\propto T$. Though the slope of $1/\ell_1(T)$ does not match exactly with $1/\ell_1(T)$ obtained from the scaling analysis [Figs.\ref{fig:SA2_Collpase_1DYSYK}(c),(d), insets], $\ell_0$ extracted from the two methods agree very well. The deviation in the temperature dependence of $\ell(T)$ between direct CFT fit [Figs.\ref{fig:SA2_CFTFit}] and CFT fit to the scaling curves [Fig.\ref{fig:SA2_Collpase_1DYSYK}] can be rationalized from the fact that all the parameters, $c_{eff}(T)$, $\ell(T)$ and $b(T)$, become $T$ dependent in the former, whereas fitting to the universal scaling curves $f_L(x=L_A/\ell(T))=S_{A_f}^{(2)}(L_A,T)-d_L(T)$ [Eq.\eqref{eq:eff_cft_fit}] using $c_{eff}$, $s_\ell$ and $b$ does not involve any temperature dependent quantities. Indeed, we see in Fig.\ref{fig:SA2_EntropyLocalEnt}(b), that the combind quantity $c_{eff}(T)/\ell(T)$ matches very well between the two methods, even though $c_{eff}$ and $\ell(T)$ do not match separately.

\section{System-size dependence of $c_{\mathrm{eff}}$ and  $\ell_0$ }\label{app:system_size_c_l_0}
\begin{figure}
    \centering
    \includegraphics[width=\linewidth]{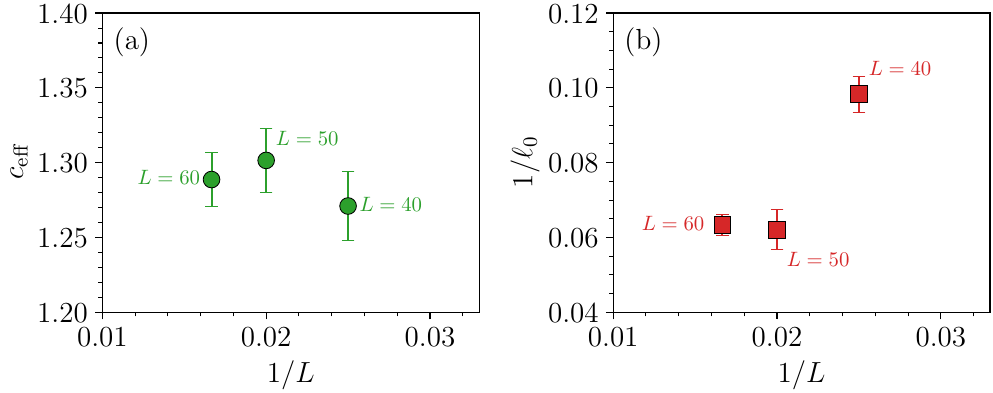}
\caption{\textbf{System-size dependence of $c_{\mathrm{eff}}$ and $1/\ell_{0}$.} (a) Effective central charge $c_{\mathrm{eff}}$, extracted from the universal scaling curves and their fits to the CFT-like crossover formula (see, e.g., Figs.\ref{fig:SA2_Collpase_1DYSYK}),
plotted as a function of the inverse system size $1/L$. (b) Inverse fermion--boson entanglement length $1/\ell_{0}$ [Eq.\eqref{eq:length_ysyk}], plotted as a function of $1/L$. Both panels correspond to the Fermi liquid regime ($\gamma/\gamma_{c}=1.84$); error bars denote the uncertainties of the corresponding fits.}
\label{fig:system_size_c_l_0}
\end{figure}

In this section, we discuss the system-size ($L$) dependence of the effective
central charge and the fermion--boson entanglement length, e.g., for the Fermi
liquid regime at $\gamma/\gamma_{c}=1.84$. Similar results can be obtained for other values of $\gamma\geq \gamma_c$ (not shown). We compute the subsystem R\'{e}nyi
entropy $S^{(2)}_{A_{f}}(L_{A},T)$ as a function of the subsystem size $L_{A}$
for the system sizes $L=40$, $50$ and $60$ by solving the self-consistency
equations in Eq.~\eqref{eq:self_consis_entan_replica} (see Sec.~\ref{sec:Ent_LargeN}) with the recursive Green's function method (see Supplemental Material, Sec. \ref{app:renyi_numerical_saddle}). Following the scaling ansatz of Eq.~\eqref{eq:scaling_ansatz}, the data for
$S^{(2)}_{A_{f}}(L_{A},T)$, obtained at different subsystem sizes $L_A\lesssim L/2$ and the temperature range $T=0.057-0.12$, are collapsed onto a single universal curve for each of these system sizes, as shown in Fig.\ref{fig:SA2_Collpase_1DYSYK} for $L=50$. The thermal length
$\ell(T)$ extracted from this collapse exhibits the same behavior as in Eq.\eqref{eq:length_ysyk} for all system sizes considered, and the
corresponding inverse fermion-boson entanglement length $1/\ell_{0}$ for $L=40$-$60$
is shown in Fig.\ref{fig:system_size_c_l_0}(b). Finally, fitting the universal curve obtained for each system size to Eq.\eqref{eq:eff_cft_fit} yields the
effective central charge $c_{\mathrm{eff}}$, which is plotted as a function of $1/L$ in Fig.\ref{fig:system_size_c_l_0}(a). Within the range of system sizes considered, we find no strong dependence of either $c_{\mathrm{eff}}$ or $1/\ell_{0}$ on $L$ to within the error bars
[Fig.\ref{fig:system_size_c_l_0}]. 

\section{Large-$N$ model with on-site flavor-hopping disorder for a diffusive metal}\label{app:diffusive_metal}
Here we discuss how elastic scatterings due to on-site disorder in a diffusive metal lead to elastic length scale and cut off to the logarithmic growth of entanglement entropy with subsystem size. Such an effect of elastic scatterings has been previously discussed in refs.\cite{potter,Haldar,Pouranvari2015AreaLaw}. We consider large-$N$ model of non-interacting fermions in 1D by adding an on-site ($r=1,\cdots,L$) random flavor ($i=1,\cdots,N$) hopping terms to $\mc{H}_f$ in Eq.\eqref{eq:H_f}, namely
\begin{align}
\mc{H}_f&=-\sum_{irr'}t_{rr'}c_{ir}^\dagger c_{ir'}-\mu\sum_{ir}c_{ir}^\dagger c_{ir}+\frac{1}{\sqrt{N}}\sum_{ijr}v_{ijr}c_{ir}^\dagger c_{jr}, \label{eq:RandomHopping_Hf}
\end{align}
where $v_{ijr}=v_{jir}^*$ is a spatially random complex flavor hopping at a site $r$ with zero mean ($\overline{v_{ijr}}=0$) and standard deviation $v$, i.e., $\overline{v_{ijr}v^*_{klr'}}=v^2\delta_{ik}\delta_{jl}\delta_{rr'}$. In $1D$ such a disordered system is Anderson localized for finite $N$. However, in the large-$N$ limit ($N\to\infty$), the model leads to a diffusive metallic phase \cite{Haldar}. 

Following the standard thermal field theory of Yukawa-SYK model in Sec. \ref{sec:SaddlePoint}, we write the disorder averaged partition function for the model in Eq.\eqref{eq:RandomHopping_Hf} in terms of large-$N$ bilocal fields $G(\tau,\tau')$ and $\Sigma(\tau,\tau')$ as
\begin{align}\label{eq:diffusive_Z}
Z&=\int \mathcal{D}\Sigma(\tau,\tau')\mathcal{D}G(\tau,\tau')\exp{(-\mathcal{S}_v)}
\end{align}
with
\begin{align}\label{eq:diffusive_action}
\mathcal{S}_{v}&=  -N\mathrm{Tr}\ln\left[-\left({G}_{0}^{-1}-\Sigma\right)\right]\nonumber \\& \nonumber +N\int d\tau d\tau'\sum_{r}\Bigg[-\Sigma_{r}(\tau',\tau)G_{r}(\tau,\tau')\\&
+\dfrac{v^2}{2}G_{r}(\tau',\tau)G_{r}(\tau,\tau')\Bigg].
\end{align}
For $N \to \infty$, the Green's function is obtained from the saddle point of the effective action $\mathcal{S}_v$ [Eq.\eqref{eq:diffusive_action}]
\begin{subequations}\label{eq:diffusive_case_self_consis}
\begin{align}
G_{rr'}^{-1}(\tau,\tau')= & {G}_{0,rr'}^{-1}(\tau,\tau')-\Sigma_{r,}(\tau,\tau')\delta_{rr'}, \label{eq:diff_InvG}\\
\Sigma_{r}(\tau,\tau')= & v^{2}G_{r}(\tau,\tau').\label{eq:diff_Sigma}
\end{align}
\end{subequations}
The above equilibrium saddle-point self-energy [Eq.\eqref{eq:diff_Sigma}] generated by the random flavour hopping $v$ [Eq. \eqref{eq:RandomHopping_Hf}] gives rise to a finite elastic scattering rate, and hence to a finite mean free path $\ell_{{el}}$, beyond which transport is diffusive rather than ballistic. This elastic length scale, emerging from the random hopping, offers a potential signature in the subsystem entanglement entropy: for $L_{A}\gg\ell_{{el}}$
the logarithmic growth of $S^{(2)}_{A}$ is cut off by $\ell_{{el}}$, and the entanglement entropy is expected to saturate to an area law. Following the large-$N$ path integral formulation of the subsytem second R\'{e}nyi entropy developed in section \ref{sec:Ent_LargeN}, we write the disorder-averaged action with entanglement replicas $\alpha,\beta=1,2$ for this quadratic model with random hopping [Eq.\eqref{eq:RandomHopping_Hf}] 
\begin{align}
\mathcal{S}&_{A}=  -N\mathrm{Tr}\ln\left[-\left(\tilde{G}_{0}^{-1}-\Sigma\right)\right]\nonumber \\&+N\int d\tau d\tau'\sum_{r,\alpha\beta}\Bigg[\Sigma_{r,\beta\alpha}(\tau',\tau)G_{r,\alpha\beta}(\tau,\tau')\nonumber\\ &
+\dfrac{v^2}{2}G_{r,\beta\alpha}(\tau',\tau)G_{r,\alpha\beta}(\tau,\tau')\Bigg].
\label{eq:diffusive_action_entan_replica}
\end{align}
For $N \to \infty$, Green's function is determined by the saddle point of the effective action $\mathcal{S}_{A}$ [Eq.\eqref{eq:diffusive_action_entan_replica}]
\begin{subequations}\label{eq:diffusive_case_self_consis_entan_replica}
\begin{align}
G_{r\alpha,r'\beta}^{-1}(\tau,\tau')= & \tilde{G}_{0,r\alpha,r'\beta}^{-1}(\tau,\tau')-\Sigma_{r,\alpha\beta}(\tau,\tau')\delta_{rr'}, \label{eq:diff_InvG_Ent}\\
\Sigma_{r,\alpha\beta}(\tau,\tau')= & v^{2}G_{r,\alpha\beta}(\tau,\tau'),\label{eq:diff_Sigma_Ent}
\end{align}
with $\tilde{G}_{0,r\alpha,r'\beta}^{-1}(\tau,\tau')$ is defined in Eq.\eqref{eq:fermion_greenfn_kick}.
Finally, the disorder averaged subsystem second R\'{e}nyi entropy $\overline{S_{A}^{(2)}}$ obtain in terms of the difference between large-$N$ second R\'{e}nyi [Eq.\eqref{eq:diffusive_action_entan_replica}] and equilibrium saddle-point [Eq.\eqref{eq:diffusive_action}] actions,
\begin{equation}
\overline{S_{A}^{(2)}}=\frac{\mathcal{S}_{A}-2\mathcal{S}_v}{N}. 
\label{eq:diffusive_final_renyi_entropy}
\end{equation}
\end{subequations}

\begin{figure}[h!]
    \centering
    \includegraphics[width=\linewidth]{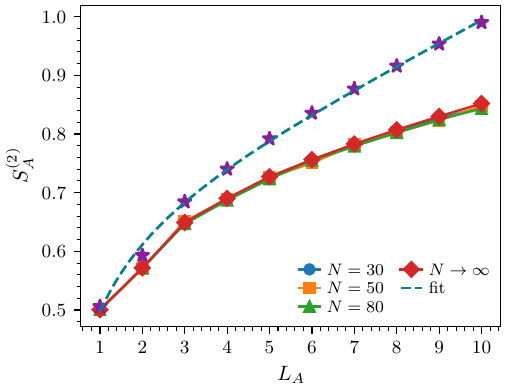}
   \caption{\textbf{Comparison of $S^{(2)}_{A}$ between the correlation-matrix and
large-$N$ results for the random hopping model:} Second R\'{e}nyi entropy $S^{(2)}_{A}$ as a function of subsystem size $L_{A}$ for the on-site random
flavour hopping model [Eq.\eqref{eq:RandomHopping_Hf}], computed with the correlation-matrix method at finite $N=30$, $50$ and $80$, and with the large-$N$ path-integral method in the
continuum limit $\delta\tau\to 0$, for system size $L=30$, $T=0.05$ and $v=1$. Also shown is the large-$N$ result for $S^{(2)}_{A}(L_{A})$ (magenta stars) ($L=30,~T=0.05$) for the 1D Yukawa-SYK model [Eq.~\eqref{eq:Model}] with the random hopping term
in Eq.~\eqref{eq:RandomHopping_Hf} added to $\mc{H}_f$ [Eq.\eqref{eq:H_f}] at $g=1$, $v=1$ and $m_{b}=1$ (Fermi liquid regime). Fitting (dashed line) these data to the CFT form [Eq.~\eqref{eq:finite_T_cft}]
yields $c_{\mathrm{eff}}=0.63\pm 0.03$ (dark cyan).}
\label{fig:large_N_vs_corr_random_hop}
\end{figure}

Since, the model with random flavour hopping is non-interacting, we can also compute the disorder-averaged subsystem second R\'{e}nyi entropy $\overline{S_{A}^{(2)}}$ using the correlation matrix method, but with finite $N$ [Eq.\eqref{eq:SA2_NonInt_CorrMat}].
We consider nearest-neighbour inter-site hopping $(t)$ with the hopping strength set to unity in Eq.\eqref{eq:RandomHopping_Hf} and a chemical potential $\mu=0$, as in the 1D Yukawa-SYK model [Eq.\eqref{eq:Model}]. We first compute $S_A^{(2)}(L_A)$ in the large-$N$ limit using the path integral method [Eq.\eqref{eq:diffusive_final_renyi_entropy}] by extrapolating to $\delta\tau \to 0$ for a system with $L=30$ at $\beta=20$ and $v=1$, as shown in Fig.\ref{fig:large_N_vs_corr_random_hop}. We compare this result with $S_A^{(2)}$ computed using the correlation matrix method [Eq.\eqref{eq:SA2_NonInt_CorrMat}] with finite but large $N$ values, in the range $N=30-80$ for system size $L=30$ at $T=0.05$. In the correlation-matrix
calculation, $S_{A}^{(2)}$ is averaged over many independent realizations $(\sim 20)$ of the random hopping $v_{ijr}$ [Eq.\eqref{eq:RandomHopping_Hf}]. We see that the results of the two methods agree very well [Fig.\ref{fig:large_N_vs_corr_random_hop}], implying that $N\gtrsim 30$ is sufficient to approach the large-$N$ limit for such system sizes $L$. 

\begin{figure}[h!]
    \centering
    \includegraphics[width=\linewidth]{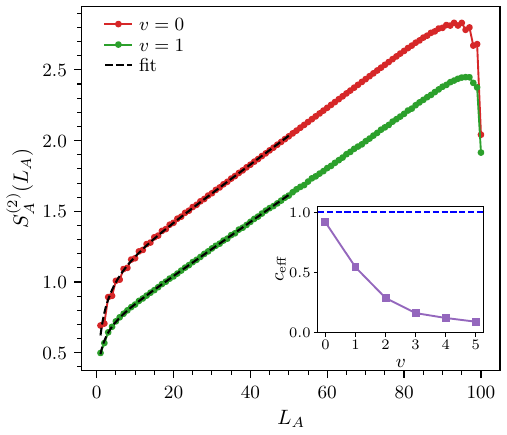}
    \caption{\textbf{The second R\'{e}nyi entropy for the large-$N$ random hopping model:} $S^{(2)}_{A}(L_{A})$ as a function of subsystem size $L_{A}$, obtained with the correlation-matrix method [Eq.~\eqref{eq:SA2_NonInt_CorrMat}] for the
clean system ($v=0$) and for $v=1$, at $L=100$, $\beta=20$, and $N=30$. The data are fitted to the CFT form [Eq.\eqref{eq:finite_T_cft}], giving $c=0.92\pm 0.02$ for $v=0$ and
$c_{\mathrm{eff}}=0.54\pm 0.01$ for $v=1$. Inset: $c_{\mathrm{eff}}$ as a function of the on-site disorder strength $v$, with the CFT value $c=1$ (clean system) shown for reference (blue dotted line).}
\label{fig:diffusive_metal_finite_T}
\end{figure}

\begin{figure}[h!]
    \centering
    \includegraphics[width=\linewidth]{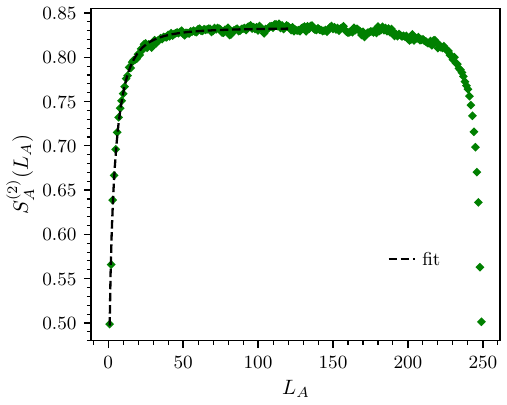}
\caption{\textbf{Area-law entanglement entropy in the diffusive Fermi liquid:} Scaling of the subsystem second R\'{e}nyi entropy $S^{(2)}_{A}$ as a function of subsystem size $L_{A}$, obtained with the correlation-matrix method for system size $L=250$ with $N=30$ at $T=0$. The data are fitted to the modified scaling form (black dotted line) $S_{A}^{(2)}(L_{A})\simeq(c_{\mathrm{eff}}/4)\ln\!\big[\left(L_{A}^{-2}+\ell_{\mathrm{el}}^{-2}(v)\right)^{-1/2}\big]+b$, yielding an effective
central charge $c_{\mathrm{eff}}=0.50\pm 0.01$ and an elastic length $\ell_{\mathrm{el}}=15.0\pm 0.3$.}
    \label{fig:area_law_T_0}
\end{figure}

We further obtain $S_A^{(2)}(L_A)$ for larger system sizes $L=100$ at $T=0.05$ with $N=30$ using the correlation matrix method [Eq.\eqref{eq:SA2_NonInt_CorrMat}] with and without random flavor hopping for $v=1$ and $v=0$, respectively, as shown in Fig.\ref{fig:diffusive_metal_finite_T}. We extract an effective central charge $c_{eff}$ by fitting the CFT formula in Eq.\eqref{eq:finite_T_cft} for second R\'{e}nyi entropy to the data in Fig.\ref{fig:diffusive_metal_finite_T} with parameters $c=c_{eff}$, $\tilde{v}$ [$\ell(T)=\tilde{v}\beta/\pi$] and $b$. The $S_{A}^{(2)}$ yields a much smaller value of $c_{eff}\approx 0.54$ for $v=1$ compared to $c_{eff}\approx0.92$ for $v=0$, without the on-site disorder. We similarly extract $c_{eff}$ for a range of values of $v$ and plot $c_{eff}$ as a function of $v$ in the inset of Fig.\ref{fig:diffusive_metal_finite_T}. The reduction of $c_{eff}$ can be rationalized by a modified subsystem size scaling \cite{potter,Haldar,Pouranvari2015AreaLaw},
\begin{align} \label{eq:area_law_random_hop}
S_{A}^{(2)}(L_A)&\simeq (c_{\mr{eff}}/4)\ln[(L_A^{-2}+\ell_{el}^{-2}(v))^{-1/2}]+\cdots,
\end{align}
at $T=0$, where the logarithmic growth of the clean system with a Fermi surface (points) is cut off by an elastic length scale $\ell_{el}(v)$ for $L_A>\ell_{el}(v)$. To validate this, we also show $S_A^{(2)}(L_A)$ computed at zero temperature ($T=0$) for $L=250$, $N=30$ and $v=1$ using the correlation matrix method in Fig.\ref{fig:area_law_T_0}. The fit of the modified entanglement growth law [Eq.\eqref{eq:area_law_random_hop}] leads to $\ell_{el}=15.0 \pm 0.3$ and $c_{\mr{eff}}=0.50±0.01$. 

We further compute the second R\'{e}nyi entropy of the 1D Yukawa--SYK model [Eq.\eqref{eq:Model}] in the presence of random on-site flavour hopping, employing the large-$N$ second R\'{e}nyi path-integral method developed in Sec.~\ref{sec:Ent_LargeN}. To this end, we add the random hopping term
in Eq.~\eqref{eq:RandomHopping_Hf} to $\mc{H}_f$ [Eq.\eqref{eq:H_f}]. This adds additional self-energies of Eq.\eqref{eq:diff_Sigma} and Eq.\eqref{eq:diff_Sigma_Ent} to Eq.\eqref{eq:SaddleSigma} and Eq.\eqref{eq:Sigma_Ent}, respectively, while the rest of formalism remaining the same. The calculation is
performed for the Yukawa coupling $g=1$, random hopping strength $v=1$ and bare bosonic mass $m_{b}=1$, with $t=1,~\mu=0$ and $K=1$ [Eq.\eqref{eq:Model}] (Fermi liquid regime), at system size $L=30$ and temperature $T=0.05$. The resulting $S^{(2)}_{A_{f}}$ is shown in Fig.~\ref{fig:large_N_vs_corr_random_hop} (magenta stars). A fit to the CFT form yields an effective central charge $c_{\mathrm{eff}}\simeq0.6$, which lies
below the corresponding clean, non-interacting value for the same system size. This suppression is consistent with the reduction of entanglement anticipated
once the random hopping establishes a finite elastic mean free path. Thus, the presence of such on-site disorder is expected to drive the 1D Yukawa--SYK model for all $\gamma$ into diffusive metals characterized by an elastic length scale $\ell_{{el}}$, associated with a finite elastic scattering rate. This gives rise to a nonzero residual resistivity at $T=0$. This elastic length scale cuts off the growth of the entanglement entropy even in the presence of a finite density of states at the Fermi surface [Fig.~\ref{fig:area_law_T_0}], in contrast to the clean limit, where the gapless Fermi surface produces the logarithmic violation of the boundary law. Hence, in the present work for the entanglement properties of the strange metal and Fermi liquid phase [Fig.\ref{fig:PhaseDiagram}], we have not considered the on-site disorder, in order to study the growth of the entanglement
entropy as a function of subsystem size.


\bibliography{references.bib} 



\clearpage
\newpage
\def\makeSM{1}
\ifdefined\makeSM

\appendix
\renewcommand{\appendixname}{}
\renewcommand{\thesection}{{S\arabic{section}}}
\renewcommand{\theequation}{\thesection.\arabic{equation}}
\renewcommand{\thefigure}{S\arabic{figure}}
\setcounter{page}{1}
\setcounter{figure}{0}
\setcounter{equation}{0}

\widetext

\centerline{\bf Supplemental Material}
\centerline{\bf for}
\begin{center}
\bf Entanglement entropy of fermions in a strange metal
\end{center}
\centerline{ \authorf, \authors, \authort, \authorfo, \authorfi}
\centerline{\affiliationf}
\centerline{\affiliations}
\centerline{\affiliationt}
\centerline{\affiliationfo}
\fi 

\def  \qinv{Q^{-1}}
\def  \q0{\frac{\w_k^2}{\G}+z}

\newcommand{\lin}{linspace}
\newcommand{\m}{\Delta_0}
\newcommand{\M}{\Delta}
\newcommand{\g}{Q}
\newcommand{\ginv}{\big(Q^{-1}\big)_{ab}}
\newcommand{\qr}{q_{\text{reg}}} 
\newcommand{\sr}{\Sigma_{\text{reg}}}
\newcommand{\qt}{\tilde{q}_{{EA}}}
\newcommand{\bfig}{\begin{figure}[H]\centering}
\newcommand{\efig}{\end{figure}}
\newcommand{\s}{\hspace{0.5cm }} 
\newcommand{\sh}{\hspace{0.25cm }} 

\section{Numerical solutions of the large-$N$ saddle point equations [Eqs.\eqref{eq:SaddlePoint}]}\label{app:numerical_saddle}
\subsection{Solution of imaginary-time large-$N$ saddle-point equations}
We solve the large-$N$ saddle-point Eqs.\eqref{eq:SaddlePoint} through numerical iterations. For each value of $\gamma$ in the range of interest, we scan over the temperature $T\neq 0$, from high to low. At each temperature, we start with an initial guess for the local Green's functions $G(\mathrm{i}\omega_n)$ and $D(\mathrm{i}\Omega_m)$ based on the converged $G,D$ from the preceding higher temperature. Using Fast Fourier Transform (FFT), $G(\tau)$ and $D(\tau)$ are obtained by evaluating the Matsubara summations,
\begin{subequations}
    \begin{align}\label{eq:FFT}
    G(\tau)=\dfrac{1}{\beta}\sum_{n} G(\mathrm{i}\omega_n) e^{-\mathrm{i}\omega_n \tau},\\
    D(\tau)=\dfrac{1}{\beta}\sum_{m} D(\mathrm{i}\Omega_m) e^{-\mathrm{i}\Omega_m \tau},
\end{align}
\end{subequations}
which are then used to compute the self-energies $\Sigma(\tau)$ and $\Pi(\tau)$ from Eq.\eqref{eq:SaddleSigma} and, Eq.\eqref{eq:Pi_tau}, respectively. We get the self-energies in Matsubara-frequency space by an another FFT,
\begin{subequations}
  \begin{align}\label{eq:inv_FFT}
    \Sigma(\mathrm{i}\omega_n)&=\int_{0}^{\beta}d\tau e^{\mathrm{i}\omega_n \tau} \Sigma(\tau)\\
    \Pi(\mathrm{i}\Omega_m)&=\int_{0}^{\beta}d\tau e^{\mathrm{i}\Omega_m \tau} \Pi(\tau)
\end{align}  
\end{subequations}
These yield a new $G(\mathrm{i}\omega_n)$ and $D(\mathrm{i}\Omega_m)$, and the process is
iterated till a desired convergence is achieved. For a given $\gamma$ and $T$, we determine $M^2(T)$ using the bisection root finding method and at each bisection step we run the entire iterative process for $G$ and $D$, as described above, until a converged solution for $M^2(T)$ satisfies the constraint Eq.\eqref{eq:SaddleConstraint}. Finally, the bare boson mass is obtained as $m_b^2(T)=M^2(T)+\Pi(\mathrm{i}\Omega_m{=}0,T)$.

\subsection{Solution of the real-frequency large-$N$ saddle-point equations}

We use $M^2(T)$ (and hence $m_b^2(T)$) determined self-consistently from the above solution of the imaginary-time saddle-point Eqs.\eqref{eq:SaddlePoint} for each $\gamma$ and $T$, to solve the real-frequency saddle-point equations [Eqs.~\eqref{eq:G_iwn}, \eqref{eq:D_iwn}, \eqref{eq:SaddleSigma}, \eqref{eq:Pi_tau}] for the retarded functions $G_R(\omega)$ and $D_R(\Omega)$, obtained through analytical continuations, $\ci \omega_n\to\omega+\ci 0^+$ and $\ci \Omega_m\to \Omega+\ci 0^+$. To this end, the retarded self-energies are given by 
\begin{subequations}\label{eq:retarded_self}
\begin{align} 
        \Sigma^R(\omega) & =\mathrm{i}g^2 \int^{\infty}_{0} dt \big[n_1(t)n_2(t)-n_3(t)n_4(t)\big] e^{\mathrm{i}\omega t}\\
        \Pi^R(\Omega) & =-\mathrm{i}g^2 \int^{\infty}_{0} dt \big[n_1(t)n_5(t) -n_6(t)n_3(t)\big]e^{\mathrm{i}\Omega t}
\end{align} 
\end{subequations}
where
\begin{subequations} \label{eq:convo_int}
   \begin{align}
    n_1(t) &=\int^{\infty}_{-\infty} d\omega \rho_f(\omega) n_{F}(\omega) e^{-\mathrm{i}\omega t} \label{eq:n1},\\
    n_2(t) &=\int^{\infty}_{-\infty} d\omega \rho_b(\omega) n_{B}(\omega) e^{-\mathrm{i}\omega t} \label{eq:n2},\\
    n_3(t) &=\int^{\infty}_{-\infty} d\omega \rho_f(-\omega) n_{F}(\omega) e^{\mathrm{i}\omega t} \label{eq:n3},\\
    n_4(t) &=\int^{\infty}_{-\infty} d\omega \rho_b(-\omega) n_{B}(\omega) e^{\mathrm{i}\omega t} \label{eq:n4},\\
    n_5(t) &=\int^{\infty}_{-\infty} d\omega \rho_f(-\omega) n_{F}(\omega) e^{-\mathrm{i}\omega t}\label{eq:n5},\\
    n_6(t) &=\int^{\infty}_{-\infty} d\omega \rho_f(\omega) n_{F}(\omega) e^{\mathrm{i}\omega t}\label{eq:n6}.
\end{align} 
\end{subequations}
The $n_F(\omega)$ and $n_B(\omega)$ are the Fermi and Bose distribution functions, respectively, at temperature $T$.
The equations for the local retarded Green's functions [Eqs.~\eqref{eq:G_iwn} and \eqref{eq:D_iwn}] are written as,
\begin{subequations}\label{eq:retarded_G}
\begin{align}
G_{R}(\omega) &= \dfrac{1}{\big[(\omega+\mathrm{i}\eta+\mu-\Sigma^{R}(\omega))^2-4t^2\big]^{1/2}},\\
D_{R}(\Omega) &= \dfrac{1}{\big[(-(\Omega+\mathrm{i}\eta)^2+M^2+2K-\tilde{\Pi}^{R}(\omega))^2-4K^2\big]^{1/2}},
\end{align}
\end{subequations}
after performing the momentum summations through integrations for the 1D dispersions $\epsilon_k=-2t\cos k$ and $\omega^2_q=2K(1-\cos q)$ of the fermions and bosons.
The local single-particle spectral functions or the density of states (DOS) are obtained from the retarded Green's functions as
\begin{subequations}\label{eq:local_spectral}
\begin{align}
     \rho_{f}(\omega)&=-\dfrac{1}{\pi} \mathrm{Im} G_{R}(\omega)\\
     \rho_{b}(\Omega)&=-\dfrac{1}{\pi} \mathrm{Im} D_{R}(\Omega)
\end{align}
\end{subequations}
Using Eqs.~\eqref{eq:retarded_self}, \eqref{eq:convo_int}, and \eqref{eq:retarded_G}, we iteratively solve for the local Green's functions using the scheme similar to the preceding section.

\section{Numerical solution of the large-$N$ second R\'{e}nyi saddle-point equations [Eqs.\eqref{eq:self_consis_entan_replica}]} \label{app:renyi_numerical_saddle}

The large-$N$ saddle point Eqs.\eqref{eq:self_consis_entan_replica} for computing $S^{(2)}_{A_f}$ [Eq.\eqref{eq:second_renyi}] are numerically much more challenging compared to the solution of the equilibrium saddle-point Eqs.\eqref{eq:SaddlePoint}, since the Green's functions $G_{r\alpha,r'\beta}(\tau,\tau')$ and $D_{r\alpha,r'\beta}(\tau,\tau')$ are matrix in space, time and entanglement replicas $\alpha,\beta=1,2$. (Here the replica index $\beta$ should not be confused with the inverse tempearture $\beta$). In the presence of the self-energy kick term $\propto\delta(\tau-\tau_{0}^{+})\delta(\tau'-\tau_{0})\delta_{r\in A}\delta_{rr'}$ which breaks space and time-translation invariance in Eq.\eqref{eq:fermion_greenfn_kick}, we cannot solve the second R\'{e}nyi saddle-point equations [Eqs.\ref{eq:self_consis_entan_replica}] using Matsubara frequency and the momentum space representation. We solve the saddle-point equations on a finite space-time lattice by discretizing in imaginary time. The main challenging part of the iterative solution of the saddle-point equations is the matrix inversions of $G^{-1}$ and $D^{-1}$ in
Eqs.\eqref{eq:InvG_Ent}) and \eqref{eq:InvD_Ent}, respectively. Since the iteration only requires the local or the on-site Green's function, we use an efficient recursive Green's function inversion \cite{bera_PRB}, which only calculates the on-site Green's function for each iteration step, instead of inverting the entire matrices $G^{-1}$ and $D^{-1}$. The
on-site Green's functions provide the self-energies in Eq.\eqref{eq:Sigma_Ent} and Eq.\eqref{eq:Pi_Ent}.
These self-energies are fed back into Eqs.\eqref{eq:InvG_Ent}, and \eqref{eq:InvD_Ent} for the next iteration till convergence.

\subsection{Imaginary-time discretization}\label{app:discretization}
To solve the saddle-point Eqs.\eqref{eq:self_consis_entan_replica} in imaginary time $\tau$ without the invariance of the time-translation, we discretize the saddle-point equations in imaginary time. We divide the time interval $[0,\beta)$ into $N_{\tau}$ segments such that
$\beta=N_{\tau}\delta\tau$. However, while discretizing the equations, we have to ensure the appropriate  antiperiodic boundary condition on the fermionic Green's functions and periodic boundary condition on the bosonic Green's functions,
\begin{subequations}
\begin{align}
G(\tau+\beta,\tau')&=-G(\tau,\tau'),\\
D(\tau+\beta,\tau')&=D(\tau,\tau'),
\end{align}
\end{subequations}
which are equivalent to the antiperiodic boundary conditions on Grassmann variables $c_{N_{\tau}}=-c_0$, $\bar{c}_{N_{\tau}}=-\bar{c}_0$, and periodic boundary conditions on the bosonic scalar fields $\phi_{N_{\tau}}=\phi_{0}$, in the time-discretized form. Here we have suppressed the space, and
replica indices for brevity. We use the indices $(n,m)$ running
from $n,m=0~ \text{to}~ N_{\tau}-1$ for $(\tau,\tau')$. We also write the following useful discretization rules
\begin{subequations}
 \begin{align}
\bar{c}(\tau)\partial_{\tau}c(\tau) &= \frac{\bar{c}_{n+1}(c_{n+1}-c_{n})}{\delta\tau}\\
\bar{c}(\tau)c(\tau) &= \bar{c}_{n+1}c_{n}\\
\left(\partial_{\tau}\phi(\tau)\right)^{2}&=\frac{\left(\phi_{n+1}-\phi_{n}\right)^{2}}{\delta\tau^{2}}\\
\phi(\tau)\phi(\tau) &= \phi_{n}\phi_{n}
\end{align}   
\end{subequations}
The above rules arise since the creation operator $c^\dagger$ always
appears slightly later in time than the annihilation operator
$c$ in the path integral. Using the above rules, we write the action in Eq.(\ref{eq:final_action_entan_replica}) a
\begin{align}
\frac{\mathcal{S}_{A_{f}}}{N}= & \,\delta\tau^{2}\,\bar{c}_{r\alpha m}\left[-\left(\tilde{G}_{0,r\alpha,r'\beta,mn}^{-1}-\Sigma_{r,\alpha\beta,mn}\delta_{rr'}\right)\right]c_{r'\beta n}+\delta\tau^{2}\,\frac{1}{2}\,\phi_{r\alpha m}\left[D_{0,r\alpha,r'\beta,mn}^{-1}-\Pi_{r,\alpha\beta,mn}\delta_{rr'}\right]\phi_{r'\beta n}\nonumber\\
 & +\delta\tau^{2}\Big[-\Sigma_{r,\beta\alpha,nm}\,G_{r,\alpha\beta,mn}+\frac{1}{2}\Pi_{r,\beta\alpha,nm}\,D_{r,\alpha\beta,mn} +\frac{g^{2}}{2}\,G_{r,\alpha\beta,mn}\,G_{r,\beta\alpha,nm}\,D_{r,\alpha\beta,mn}\Big].
\end{align}
In the above action, repeated indices are understood to be summed over according to the Einstein summation convention, with $r=1\cdots L$ and $m,n=0 \cdots N_{\tau}-1$ and $\alpha,\beta=1,2$.
The inverse of the fermionic Green's function appearing above is given by 
\begin{align}
-\Tilde{G}_{0,r\alpha,r'\beta,mn}^{-1}= & g^{-1}_{\alpha m,\beta n}\delta_{rr'}+\dfrac{1}{\delta\tau}t_{rr'}\zeta_{mn}\delta_{\alpha\beta}+M_{\alpha\beta}\delta_{r\in A}\delta_{rr'}\frac{\delta_{m,p+1}\delta_{n,p}}{\delta\tau^{2}}
\end{align}
where the index $p\in[0,N_{\tau}]$ is arbitrary depending on $\tau_0$. Here
\begin{subequations}
    \begin{align}
    g^{-1}_{\alpha m,\beta n}=\dfrac{1}{\delta\tau^2}(\zeta_{mn}-\delta_{mn})\delta_{\alpha\beta}-\mu\dfrac{1}{\delta\tau}\zeta_{mn}\delta_{\alpha\beta}
\end{align}
\text{with}
\begin{align}
\zeta_{mn}= & \delta_{m,n+1}\hspace{1em}\hspace{1em}n<N_{\tau}-1 \nonumber\\
= & -1\hspace{1em}\hspace{1em}m=0,n=N_{\tau}-1
\end{align}
\end{subequations}
Similarly, the bosonic Green's function is given by
\begin{subequations}
\begin{align}
    D^{-1}_{0,r\alpha,r'\beta,mn}&=D^{-1}_{0,rr',mn}\delta_{\alpha\beta}\\
    D^{-1}_{0,rr',mn}&=\dfrac{2\delta_{mn} - \delta_{m,n+1} - \delta_{m,n-1}}{\delta \tau^{3}}\delta_{rr'}+ \dfrac{m^{2}_{b}\delta_{rr'} + K_{rr'}}{ \delta \tau }\delta_{mn}, \hspace{1em} N_{\tau}\equiv0
\end{align}
\end{subequations}
The saddle point equations become 
\begin{subequations}\label{eq:discretized_saddle_point_eqs}
    \begin{align}
G_{r\alpha,r'\beta,mn}^{-1}= & G_{0,r\alpha,r'\beta,mn}^{-1}-\Sigma_{r,\alpha\beta,mn}\delta_{rr'}\label{eq:invG-1}\\
D_{r\alpha,r'\beta,mn}^{-1}= & D_{0,r\alpha,r'\beta,mn}^{-1}-\Pi_{r,\alpha\beta,mn}\delta_{rr'}\label{eq:invD-1}\\
\Sigma_{r,\alpha\beta,mn}= & g^{2}G_{r,\alpha\beta,mn}D_{r,\alpha\beta,nm}\label{eq:fermionSelfEnergy-1}\\
\Pi_{r,\alpha\beta,mn}= & -g^{2}G_{r,\alpha\beta,mn}G_{r,\beta\alpha,nm}\label{eq:bosonSelfEnergy-1}
\end{align}
\end{subequations}

The on-site Green's functions
are obtained by the following matrix inversion, using a recursive
method (discussed below) 
\begin{subequations}\label{eq:inversion}
   \begin{align}
\sum_{r_{1}\gamma n_{1}}G_{r\alpha,r_{1}\gamma,mn_{1}}^{-1}G_{r_{1}\gamma,r'\beta,n_{1}n}= & \delta_{rr'}\delta_{\alpha\beta}\frac{\delta_{mn}}{\delta\tau^{2}}\label{eq:inversionG}\\
\sum_{r_{1}\gamma n_{1}}D_{r\alpha,r_{1}\gamma,mn_{1}}^{-1}D_{r_{1}\gamma,r'\beta,n_{1}n}= & \delta_{rr'}\delta_{\alpha\beta}\frac{\delta_{mn}}{\delta\tau^{2}}\label{eq:inversionD}
\end{align} 
\end{subequations}
\subsection{Recursive Green's function method}\label{app:recursive}
The most computationally demanding part of solving the entanglement large-$N$ equations is the inversion of $G^{-1}$ and $D^{-1}$, matrices of dimension $\sim L N_\tau \times L N_\tau$, to obtain $G$ and $D$ via Eq.(\ref{eq:inversion}) for a system size $L$. A direct inversion with the large-$N$ self-consistency loop [Eq.(\ref{eq:discretized_saddle_point_eqs}] is only practically feasible for relatively small systems of size $L\lesssim10$ with $N_{\tau}\lesssim1000$. For larger system sizes, we use an efficient recursive Green's
function method \cite{bera_PRB} along the lattice direction. The recursive method can
be implemented for the open boundary condition (OBC) as well as for
the periodic boundary condition (PBC) \cite{bera_PRB}. To describe the method, we will use the symbol $G$ for both $G$ and $D$ in Eqs.\ref{eq:inversionG}, and \ref{eq:inversionD}.
We rewrite Eq.(\ref{eq:inversion}) as a
\begin{subequations}\label{eq:identity}
\begin{align} 
\boldsymbol{G}^{-1}\boldsymbol{G}= & \boldsymbol{I}\label{eq:identity_G}\\
\boldsymbol{I}_{r\alpha m,r'\beta n}= & \delta_{rr'}\delta_{\alpha\beta}\delta_{mn}
\end{align}
\end{subequations}
where,
\begin{subequations}
\begin{align}
    \boldsymbol{G}_{r\alpha m,r'\beta n}= & \delta\tau G_{r\alpha m,r'\beta n}\\
\boldsymbol{G}_{r\alpha m,r'\beta n}^{-1}= & \delta\tau G_{r\alpha m,r'\beta n}^{-1}
\end{align}
\end{subequations}
We separate the system spatially into two parts $S$ and $R$, and
write 
\begin{align}
\begin{pmatrix}(\boldsymbol{G}_{R})^{-1} & -\boldsymbol{T}_{RS}\\
-\boldsymbol{T}{}_{RS} & (\boldsymbol{G}_{S})^{-1}
\end{pmatrix}\begin{pmatrix}\boldsymbol{G}_{R}^{R+S)} & \boldsymbol{G}_{RS}^{(R+S)}\\
\boldsymbol{G}_{SR}^{(R+S)} & \boldsymbol{G}_{S}^{(R+S)}
\end{pmatrix} & =\boldsymbol{I}\label{eq:RecursiveInverse}
\end{align}

Here $\boldsymbol{G}_{R}$ and $\boldsymbol{G}_{S}$ are Green's functions
of the system and the rest in the absence of any coupling between
them. $\boldsymbol{T}$ connects the systems and the rest, and $\boldsymbol{G}^{(R+S)}$
is the full Green's function of the combined system. From the above,
we get 
\begin{subequations}
 \begin{align}
\boldsymbol{G}_{R}^{(R+S)}= & \boldsymbol{G}_{R}+\boldsymbol{G}_{R}\boldsymbol{T}_{RS}\boldsymbol{G}_{SR}^{(R+S)}\\
\boldsymbol{G}_{RS}^{(R+S)}= & \boldsymbol{G}_{R}\boldsymbol{T}_{RS}\boldsymbol{G}_{s}^{(R+S)}\\
\boldsymbol{G}_{SR}^{(R+S)}= & \boldsymbol{G}_{S}\boldsymbol{T}_{RS}^{\dagger}\boldsymbol{G}_{R}^{(R+S)}\\
\boldsymbol{G}_{S}^{(R+S)}= & \boldsymbol{G}_{S}+\boldsymbol{G}_{S}\boldsymbol{T}_{RS}^{\dagger}\boldsymbol{G}_{RS}^{(R+S)}
\end{align}   
\end{subequations}
\subsubsection{Recursive solution of Eq.~(\ref{eq:identity_G})}
We imagine successively building the system from the left, starting
from the first layer at $r=l=1$ and then adding successive layers
till $l=L$. Imagine that at the $l$-th step of recursion, we have
only left $l+1$ layers, and we separate the system into left $l$
layers (``$L$''), i.e., $R$ of the preceding section, and add
one layer (system ``$S$'') more. We denote the Green's function
of the left $l$ layers as $\boldsymbol{G}^{(l)}$ and that of $l+1$
layers as $\boldsymbol{G}^{(l+1)}$. From Eq.~(\ref{eq:RecursiveInverse})
\begin{align}
\begin{pmatrix}(\boldsymbol{G}_{L}^{(l)})^{-1} & -\boldsymbol{T}_{LS}\\
-\boldsymbol{T}_{LS}^{\dagger} & (\boldsymbol{G}_{S})^{-1}
\end{pmatrix}\begin{pmatrix}\boldsymbol{G}_{L}^{(l+1)} & \boldsymbol{G}_{LS}^{(l+1)}\\
\boldsymbol{G}_{SL}^{(l+1)} & \boldsymbol{G}_{S}^{(l+1)}
\end{pmatrix} & =\boldsymbol{I}\label{recursiveinv_1}
\end{align}
where the coupling between $L$ and $S$ for fermionic Green's function
Eq.~(\ref{eq:inversionG}) is given by 
\begin{align*}
\boldsymbol{T}_{rr'}= & \boldsymbol{\zeta}t_{rr'}\hspace{1em}\hspace{1em}r\leq l,r'=l+1\\
= & 0\hspace{1em}\hspace{1em}\mathrm{otherwise}\\
\boldsymbol{\zeta}_{\alpha m,\beta n}= & \zeta_{mn}\delta_{\alpha\beta}
\end{align*}
and for bosonic Green's function in Eq.~(\ref{eq:inversionD})
\begin{align*}
\boldsymbol{T}_{rr'}= & \boldsymbol{\delta}K_{rr'}\hspace{1em}\hspace{1em}r\leq l,r'=l+1\\
= & 0\hspace{1em}\hspace{1em}\mathrm{otherwise}\\
\boldsymbol{\delta}= & \delta_{mn}\delta_{\alpha\beta}
\end{align*}

Here $r\leq l$ indicates that the site $r$ belongs to a layer from
$1$ to $l$. The inverse Green's functions of $L$ and $S$ in the
absence of any coupling between them are 
\begin{align*}
\left(\boldsymbol{G}_{L}^{(l)}\right)_{r\alpha m,r'\beta n}^{-1}= & \boldsymbol{G}_{r\alpha m,r'\beta n}^{-1}\hspace{1em}\hspace{1em}r,r'\leq l\\
\left(\boldsymbol{G}_{S}\right)_{r\alpha m,r'\beta n}^{-1}= & \boldsymbol{G}_{r\alpha m,r'\beta n}^{-1}\hspace{1em}\hspace{1em}r,r'=l+1
\end{align*}
Here $r\leq l$ indicates that the site $r$ belongs to a layer from
$1$ to $l$. The full Green's function we want to compute is $\boldsymbol{G}=\boldsymbol{G}^{(L)} $. We can write Eqs.~(\ref{recursiveinv_1}) as 
\begin{align}
\boldsymbol{G}_{L}^{(l+1)}= & \boldsymbol{G}_{L}^{(l)}+\boldsymbol{G}_{L}^{(l)}\boldsymbol{T}_{LS}\boldsymbol{G}_{SL}^{(l+1)}\label{eq:recursion_1}\\
\boldsymbol{G}_{LS}^{(l+1)}= & \boldsymbol{G}_{L}^{(l)}\boldsymbol{T}_{LS}\boldsymbol{G}_{s}^{(l+1)}\label{eq:recursion_2}\\
\boldsymbol{G}_{SL}^{(l+1)}= & \boldsymbol{G}_{S}\boldsymbol{T}_{SL}^{\dagger}\boldsymbol{G}_{L}^{(l+1)}\label{eq:recursion_3}\\
\boldsymbol{G}_{S}^{(l+1)}= & \boldsymbol{G}_{S}+\boldsymbol{G}_{S}\boldsymbol{T}_{LS}^{\dagger}\boldsymbol{G}_{LS}^{(l+1)}\label{eq:recursion_4}
\end{align}
We now rewrite Eq.~(\ref{eq:recursion_1}) keeping only the index $r$
for the layers, where all other indices $\alpha$ and $m$, and contracted
for matrix multiplications. 
\begin{align}
\boldsymbol{G}_{r,r'}^{(l+1)} & =\boldsymbol{G}_{r,r'}^{(l)}+\sum_{r_{1}\leq l}\boldsymbol{G}_{r,r_{1}}^{(l)}\boldsymbol{T}_{r_{1},l+1}\boldsymbol{G}_{l+1,r}^{(l+1)}\hspace{0.4cm}r,r'\leq l\label{eq:Gxx'}
\end{align}
From Eq.~(\ref{eq:recursion_2}), we get 
\begin{align}
\boldsymbol{G}_{r,l+1}^{(l+1)} & =\sum_{r_{1}\leq l}\boldsymbol{G}_{r,r_{1}}^{(l)}\boldsymbol{T}_{r_{1},l+1}\boldsymbol{G}_{l+1,l+1}^{(l+1)}\hspace{0.4cm}r\leq l\label{eq:G_xlp1}
\end{align}
Since $G_{r\alpha m,r'n}^{*}=G_{r'\beta n,r\alpha m}$, we can obtain
from the above 
\begin{align}
\boldsymbol{G}_{l+1,r}^{(l+1)} & =\sum_{r_{1}\leq l}\boldsymbol{G}_{l+1,l+1}^{(l+1)}\boldsymbol{T}_{l+1,r_{1}}\boldsymbol{G}_{r_{1},r}^{(l)}\hspace{0.4cm}r\leq l\label{eq:G_lp1x}
\end{align}
Using the above in Eq.~(\ref{eq:Gxx'}), we obtain for $r,r'\leq l$
\begin{align}
\boldsymbol{G}_{r,r'}^{(l+1)} & =\boldsymbol{G}_{r,r'}^{(l)}+\sum_{r_{1},r_{2}\leq l}\boldsymbol{G}_{r,r_{1}}^{(l)}\boldsymbol{T}_{r_{1},l+1}\boldsymbol{G}_{l+1,l+1}^{(l+1)}\boldsymbol{T}_{l+1,r_{2}}\boldsymbol{G}_{r_{2},r'}^{(l)}\label{eq:Gxx'_final}
\end{align}
In the above equation, the only unknown quantity is $\boldsymbol{G}_{l+1,l+1}^{(l+1)}$.
This can be obtained as follows. From Eqs.~(\ref{eq:recursion_2}) and (\ref{eq:recursion_4}),
we get 
\begin{align}
\boldsymbol{G}_{S}^{(l+1)}= & \boldsymbol{G}_{S}+\boldsymbol{G}_{S}\boldsymbol{T}_{LS}^{\dagger}\boldsymbol{G}_{L}^{(l)}\boldsymbol{T}_{LS}\boldsymbol{G}_{S}^{(l+1)}
\end{align}
The above can be written in the form of a Dyson equation, 
\begin{align}
\big(\boldsymbol{G}_{S}^{l+1}\big)^{-1} & =(\boldsymbol{G}_{l+1,l+1}^{(l+1)}\big)^{-1}=\boldsymbol{G}_{S}^{-1}-\boldsymbol{\Sigma}^{(l)}\label{eq:Dyson_eq1}
\end{align}
with the self-energy 
\begin{align}
\boldsymbol{\Sigma}^{(l)} & =\sum_{r_{1},r_{2}\leq l}\boldsymbol{T}_{l+1,x_{1}}\boldsymbol{G}_{r_{1},r_{2}}^{(l)}\boldsymbol{T}_{r_{2},l+1}\label{eq:selfE_dyson}
\end{align}
Hence $\boldsymbol{G}_{l+1,l+1}^{(l+1)}$ can be obtained using Eq.~(\ref{eq:Dyson_eq1}).
Thus from Eqs.~(\ref{eq:G_xlp1})--(\ref{eq:Gxx'_final}), (\ref{eq:Dyson_eq1}) and (\ref{eq:selfE_dyson}),
we can construct the complete Green's function matrix $\boldsymbol{G}_{r\alpha m,r'\beta n}^{(l+1)}$
($r\leq l+1$) of the system of $l+1$ layers from that of $l$ layers.
The process can be applied recursively, starting with $l=1$ and continuing
till $l=L-1$, which will yield us the Green's function of system
size $L$, i.e., $\boldsymbol{G}^{(L)}$. However, for the large-$N$ self-consistency [Eq.\ref{eq:discretized_saddle_point_eqs}] one does not need the full Green's function at each iteration, but only
the on-site Green's functions which provide the local self-energy.

For periodic boundary condition (PBC), we need to incorporate the
hopping matrix element between site $1$ and site $L$. We can implement
this in the recursive procedure by changing the hopping coupling matrix
in the last iteration accordingly when we add the $l=L-1$ layer with
a single layer system $S$ to form the required system size $l+1=L$.
In particular, we can explicitly write the hopping coupling matrix
for the nearest neighbor for PBC below. For fermions 
\begin{align}
\boldsymbol{T}_{r,r'} & =\boldsymbol{\zeta}t_{rr'}\hspace{0.5cm}r=l;r'=l+1\hspace{0.25cm}\text{if}\hspace{0.25cm}l\leq L-2\nonumber \\
 & =\boldsymbol{\zeta}t_{rr'}\hspace{0.5cm}r=1,l;r'=l+1\hspace{0.25cm}\text{if}\hspace{0.25cm}l=L-1\nonumber \\
 & =0\hspace{0.5cm}\text{otherwise},
\end{align}

For bosons 
\begin{align}
\boldsymbol{T}_{r,r'} & =\boldsymbol{\delta}K_{rr'}\hspace{0.5cm}r=l;r'=l+1\hspace{0.25cm}\text{if}\hspace{0.25cm}l\leq L-2\nonumber \\
 & =\boldsymbol{\delta}t_{rr'}\hspace{0.5cm}r=1,l;r'=l+1\hspace{0.25cm}\text{if}\hspace{0.25cm}l=L-1\nonumber \\
 & =0\hspace{0.5cm}\text{otherwise},
\end{align}







\end{document}